\documentclass{article}
\newif\ifclean
\cleantrue  % hide extra markup 
\cleanfalse % show extra markup
\usepackage{arxiv}
\usepackage[square,sort,comma,numbers]{natbib}
\usepackage{graphicx}	
\usepackage{amsmath,amssymb,amsthm}
\usepackage{bm}
\usepackage[usenames,dvipsnames]{xcolor}
\usepackage[normalem]{ulem}
\newcommand{\COMMENT}[1]{\textcolor{cyan}{{[ \sc{#1} ]}}} % comments

\newcommand{\QUESTION}[1]{\textcolor{ForestGreen}{{#1}}}
\newcommand{\red}[1]{\textcolor{red}{{#1}}}

\newcommand{\tr}{\mathrm{tr}}

\newlength{\figwidth}
\newlength{\figwidthtwo}
\newlength{\figwidththree}
\newcommand{\aref}[1]{App.\,\ref{#1}}

\newcommand{\cref}[1]{Ref.\,\cite{#1}}
\newcommand{\crefs}[1]{Refs.\,\cite{#1}}

\newcommand{\Ac}{\mathcal{A}}
\newcommand{\Bc}{\mathcal{B}}
\newcommand{\Ic}{\mathcal{I}}

\newcommand{\varepsilonb}{\boldsymbol{\varepsilon}}

\newcommand{\eb}{\mathbf{e}}

\newcommand{\nb}{\mathbf{n}}

\newcommand{\pb}{\mathbf{p}}

\newcommand{\Ab}{\mathbf{A}}

\newcommand{\Cb}{\mathbf{C}}

\newcommand{\Gb}{\mathbf{G}}

\newcommand{\Fb}{\mathbf{F}}

\newcommand{\Nb}{\mathbf{N}}
\newcommand{\Pb}{\mathbf{P}}
\newcommand{\Sb}{\mathbf{S}}
\newcommand{\Ib}{\mathbf{I}}
\newcommand{\Rb}{\mathbf{R}}
\newcommand{\Qb}{\mathbf{Q}}

\usepackage{booktabs}
\usepackage{tabularx}
\usepackage[letterpaper]{geometry}
\usepackage{subcaption}
\graphicspath{{./Figures/}}

\usepackage{comment}
\newcommand{\caution}{\red{\bf Draft: \today. Do not distribute.}}
\ifclean
\renewcommand{\COMMENT}[1]{{}}
\renewcommand{\QUESTION}[1]{{}}
\else
\fi
\DeclareMathOperator*{\argmin}{arg\,min}
\usepackage[utf8]{inputenc}
\usepackage[english]{babel}
\usepackage{natbib}
\title{\bf Inverse design of anisotropic microstructures using physics-augmented neural networks}
\author{
Asghar A. Jadoon$^a$, Karl A. Kalina$^b$, Manuel K. Rausch$^a$, Reese Jones$^c$,
Jan N. Fuhg$^a$\footnote{corresponding: jan.fuhg@utexas.edu}  \\[0.1in]
$^a${\it The University of Texas at Austin,  Austin TX, USA}\\[0.05in]
$^b${\it
TU Dresden, Dresden, Germany}\\[0.05in]
$^c${\it Sandia National Laboratories,  Livermore CA, USA}
}
\date{\today}

\begin{document}
\maketitle

\begin{abstract}
Composite materials often exhibit mechanical anisotropy owing to the material properties or geometrical configurations of the microstructure. This makes their inverse design a two-fold problem. First, we must learn the type and orientation of anisotropy and then find the optimal design parameters to achieve the desired mechanical response. In our work, we solve this challenge by first training a forward surrogate model based on the macroscopic stress-strain data obtained via computational homogenization for a given multiscale material. To this end, we use partially Input Convex Neural Networks (pICNNs) to obtain a representation of the strain energy in terms of the invariants of the Cauchy-Green deformation tensor which is polyconvex with respect to the deformation gradient whereas it can have an arbitrary form with respect to the design parameters. The network architecture and the strain energy function are further modified to incorporate, by construction, physics and mechanistic assumptions into the framework. While training the neural network, we find the type of anisotropy, if any, along with the preferred directions. Once the model is trained, we solve the inverse problem using an evolution strategy to obtain the design parameters that give a desired mechanical response. We test the framework against synthetic macroscale and also homogenized data. For cases where polyconvexity might be violated during the homogenization process, we present viable alternate formulations. The trained model is also integrated into a finite element framework to invert design parameters that result in a desired macroscopic response. We show that the invariant-based model is able to solve the inverse problem for a stress-strain dataset with a different preferred direction than the one it was trained on and is able to not only learn the polyconvex potentials of hyperelastic materials but also recover the correct parameters for the inverse design problem.

\end{abstract}

\section{Introduction}
Most natural and engineered composite materials \cite{nguyen2007modeling,chen1970mechanical,mortazavian2015effects,yang1998anisotropic,pinsky2005computational,datta2019anisotropy} exhibit anisotropy, i.e., the material exhibits a direction-dependent response \cite{courtney2005mechanical,saeb2016aspects} which usually stems from the geometric and material properties of the microstructure. Due to the advances in additive manufacturing (AM), it is now possible to create complex, multi-material geometries with specific microstructures \cite{Schumacher2015MicrostructuresTC}. Therefore, careful design of materials and microstructures can lead to desired macro-mechanical responses different from the source material \cite{surjadi2019mechanical}. 
Computational modeling could help in improving this design process by avoiding costly tests and experiments. However, this approach faces two challenges: (i) Even if the individual materials may be isotropic, the overall response of the continuum can be anisotropic based on the microscopic geometry or the spatial distribution of these materials \cite{kok2018anisotropy}; (ii) The material response can exhibit different anisotropic characteristics \cite{newnham2005properties}, the type and orientation of which are generally not easily deducible before material testing, even from microscopic imaging of the material \cite{guild1993microstructural,trimby1999microstructural}. In this work, we present a computational approach for the design of (hyperelastic) microstructures under varying anisotropies with targeted stress responses.

 Herein, we consider only two length scales -- the micro- and the macro-scale \cite{horstemeyer2010multiscale, weinan2011principles}. When the length scales are well-separated and there exists no coupling between the scales, we can use homogenization methods \cite{schroder2014numerical}. The idea of homogenization is to replace the heterogeneous structure with an equivalent homogeneous one, with macroscopic properties that yield the same response as the multiscale material. The first works on homogenization of mechanical response found bounds \cite{Voigt1889,Reuss1929,taylor1938plastic,Sachs1929,hashin1963variational} on effective moduli which were later extended to anisotropic composites \cite{milton1988variational}. However, these bounds do not take the morphology of the microstructure into account and are generally quite loose \cite{babuvska1976homogenization}. Assuming that there are no interactions between the inhomogeneities, Eshelby presented an equivalent inclusion method \cite{eshelby1957determination} where the inclusions are replaced by a single equivalent inclusion. However, the assumption of no interaction is unrealistic for materials with randomly dispersed inclusions and does not even hold for materials with a few percent volume fraction of inclusions \cite{Zohdi2001}. Nevertheless, Eshelby's work paved the way for more rigorous analytical homogenization schemes (see e.g. \cite{MORI1973571, LUO1987347, budiansky1965elastic, WILLIS1977185}) that took into account the interaction between different material phases, which were also then extended to nonlinear composites and finite deformations \cite{hill1972constitutive, OGDEN1974541}. In the last couple of decades, researchers have shifted their focus to computational homogenization instead \cite{GEERS20102175, Geers2017}. A main component of computational homogenization is the Representative Volume Element (RVE) which, as the name suggests, is the smallest volume element with a detailed model of the microstructure whose response is representative of the entire volume. 
To this end, concurrent nonlinear simulations can be used to resolve the strong coupling between a macroscopic problem and a microscopic RVE \cite{ghosh1996two,geers2016multiscale}. For two-scale problems, a popular approach is the FE$^{2}$ method \cite{feyel1999multiscale,raju2021review}, also called the coupled multiscale scheme, which transfers information on the deformation gradient and the effective stress and material tangent back-and-forth between quadrature points at the microscale and the RVE. While this approach is accurate and versatile it suffers from high computational costs \cite{fuhg2022local}, especially in the context of inverse microstructure design. Surrogate modeling techniques have therefore been introduced that aim to learn the effective mechanical response of the RVE \cite{fuhg2024review} and are classified as decoupled multiscale schemes. Our work falls into this category.

In this current work, we restrict ourselves to hyperelastic materials. We assume that a free energy potential can be employed to derive the mechanical response of a material and that anisotropy can be introduced through a modification of the free energy potential.
This is usually done through structural tensors that can impart directional information into the free energy potential \cite{boehler1987representations,itskov2004class}. We furthermore assume that the potential is parameterized by a set of material and geometric design parameters that describe the RVE.
The universal approximation capabilities of neural networks (NNs) \cite{hornik1989multilayer} make them a tempting choice as mapping functions to replace the free energy potential, even though other regression techniques have also been employed \cite{frankel2020tensor,fuhg2022physics}. Through the derivatives of these neural networks, we can find the internal stresses and the tangent moduli. However, the free energy potential cannot be arbitrary and must not violate established mechanical principles \cite{linden2023neural}. To this end, modified neural network architectures have been used that allow formulating potentials that respect these principles \cite{klein2022polyconvex, fuhg2022learning, linka2023new}. Further modifications to the potential itself are also usually made to ensure the existence of a stress-free state and to introduce coercivity. One of the most important constraints is the polyconvexity of the free energy function. Polyconvextiy is sufficient for quasiconvexity, which in turn guarantees rank-one convexity and thus finally guarantees that the Euler equations of the associated functional are elliptic. This can be ensured by enforcing that the potential is convex in the minors of the deformation gradient. However, the polyconvexity requirements only exist with respect to the deformation gradient while there are no general constraints with regard to RVE design parameters \cite{Klein_Roth_Valizadeh_Weeger_2023}. 

We propose to rely on parameterized physics-augmented NNs to capture all these requirements \cite{Klein_Roth_Valizadeh_Weeger_2023}. We then generate a response dataset of parameterized RVEs and leverage these flexible NNs to derive design-parameter-dependent material responses that can infer the anisotropy type and existing preferred directions during training. After the training process, the trained models can be used to solve the inverse problem of finding a microstructure (via its design variables) with a targeted material response. 

The paper is organized as follows. A theoretical background of (second-order) anisotropic hyperelasticity is presented in Section 2. Some constraints on the free energy potential and how they can be introduced in a neural network setting are explained in Section 3 along with a detailed framework for the forward problem. The optimization framework for solving the inverse problem is formulated in Section 4 while Section 5 focuses on data generation for both macroscale and microscale problems. The results for training our framework and the solutions of the inverse problems are presented in Section 6. The paper ends with concluding remarks in Section 7.  

\section{Theory}
For hyperelastic materials, we derive the mechanical properties by employing a strain (or free) energy potential $\Psi (\Fb, \mathcal{D})$ where $\Fb$ is the deformation gradient and $\mathcal{D}$ is a set of design variables. A more convenient representation of the free energy is in terms of the right Cauchy-Green deformation tensor $\Cb = \Fb^T \Fb$ as it ensures frame indifference since $\Cb(\Fb)=\Cb(\Qb\Fb)$ for every $\Qb \in \text{Orth}^+$. Through thermodynamic considerations, we can define the second Piola-Kirchhoff stress as:
\begin{equation} \label{eq:2PK}
\Sb = 2\partial_\Cb \Psi (\Cb, \mathcal{D}) \ .
\end{equation}
The free energy potential must satisfy the material symmetry conditions\citep{ogden1997non}, i.e.,
\begin{equation} \label{eq:mat_sym}
\Psi(\Cb) = \Psi(\Gb\Cb\Gb^T, \mathcal{D}) \ ,
\end{equation}
 for every $\Gb \in \mathcal{G} \subseteq \text{Orth}$. To account for the symmetry group $\mathcal{G}$, we introduce a structure tensor, $\Ab$, or a set of structure tensors, $\Ac = \{ \Ab_i \}$, as additional arguments \cite{zhang1990structural,svendsen1994representation} to the free energy potential to get an equivalent representation:
\begin{equation} \label{eq:aug_arg}
\Psi = \Psi(\Cb, \Ac, \mathcal{D}) \ .
\end{equation}

In this work, we only consider transversely isotropic and orthotropic anisotropies which can be represented using second-order structure tensors. Whereas, for some symmetry groups, higher-order structure tensors are needed \cite{kalina2024neuralnetworksmeetanisotropic}. Setting $\Nb_i = \nb_i \otimes \nb_i$, where $\nb_i$ is a unit vector, we can define a set of second-order structure tensors for the selected anisotropy groups. As in \citep{fuhg2022learning}, we take $\Ac = \{ \Ib = \sum_i \Nb_i \}$ for isotropy, $\Ac = \Nb_1$ for transverse isotropy and $\Ac = \{ \Nb_1, \Nb_2, \Ib \}$ for orthotropy. Note there is no need for $\Nb_3$ since it is already in the span of $\{ \Nb_1, \Nb_2, \Ib \}$. Exploiting the assumption that preferred directions of anisotropy groups under consideration must be orthogonal to each other, we can define the structural tensors through rotations of the Cartesian basis as:
\begin{equation} \label{eq:rot_G}
\Nb_i(\Rb) = \nb_i \otimes \nb_i 
      = \Rb \eb_i \otimes \Rb \eb_i 
%      = \Rb \boxtimes ( \eb_i \otimes \eb_i ) 
      \ ,
\end{equation}
where $\nb_i$ are the orthogonal bases aligned with the preferred directions of anisotropy. We can parameterize the rotation tensor $\Rb$ in terms of a unit vector $\pb$ and a rotation angle $\varphi\in [0,2\pi]$ using the Rodrigues' formula as:
\begin{equation} \label{eq:rodrigues}
\Rb(\varphi\pb) = \exp(\varphi \Pb) 
= \Ib + (\sin \varphi) \Pb + (1-\cos \varphi) \Pb^2 \ ,
\end{equation}
where $\Pb \equiv \varepsilonb \pb$ and $\varepsilonb$ is the third order permutation tensor. 

We ensure objectivity by expressing the potential $\Psi$ in terms of invariants~\citep{gurtin1982introduction} of $\Cb$ and $\Ac = \{ \Ab_i \}$. A complete set of invariants can be expressed as $\Ic = \{\Ic_\text{iso}, \Ic_\text{trans}, \Ic_\text{ortho} \}$, with the subscripts denoting isotropic, transversely isotropic and orthotropic invariants respectively. Thus, we can write the potential as:
\begin{equation}
\Psi = \Psi(\Cb, \Ac, \mathcal{D} ) = {\Psi}( \Ic, \mathcal{D}) \ .
\end{equation}
From this potential representation, following \citep{fuhg2022learning}, we can then write Eq. \eqref{eq:2PK} in terms of a tensor bases $\Bc$ as:
\begin{equation}\label{eq:stress}
\Sb  = 2 \partial_\Cb \Psi 
= 2 \sum_i \partial_{\Ic_i} {\Psi} \ \partial_\Cb \Ic_i
= \sum_i c_i(\Ic, \mathcal{D}) \, \Bc_i \ ,
\end{equation}
where $\Bc_i \equiv \partial_\Cb \Ic_i$ are \textit{a priori} known for chosen invariants and $c_i \equiv \partial_{\Ic_i} \Psi$ depend on the potential function itself. We define the isotropic invariants as:
\begin{equation}\label{eq:inv_iso}
\Ic_\text{iso} = \{\Ic_1 = \tr (\Cb), \Ic_2 = \tr ({\text{Cof}\Cb}), \Ic_3 = J, \Ic_4 = -2J
\} \ ,
\end{equation}
where $\text{Cof}$ denotes the cofactor operator and $J=\sqrt{\det \Cb}$. The first three isotropic invariants are sufficient for complete representation, but the fourth one is added to tackle an issue with NN-based potential modeling as will be explained in the next section. Note that the second invariant $\tr ({\text{Cof}\Cb})$ is equivalent to $\frac{1}{2}\left({(\tr \Cb)}^2 - \tr {(\Cb^2)} \right)$. The tensor bases for isotropic invariants can then be represented as:
\begin{equation}\label{eq:basis_iso}
\Bc_\text{iso} = \{\Ib, \tr(\Cb) \Ib - \Cb, \frac{1}{2}J \Cb^{-1}, -J \Cb^{-1}
\}.
\end{equation}
Substituting this basis into Eq. \eqref{eq:stress}, we obtain the isotropic stresses as:
\begin{equation} \label{eq:stress_iso}
\Sb_\text{iso} = 2 \left[( \partial_{I_1}  \Psi ) \, \Ib + ( \partial_{I_2} \Psi ) \, (\tr(\Cb) \Ib - \Cb) + \frac{1}{2}( \partial_{I_3} \Psi )\, J \Cb^{-1} - ( \partial_{I_4} \Psi )\, J \Cb^{-1} \right].
\end{equation}

Normally, we do not know which anisotropy class the material belongs to or what the preferred directions are. In this case, we can employ the methodology presented in \citep{fuhg2022learning} to learn the anisotropy class and preferred directions during the forward process. To this end, we introduce two parameters, $\bar{\alpha}_{1}$ and $\bar{\alpha}_{2}$, which determine the degree of anisotropy. These are further placed inside the logistic function, i.e.,

\begin{equation}\label{eq:alpha_sig}
\alpha_i = \frac{1}{1+ e^{-\bar{\alpha_i}}} \ .
\end{equation}

The parameters $\alpha_{1}$ and $\alpha_{2}$ will be zero if the material is isotropic and the stresses can be represented using only the isotropic invariants. Whereas, one of them will be nonzero for the case of transverse isotropy and both will be nonzero for the orthotropic case. Since all the free energy representations contain some isotropic part, we only use these parameters for anisotropic components. Details on the learning process of these parameters during training will be presented in subsequent sections. For now, we introduce them with the anisotropic invariants. The invariant set for transversely isotropic material can be defined as:
\begin{equation}\label{eq:inv_trans}
\Ic_\text{trans} = \Ic_\text{iso} \cup  \{
\Ic_5 = \alpha_1 \tr (\Cb \Nb_1), \Ic_6=\alpha_1 \tr ((\text{Cof}\Cb) \Nb_1)
\} \ ,
\end{equation}
with the basis:
\begin{equation}\label{eq:basis_trans}
\Bc_\text{trans} = \Bc_\text{iso} \cup   \{
     \alpha_1 \Nb_1,
     \alpha_1 [\tr ((\text{Cof}\Cb) \Nb_1)\Cb^{-1} - (\text{Cof}\Cb) \Nb_1\Cb^{-1}]
\} \ .
\end{equation}
Therefore, the stress can be evaluated as:
\begin{equation} \label{eq:stress_trans}
\Sb_\text{trans} = \Sb_\text{iso} + 2 \left[
  ( \partial_{I_5}  \Psi ) \, \alpha_1 \Nb_1
+ ( \partial_{I_6}  \Psi ) \, \alpha_1[\tr ((\text{Cof}\Cb) \Nb_1)\Cb^{-1} -  (\text{Cof}\Cb) \Nb_1\Cb^{-1} ] \right] \ .
\end{equation}
Finally, we can write the orthotropic invariants as:
\begin{equation}\label{eq:inv_ortho}
\Ic_\text{ortho} = \Ic_\text{trans} \cup  \{
\Ic_7=\alpha_2 \tr (\Cb \Nb_2), \Ic_8=\alpha_2 \tr ((\text{Cof}\Cb) \Nb_2)
\} \ ,
\end{equation}
with the corresponding bases:
\begin{equation}\label{eq:basis_ortho}
\Bc_\text{ortho} = \Bc_\text{trans} \cup   \{
     \alpha_2 \Nb_2,
     \alpha_2 [\tr ((\text{Cof}\Cb) \Nb_2)\Cb^{-1} - (\text{Cof}\Cb) \Nb_2\Cb^{-1}]
\} \ ,
\end{equation}
leading to a stress representation of the form:
\begin{equation} \label{eq:stress_ortho}
\Sb_\text{ortho} = \Sb_\text{trans} + 2 \left[
  ( \partial_{I_7}  \Psi ) \, \alpha_2 \Nb_2
+ ( \partial_{I_8}  \Psi ) \, \alpha_2 [ \tr ((\text{Cof}\Cb) \Nb_2)\Cb^{-1} - (\text{Cof}\Cb) \Nb_2\Cb^{-1} ] \right].
\end{equation}

%\section{Methodology}
\section{Model construction} \label{Sec:Model}

Apart from material symmetry and frame invariance, there exist some additional constraints on the form of the free energy potential which will be discussed in the this section along with details on the enforcement of these constraints in a neural network framework. A free energy potential exhibiting ellipticity ensures material stability\citep{Zee1983-ml}. However, enforcing this by construction in a neural network setting is a daunting task. An alternate approach is to ensure polyconvexity of the energy potential since Ball \cite{ball1976convexity} proved the existence of boundary value problem minimizers for polyconvex and coercive potentials. In addition, polyconvexity guarantees sequential weak lower semicontinuity (s.w.l.s.). Together with the coercivity it is sufficient for the existence of minimizers. Polyconvexity is generally a stronger condition than ellipticity \citep{Alibert1992, Aubert1987} but is simpler to enforce. The polyconvexity condition is fulfilled if and only if a function $\mathcal{P} $ exists such that
\begin{equation}
  \Psi(\Fb, \Ac, \mathcal{D}) = \mathcal{P}(\Fb, \text{Cof} \,\Fb, \det \Fb, \Ac, \mathcal{D})
\end{equation}
where $\mathcal{P}$ is convex with regard to the minors of $\Fb$, i.e., $\Fb$, \text{Cof} $\Fb$, and $\det \Fb$, which govern the deformation of line, area, and volume elements, respectively. Since we employ an invariant formulation, the invariants must be convex with respect to these arguments and thereby, be polyconvex in $\Fb$. Although all the invariants in $\Ic$ fulfill this requirement, this places an additional constraint on the form of strain energy. Specifically, the strain energy must be convex and monotonically nondecreasing with respect to these invariants in order to preserve polyconvexity\citep{Boyd_Vandenberghe_2011}. We want to model the free energy potential using a neural network such that $\Psi(\Ic, \mathcal{D}) = \Psi^{\mathcal{N\!N}}(\Ic, \mathcal{D})$ but it is evident that the direct application of a classical feedforward neural network does not provide a polyconvex potential. In a recent work \cite{amos2017input}, two neural network architectures were proposed: First, fully Input Convex Neural Networks (fICNNs) were presented that gave a convex scalar-valued output with respect to all the inputs. Second, partially Input Convex Neural Networks (pICNNs) which can be convex with respect to some inputs and can take arbitrary forms with regard to others. The latter is particularly useful for our problem since our strain energy potentials generally are also dependent on the design variables $\mathcal{D}$. 

\begin{figure}%[h]
    \centering
    \includegraphics[width=1.0\textwidth]{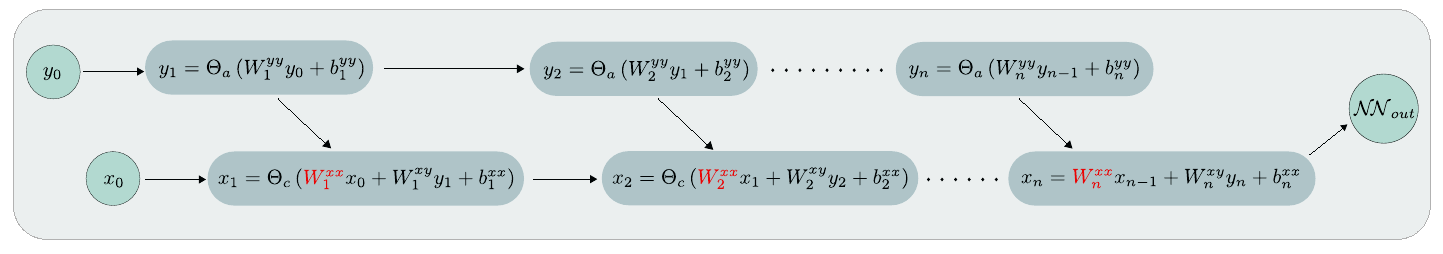}
    \caption{Implementation schematic for partially input convex neural network employed. This network is convex in $x_0$ which would be the invariants in our framework whereas it can take an arbitrary form with respect to $y_0$ which represent the design variables in our framework. Here, $\Theta_c$ represents a positive, monotonically nondecreasing, and convex activation function whereas $\Theta_a$ represents an arbitrary activation function. The red color indicates the weights constrained to be positive to ensure convexity and monotonicity with respect to inputs $x_0$.
    % \COMMENT{RJ: State that this network is convex in x0 in the caption and add an explicit output please}
    }
    \label{fig:pICNN}
\end{figure}

We can formally write the update of these pICNNs as:
\begin{subequations}
\begin{align}
                \mathbf{y}_{h+1} &= \Theta_{a} \left( W_h^{[yy]} \mathbf{y}_{h} + \mathbf{b}_h^{[y]} \right) , \quad h = 0, \ldots, H \label{y_eq} \\
        \mathbf{x}_{h+1} &= \Theta_c \left( W_h^{[xx]} \mathbf{x}_h + W_h^{[xx0]} \mathbf{x}_0 + W_h^{[xy]} \mathbf{y}_{h+1} + \mathbf{b}_h^{[x]} \right) , \quad h = 0, \ldots, H \label{x_eq}
\end{align}
\end{subequations}
This neural network is convex with respect to the inputs $\mathbf{x}_0$ given $W_h^{[xx]}$ is non-negative for $h = 1, \ldots, H$ and $\Theta_c$ is a convex, monotonically non-decreasing function. And the output can take any form with respect to the inputs $\mathbf{y}_0$. We are free to choose $\Theta_a$ with the subscript ``$a$'' denoting the arbitrary nature of this activation function. We can also constrain this network to be convex and monotonically increasing by ensuring $W_h^{[xx]}$ is non-negative for $h = 0, \ldots, H$ and $\Theta_c$ is a positive, convex, and monotonically non-decreasing function \cite{klein2022polyconvex}. The bias vector $\mathbf{b}^{[x]}$ can take on any value and contributes substantially to the representation power of the network. Unfortunately, the same cannot be said about the so-called passthrough layers for the inputs $\mathbf{x}_0$. As pointed out in \citep{Klein_Roth_Valizadeh_Weeger_2023}, the (additional) positivity constraint on the weights in the first layer curbs their benefit. Therefore, we employ a slightly modified version of pICNNs. 
The implementation schematic of our architecture is illustrated in figure \ref{fig:pICNN}. With the same constraints as before, we can formally write the modified pICNN as:
\begin{subequations}
\begin{align}
                \mathbf{y}_{h+1} &= \Theta_{a} \left( W_h^{[yy]} \mathbf{y}_{h} + \mathbf{b}_h^{[y]} \right) , \quad h = 0, \ldots, H \label{y_eq_new} \\
        \mathbf{x}_{h+1} &= \Theta_c \left( W_h^{[xx]} \mathbf{x}_h + W_h^{[xy]} \mathbf{y}_{h+1} + \mathbf{b}_h^{[x]} \right) , \quad h = 0, \ldots, H
        \label{x_eq_new}
\end{align}
\end{subequations}

The reason behind the introduction of the fourth isotropic invariant now becomes apparent. Since we have a convex, but more importantly, monotonically increasing neural network, all the $\partial_{\Ic_i}  \Psi$ terms in Eq.\ \eqref{eq:stress_iso} will be positive. Furthermore, the first three bases would also be positive, at least the diagonal terms. Therefore, in order to represent negative stresses, we need to include the additional invariant \cite{klein2022polyconvex, linden2023neural}. 

A further constraint~\citep{kruvzik2019mathematical} on the potential is that it should be coercive, i.e.,
\begin{equation}\label{eq::growth}
\Psi \rightarrow \infty \quad\text{if}\quad J \rightarrow 0 \quad\text{or} \quad J \rightarrow \infty
\end{equation}
This is also referred to in the literature as the \emph{volumetric growth} condition. It also stems from the physical consideration that the volume of an object cannot vanish or grow to infinity~\citep{holzapfel2002nonlinear}. As argued in \cite{klein2022polyconvex}, this constraint is rather theoretical in nature since most materials do not experience loading that touch on the coercivity conditions. However, to keep our framework general, we additively modify our potential to have:
\begin{equation}\label{eq::PsiplusPsig}
\Psi(\Ic, \mathcal{D}) = \Psi^{\mathcal{N\!N}}(\Ic, \mathcal{D}) + \Psi^{\text{gr}}(J)
\end{equation}
where $(\cdot)^\text{gr}$ indicates growth. We introduce a slightly modified form of the polyconvex term presented in \citep{HARTMANN20032767} to define a coercive form:
\begin{equation}\label{eq::Psi_g}
\Psi^{\text{gr}} = \gamma \left( J + \frac{1}{J} -2 \right)^2.
\end{equation}
The additional term $\gamma$ is problem-specific and can be used to control the magnitude of the stresses resulting from this term. This becomes apparent when we write out the stress contribution from $\Psi^{\text{gr}}$, i.e.,
\begin{equation}\label{eq::stress-growth}
\mathbf{S}^\text{gr} = 2 \partial_{\Cb} \Psi^\text{gr} = 2\gamma\left( J + \frac{1}{J} -2 \right)\left( 1 - \frac{1}{J^2}\right)J\Cb^{-1}.
\end{equation}

The free energy of the system must be zero at the undeformed configuration. Therefore, we add a constant term to our potential such that it becomes:

\begin{equation}
\Psi = \Psi^{\mathcal{N\!N}} + \Psi^{\text{gr}} + \Psi^{\text{n}},
\end{equation}
where $(\cdot)^\text{n}$ denotes normalization. We define it as:
\begin{equation}\label{eq::Psi_n}
\Psi^{\text{n}} = -\Psi^{\mathcal{N\!N}}(\Bar{\Ic}, \mathcal{D}),
\end{equation}

where $\Bar{\Ic}$ denotes the invariants evaluated in the undeformed configuration. In particular, the invariants read
\begin{subequations}
\begin{align}
    \Bar{\Ic}_\text{iso} &= \{3, 3, 1, -2\}
    \}, \\
    \Bar{\Ic}_\text{trans} &= \Bar{\Ic}_\text{iso} \cup  \{
    \alpha_1 \tr\Nb_1, \alpha_1 \tr\Nb_1
    \},\\
    \Bar{\Ic}_\text{ortho} &= \Bar{\Ic}_\text{trans} \cup  \{
    \alpha_2 \tr\Nb_2, \alpha_2 \tr\Nb_2
    \},
\end{align}
\end{subequations}

with the following basis tensors:

\begin{subequations}
\begin{align}
    \Bar{\Bc}_\text{iso} &= \{\Ib, 2 \Ib , \frac{1}{2}\Ib, -\Ib\}
    \}, \\
    \Bar{\Bc}_\text{trans} &= \Bar{\Bc}_\text{iso} \cup  \{
     \alpha_1 \Nb_1 ,\alpha_1\tr(\Nb_1)\Ib - \alpha_1\Nb_1
    \},\\
    \Bar{\Bc}_\text{ortho} &= \Bar{\Bc}_\text{trans} \cup  \{
    \alpha_2 \Nb_2 ,\alpha_2\tr(\Nb_2)\Ib - \alpha_2\Nb_2
    \},
\end{align}
\end{subequations}

where $\tr\Nb_1=\tr\Nb_2=1$ in our framework since it is composed of an outer product of unit vectors. In addition to the free energy being zero at the undeformed state, we also require the stresses to be zero. To this end, we further modify the free energy as:

\begin{equation} \label{completePsi}
\Psi = \Psi^{\mathcal{N\!N}} + \Psi^{\text{gr}} + \Psi^{\text{n}} + \Psi^{\text{sn}}
\end{equation}

where $\Psi^{\text{sn}}$ denotes the free energy associated with the stress normalization and is defined as:

\begin{equation} \label{Finalnormalization}
{\Psi}^{\text{sn}} = -{\mathfrak{o}}(\Ic_3 - \Bar{\Ic_3}) + \mathfrak{p}(\Ic_5-\Bar{\Ic_5})+ \mathfrak{q}(\Ic_6-\Bar{\Ic_6})+ \mathfrak{r}(\Ic_7-\Bar{\Ic_7})+ \mathfrak{s}(\Ic_8-\Bar{\Ic_8}),
\end{equation}

where,

\begin{subequations}
\begin{align}
 \nonumber {\mathfrak{o}} &= 2 \left[\partial_{\bar{I}_1}  \Psi + 2\partial_{\bar{I}_2} \Psi + \frac{1}{2}\partial_{\bar{I}_3} \Psi - \partial_{\bar{I}_4} \Psi + (\mathfrak{p}+\mathfrak{q})\tr(\alpha_1\Nb_1) + (\mathfrak{r}+\mathfrak{s})\tr(\alpha_2\Nb_2) \right] \\ 
\nonumber \mathfrak{p} &= \partial_{\bar{I}_6} \Psi , \quad
\mathfrak{q} = \partial_{\bar{I}_5} \Psi , \quad
\mathfrak{r} = \partial_{\bar{I}_8} \Psi , \quad
\mathfrak{s} = \partial_{\bar{I}_7} \Psi
\end{align}
\end{subequations}

Such a form for $\Psi^{\text{sn}}$ preserves polyconvexity of the free energy while ensuring a stress-free configuration at the undeformed state. Since it is a linear equation in $\Ic_3$, i.e., $J$, terms with $\mathfrak{o}$ do not appear in the hessian and therefore can have the negative sign with them. Whereas all the other terms involve polyconvex invariants multiplied with some positive constants since we have a monotonically increasing neural network and therefore preserve polyconvexity of the free energy. A detailed derivation for stress normalization part of the free energy is presented in \aref{app:normalization}. The applied stress normalization technique is close to the one proposed in \cite{linden2023neural}. However, we use a slightly different approach to normalize the anisotropic part.

\begin{figure}
    \centering
    \includegraphics[width=0.75\linewidth]{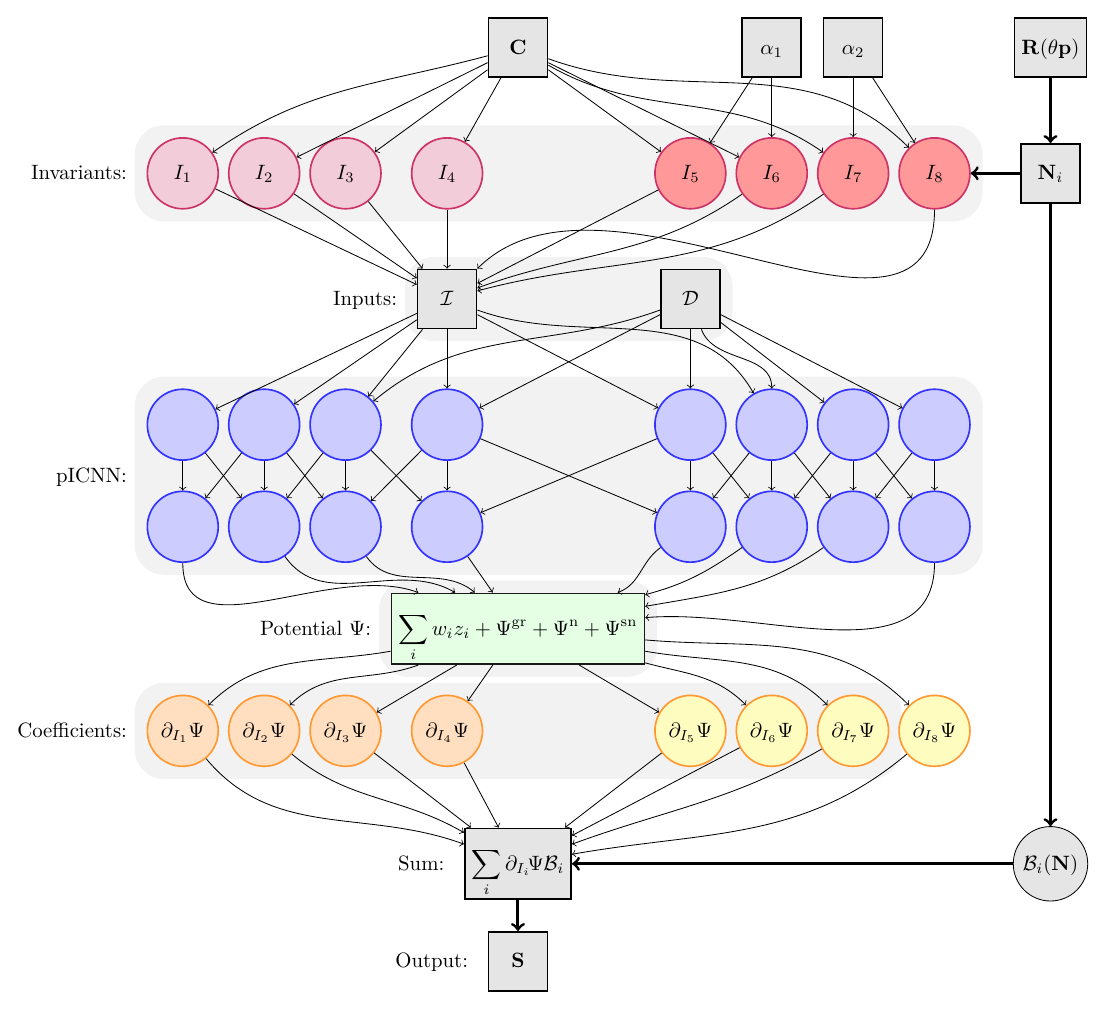}
    \caption{Physics-augmented neural network model for the solution of the forward problem.}
    \label{fig:ForwardProbSchem}
\end{figure}

\section{Optimization problem}
 Data generation will be discussed in detail in the next section but for all problems, our datasets basically contain the right Cauchy-Green deformation tensors and the stresses associated with them, along with some design variables. The training dataset denoted with the superscript $\dagger$ has the form $\lbrace \Sb^\dagger,\Cb^\dagger,\mathcal{D}^\dagger \rbrace$ where we know the associated design variables $\mathcal{D}^\dagger$. Whereas the target dataset has the form $\lbrace \mathcal{T}^\star, \cdot^\star, \mathcal{D}^\star \rbrace$ where $\mathcal{T}^\star$ denotes the target response which could be e.g. specified stresses, displacements, etc., $\cdot^\star$ represents all the other inputs for the target dataset and $\mathcal{D}^\star$ are the (unknown) target design variables. The optimization problem can be written as:
\begin{equation}
\begin{aligned}
       \argmin_{\mathcal{D}^{\star}} \, \, &\| \mathcal{T}^{\star} - \hat{\mathcal{T}}^{\star}(\cdot^\star , \mathcal{D}^{\star}) \|_2^2 \\
       \text{s.t.} \argmin_{\bm{\theta}} \, \, &\| \Sb^{\dagger} - \hat{\Sb}^{\dagger}(\Ic^{\dagger}, \mathcal{D}^{\dagger}) \|_2^2 \,
\end{aligned}
\end{equation}
 where $\bm{\theta}$ is a set of the neural network parameters (weights and biases). This problem can be solved sequentially: We first find $\bm{\theta}$ that minimizes the loss for predicted and true training stresses. We call this the forward problem. Once we have the trained neural network, we solve the inverse problem where we find the design variables $\mathcal{D}^{\star}$ that minimize the loss between true $\mathcal{T}^\star$ and predicted target response $\hat{\mathcal{T}}^\star$.

% \COMMENT{RJ: the idea of matching a target dataset is new to me, usually design optimization seeks to maximize some performance metric (subject to cost constraints) e.g. max stiffness for constant weight. Can we add more explanation here, pls? Reading further I get the idea that design variables are configuration variables for simple/easily parameterized microstructures? So there is no performance optimization per se? Ok 6.3 is close to what I would expect for design optimizaton.}
 
 First, we discuss the simple scenario where we already know the anisotropy class and the preferred directions, if any. In such cases, the forward problem reduces to training the pICNN with the respective invariant sets to get the correct free energy potential whose derivatives, when substituted into Eq.\ \eqref{eq:stress_iso}, \eqref{eq:stress_trans} or \eqref{eq:stress_ortho}, give the stresses. In this setting, the loss function looks like:
\begin{equation}
L = \| \Sb^{\dagger} - \hat{\Sb}^{\dagger} \|_2^2 ,
\end{equation}
where $\Sb$ are the predicted stresses and $\hat{\Sb}$ are the true or actual stresses. As pointed out earlier, we generally do not know the anisotropy class and preferred directions. This is where the parameters $\alpha_1$ and $\alpha_2$ come in. We modify the loss function to drive these parameters to zero using $L^p$ regularization unless it is impossible to represent the free energy (or the stresses) without them being nonzero. We write the loss function as:

\begin{equation}
L = \| \Sb^{\dagger} - \hat{\Sb}^{\dagger} \|_2^2 + \varepsilon \| \boldsymbol{\alpha} \|_{p}^{p} \ , \quad    \text{with} \quad \| \boldsymbol{\alpha} \|_{p} = \left( \alpha_{1}^{p} + \alpha_{2}^{p} \right)^\frac{1}{p} \ .
\end{equation}

% \textcolor{red}{KK: $L^p$ regularization should be $\| \boldsymbol{\alpha} \|_p = \left( \alpha_{1}^{p} + \alpha_{2}^{p} \right)^\frac{1}{p}$}

Here, we choose $p=\frac{1}{4}$ \cite{flaschel2021unsupervised}, and $\varepsilon$ is the penalty parameter which needs to be tuned for a given problem. Since we do not know the preferred directions, we parametrize the tensors $\Nb_1$ and $\Nb_2$ in terms of the rotation tensor $\Rb$ using Eq. \eqref{eq:rot_G}. We can further parameterize $\Rb$ through the Rodrigues' formula from Eq. \eqref{eq:rodrigues} such that we have four trainable parameters: the angle $\varphi$ and three components of the unit vector $\pb$.
% \COMMENT{RJ: there are only 2 independent components to p}.

Our pICNN architecture had 3 layers with 30 neurons for the layers associated with inputs $\mathcal{D}$ and 40 neurons for inputs $\Ic$ requiring convexity and monotonicity. We implemented this architecture in PyTorch \cite{NEURIPS20199015} and used the Adam optimizer \cite{kingma2014adam} to optimize the neural network parameters over $5\times10^{5}$ epochs. All datasets were input in a single batch to the network with a learning rate of $10^{-3}$. Once we have trained our pICNN, we can recover the preferred directions. The trained network gives us the parameters $\varphi$ and $\pb$, which when plugged into Eqs. \eqref{eq:rodrigues} and \eqref{eq:rot_G} give the structure tensors $\Nb_i$. We then employ the Nelder-Mead \cite{nelder1965simplex, gao2012implementing} optimization algorithm to find the unit vectors $\nb_i$ that give the same corresponding structure tensor $\Nb_i$ as the one predicted by the network. Finally, we can solve the inverse design problem where we are given a set of inputs, and we want to find the design parameters that give the targeted mechanical response. To this end, we employ Covariant Matrix Adaptation Evolutionary Strategy (CMA-ES) \cite{cmaes, Hansen16a} since it is less prone to getting stuck in local minima. It is a derivative-free evolutionary strategy that works by updating the mean and covariance matrix of a multivariate distribution based on the fitness of the previous generation to generate a new offspring. 

Since our framework employs invariant formulation and satisfies the material frame indifference and objectivity requirements by construction, we can write:
\begin{equation}
\Psi = \Psi(\Cb, \Ac_i) \ = \Psi(\bar{\Cb}, \bar{\Ac}_i) \,
\end{equation}
where
\begin{equation}
\bar{\Cb} = \Qb^T \Cb  \Qb , \quad \bar{\Ac}_i  = \Qb^T \Ac_i \Qb\ \quad \forall \Qb \in \text{Orth}^+,
\end{equation}
and the stresses read:
\begin{equation}
\bar{\Sb}  = 2 \partial_{\bar{\Cb}} \Psi 
\end{equation}
This allows us also to invert the preferred direction which may be different from the one our model was trained on. We can do so by once again employing the Rodrigues' formula from Eq. \eqref{eq:rodrigues} to get $\varphi$ and three components of the unit vector $\pb$ as additional design variables in our optimization framework which can then be inverted again to get the preferred direction of anisotropy for some stress-deformation data with different material parameters and preferred direction than the ones seen by the model during training.

\section{Data Generation}
In this section, we discuss the choice and distribution of material parameters for the data generation process. 
We distinguish between synthetic problems where we look at known parameterized constitutive laws, and example problems where actual parameterized microstructures are used to create the training data through computational homogenization under consideration of periodic boundary conditions.

\subsection{Synthetic macroscale Data}\label{sec:macrodata}
For our macroscale problems, we sample for simplicity from the deformation gradient space bounded by $\Delta$ around the undeformed configuration ($\Fb = \Ib$), i.e.,
${F}_{ij} \in \delta_{ij} \in [-\Delta,\Delta]$
\noindent with $\Delta > 0$. Despite the likelihood of generating negative stretches
with this scheme, all samples are valid since we only use $\Cb=\Fb^{T} \Fb$ in the generative model evaluations. We use Latin Hypercube Sampling \citep{stein1987large} to generate $N_F$ space-filling samples in this nine-dimensional bounded space with $\Delta= 0.2$. 
 Both isotropic and anisotropic representations of the free energies are presented along with the resulting second Piola-Kirchhoff stresses and the resulting tangent modulii.

\paragraph{Isotropic hyperelasticity}
To model compressible isotropy, we employ the well-known neo-Hookean model described by a strain energy function discussed in \crefs{ciarlet2021mathematical,bonet1998simple}:
\begin{equation}
\Psi = \frac{1}{2} c_1 (I_1 - 3) - c_1 \log J + \frac{1}{2} c_2 (J -1)^2 \ ,
\end{equation}
which yields a second Piola-Kirchhoff stress of the form:
\begin{equation}
\Sb = c_1 (\Ib - \Cb^{-1} ) + c_2 J(J-1) \Cb^{-1} \ .
\end{equation}
% leading to the following fourth-order tangent modulus:
% \begin{equation}
%     \mathbb{C} = 2 \partial_{\Sb} \Psi = \left( 2 c_1 - 2 c_2 J (J-1)\right)\left( \Cb^{-1} \odot \Cb^{-1} \right) + \left(2 c_2 J (2J-1) \right)\left(\Cb^{-1} \otimes \Cb^{-1}\right) .
% \end{equation}
%  and $\otimes$ represents the dyadic product whereas $\Cb^{-1} \odot \Cb^{-1} = -\partial_{\Cb} \Cb^{-1}$ and can be calculated as \cite{holzapfel2002nonlinear}:
% \begin{equation}
%    \left( \Cb^{-1} \odot \Cb^{-1} \right)_{ABCD} = \frac{1}{2}\left(\Cb^{-1}_{AC}\Cb^{-1}_{BD} + \Cb^{-1}_{AD}\Cb^{-1}_{BC} \right) .
% \end{equation}
Herein $I_1 = \tr \Cb$ and $J \equiv \sqrt{\det{\Cb}} = \sqrt{I_3}$. Five samples for the material parameters $c_1$ and $c_2$ are chosen uniformly from the range $[1.0, 5.0]$ and we have a dataset of the form $\lbrace {c_1}_{i}, {c_2}_{j}, \Cb_{k}, \Sb_{k} \rbrace$ with $i=1,\ldots,5$, $j=1,\ldots,5$ and $k=1,\ldots,N_F$. Hence, for all the combinations of these inputs, we get $5\times5\times N_F$ sets of input data.

\paragraph{Anisotropic hyperelasticity}
For materials belonging to anisotropic classes, namely transverse isotropy and orthotropy, we use a modified version of the strain energy function proposed in \cref{holzapfel2000new} which reads
\begin{equation}
    \Psi = c_{1} ( I_{1}- 3 ) + \frac{c_{1}}{c_{2}} (J^{-2 c_{2}}- 1)+ c_{3} \left(\exp( c_{4} (I_{4}-1)^{4})   + \exp(c_{5} (I_{6}-1)^{4})   -2 \right) \ ,
\end{equation}
where $I_4 \equiv \tr \Cb\Nb_1$ and $I_6 \equiv \tr \Cb\Nb_2$, giving the second Piola-Kirchhoff stress as
\begin{equation}
\begin{aligned}
        \Sb &= 2 c_{1} \Ib - 2 c_{1} I_{3}^{-c_{2}} \Cb^{-1} + 8 c_{3} c_{4} (I_{4}-1)^{3} \exp(c_{4}(I_{4}-1)^{4}) \Nb_{1} \\
        &+ 8 c_{3} c_{5} (I_{6}-1)^{3} \exp(c_{5}(I_{6}-1)^{4}) \Nb_{2} \ .
\end{aligned}
\end{equation}
% The tangent modulus in this case becomes:
% \begin{equation}
% \begin{aligned}
%         \mathbb{C} &= 4 c_1 {I_3}^{-c_2} \left( \Cb^{-1} \odot \Cb^{-1} \right) + 4 c_1 c_2 {I_3}^{-c_2} \left(\Cb^{-1} \otimes \Cb^{-1}\right)\\
%         &+ 16 c_{3} c_{4} \exp(c_{4}(I_{4}-1)^{4}) (3 (I_{4}-1)^{2} + 4 c_{4}(I_{4}-1)^{6}) \Nb_1 \otimes \Nb_1\\
%                 &+ 16 c_{3} c_{5} \exp(c_{5}(I_{6}-1)^{4}) (3 (I_{6}-1)^{2} + 4 c_{5}(I_{6}-1)^{6}) \Nb_2 \otimes \Nb_2
% \end{aligned}
% \end{equation}

We can toggle between anisotropy classes from this strain energy function since $c_4 \neq 0$ and $c_5 = 0$ gives a transversely isotropic behavior while $c_4 \neq 0$ and $c_5 \neq 0$ results in orthotropic material behavior. In our case, we set $c_2 = 0.75$ and $c_3=1$ and solve the inverse problem for the remaining parameters. We again get five samples for each parameter choosing uniformly for $c_1$ from $[1.0, 5.0]$, $c_4$ from $[3.0, 7.0]$ and $c_5$ from $[2.0, 6.0]$. For the transversely isotropic case, we get the dataset $\lbrace {c_1}_{i}, {c_4}_{j}, \Cb_{k}, \Sb_{k} \rbrace$ with $i=1,\ldots,5$, $j=1,\ldots,5$ and $k=1,\ldots,N_F$ whereas for orthotropic case, we have $\lbrace {c_1}_{i}, {c_4}_{j}, {c_5}_{k}, \Cb_{l}, \Sb_{l} \rbrace$ with $i=1,\ldots,5$, $j=1,\ldots,5$, $k=1,\ldots,5$ and $l=1,\ldots,N_F$. We choose the preferred directions:
$\bm{n}_{1} = (\frac{1}{\sqrt{3}},\frac{\sqrt{2}}{\sqrt{3}},0)$ and $\bm{n}_{2} = (\frac{\sqrt{2}}{\sqrt{3}},-\frac{1}{\sqrt{3}},0)$.

\subsection{Homogenized Data}\label{sec:microData}
There are multiple ways to ensure that the sampled deformation gradients have positive stretches such as rejection sampling, sampling from symmetric deformation gradients \cite{fuhg2021model}, or by relying on polar decomposition \cite{kalina2024neural}. We choose the latter for sampling $N_{F}$ deformation gradients that are applied within the following two RVE problems via periodic boundary conditions. 
%In addition, we reduce the number of samples by filtering all sampled states within the invariant space belonging to the RVE's effective behavior \cite{kalina2024neural}.
The RVE simulations have been done within an in-house Matlab FE code.

\begin{figure}%[h]
    \centering
    \includegraphics[scale = 0.3]{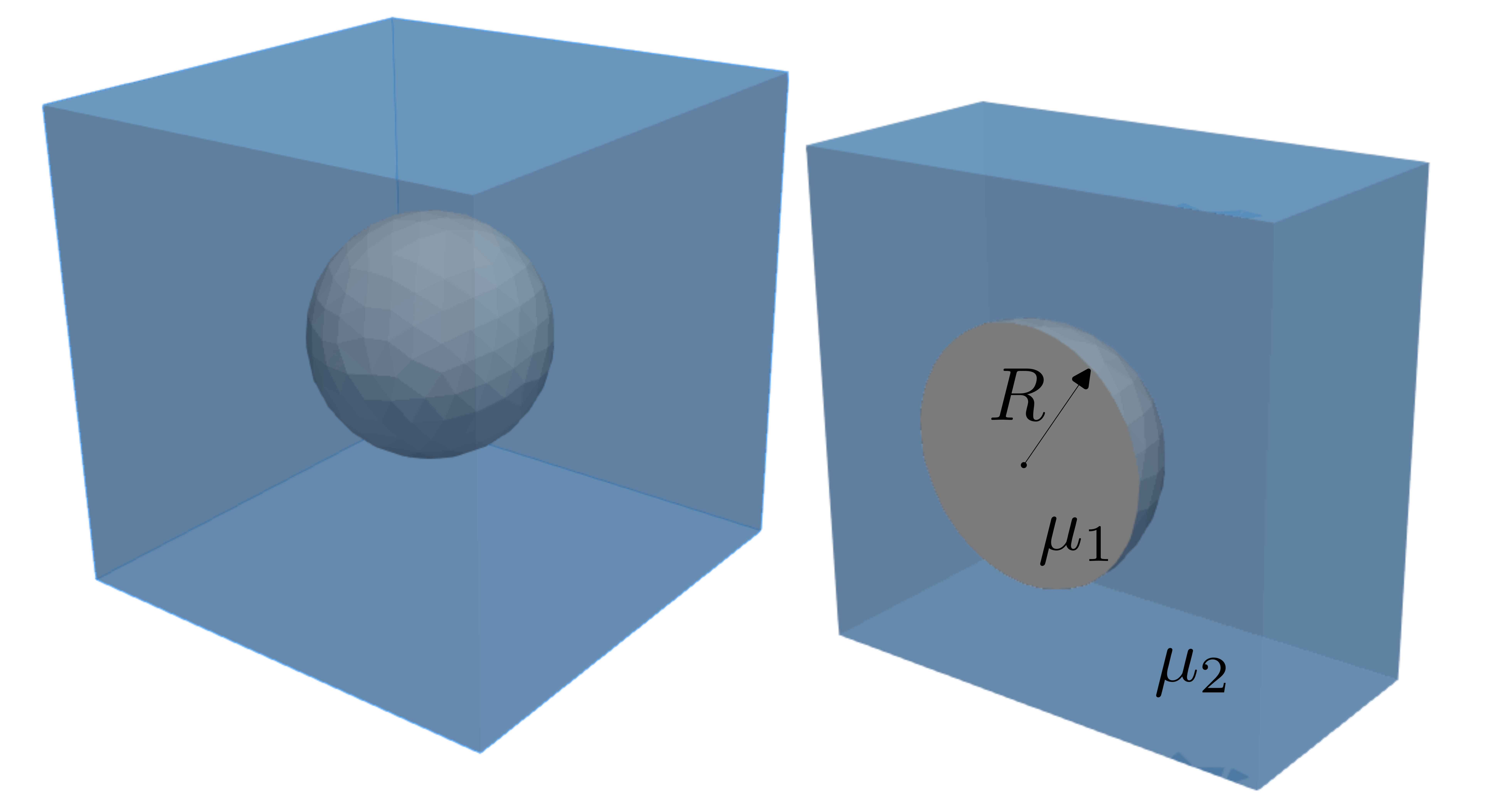}
    \caption{RVE of size 1$\times$1$\times$1 with a single spherical inclusion in the center.}
    \label{fig:RVESingleInc}
\end{figure}

\begin{figure}%[h]
    \centering
    \includegraphics{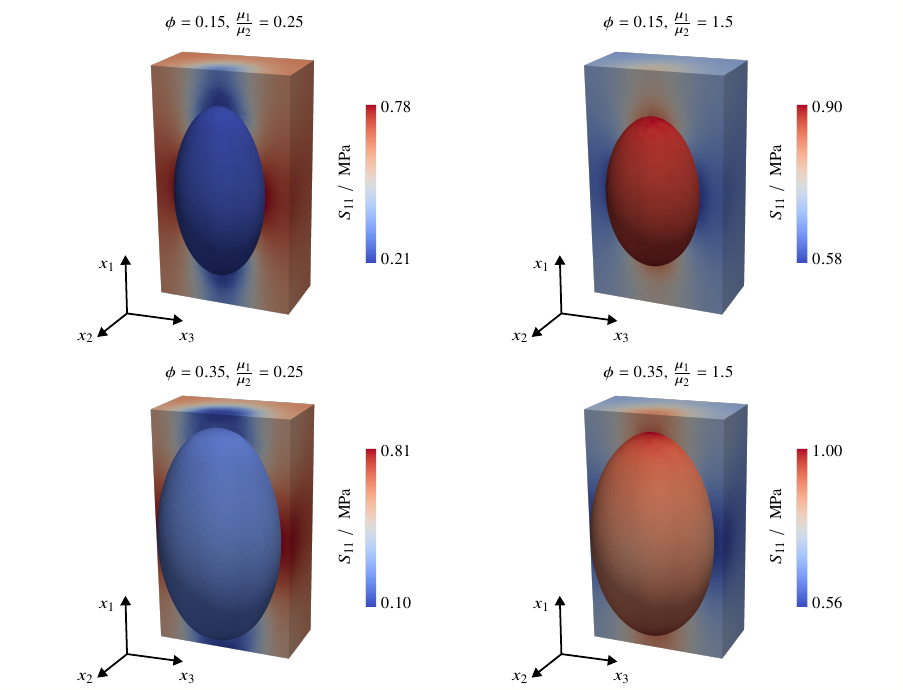}
    \caption{Local stress fields for different values of $\phi$ and $\frac{\mu_1}{\mu_2}$ under uniaxial loading of the single inclusion RVE. Shown is the component $S_{11}$ of the second Piola-Kirchhoff stress tensor for a macroscopi stretching of $\lambda_1 = 1.45$ in the $x_1$ direction applied via periodic boundary conditions.}
    \label{fig:RVE_eqStress}
\end{figure}

\paragraph{Single inclusion RVE}
First, we consider an RVE with a single spherical inclusion at the center shown in figure \ref{fig:RVESingleInc}. For both the inclusion and the matrix material, we use an isotropic hyperelastic law with a coupled form of a compressible neo-Hookean model \cite{holzapfel2002nonlinear} as:
\begin{equation}
\Psi = \frac{\mu}{2} \left(I_1 - \ln I_3^2 -3 \right) + \frac{\lambda}{4} \left(I_3^2 - \ln I_3^2 -1\right) \, ,
\label{eq:neo-Hooke_Karl}
\end{equation}
which gives the second Piola-Kirchhoff stresses as:
\begin{equation}
    \Sb = \mu(\Ib-\Cb^{-1}) + \frac{\lambda}{2} (I_3^2 \Cb^{-1} - \Cb^{-1}) \ .
\end{equation}
% and the following tangent modulus:
% % \COMMENT{RJ: odot is usually Hadamard product which is not how it's used here pls define}
% \begin{equation}
%     \mathbb{C} = 4 \beta c_1 {I_3}^{-\beta} \left(\Cb^{-1} \otimes \Cb^{-1}\right) + 4 c_1 {I_3}^{-\beta}\left( \Cb^{-1} \odot \Cb^{-1} \right) .
% \end{equation}
Here, $\mu$ denotes the shear modulus and $\lambda = \frac{2 \mu \nu}{1-2\nu}$, where $\nu$ represent the Poisson's ratio. For the generation of homogenized stress-strain data, we can control the volume fraction $\phi$ of the inclusion through its radius $R$ and the stiffness ratio by prescribing varying $\frac{\mu_1}{\mu_2}$. The Poisson's ratio is fixed to $\nu=0.44$ for both phases. 
As evident from figure \ref{fig:RVESingleInc}, $\mu_1$ represents the shear modulus of the inclusion whereas $\mu_2$ is the shear modulus of the matrix material which was chosen to $\mu_2 = 1\, \text{MPa}$. We generate a total of 50 samples from $\left(\phi, \frac{\mu_1}{\mu_2}\right) \in [0.15,0.35] \times [0.25,1.5]$ via Latin Hypercube Sampling. For each pair $\phi_i,\left(\mu_1/\mu_2\right)_i$, $N_F=13\cdot 20$ deformation states are prescribed within the RVE simulation. Thereby, the deformation gradients were sampled by first sampling 1000 deformation gradients through LHS and then sorting in the invariant space to get rid of the redundant ones since two different deformation gradients can have the same invariants. An efficient way of covering the entire deformation gradient space is to then store the Cauchy-Green tensor and the associated stress at each load step \cite{kalina2024neural}, where 20 increments have been chosen.\footnote{In the sampling technique used for the deformation states, it was assumed that the effective behavior of the RVE is isotropic, cf. \cite{kalina2024neural} for more details on the sampling technique. A discussion on the isotropy of the RVE is given in appendix~\ref{app:RVEIsotropy}.}
%Similar to the synthetic macroscopic dataset, we get five samples for $R$ from $[0.15,0.5]$ and six samples for $\frac{\mu_1}{\mu_2}$ from $[0.25,1.5]$. 
Finally, we get a dataset of the form $\lbrace {\phi}_{i}, {(\mu_1/\mu_2)}_{i}, \Cb_{k}, \Sb_{k} \rbrace$ with $i=1,\ldots,50$ and $k=1,\ldots,N_F$. 
Hence, for all the combinations of these inputs, we get $50\times 260$ data tuples. Here, $\Sb$ represents homogenized macroscopic stresses. Figure \ref{fig:RVE_eqStress} shows the equivalent stresses in the RVE for different choices of design parameters.

\begin{figure}%[h]
    \centering
    \includegraphics[scale = 0.3]{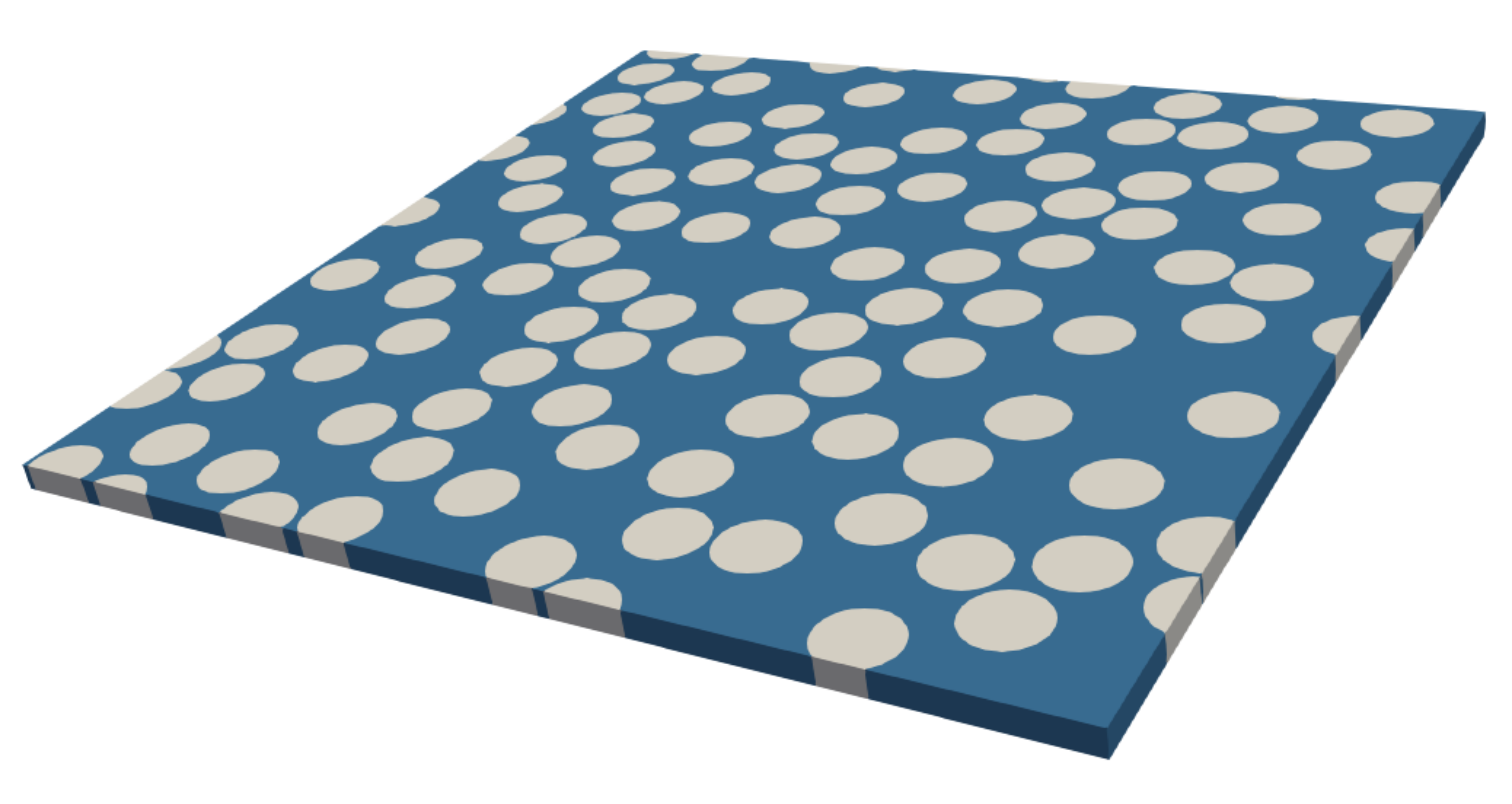}
    \caption{RVE with fibers oriented along $\bm{n} = (0,0,1)$.}
    \label{fig:RVEfiber}
\end{figure}

\paragraph{RVE with unidirectional fibers}
Now we consider a different RVE embedded with fibers, all aligned in a single direction, as shown in Figure \ref{fig:RVEfiber}. The free energy density for this RVE was formulated using the same neo-Hookean model as given in Eq.~\eqref{eq:neo-Hooke_Karl}.
% \begin{equation}
% \Psi = \sum_{n=1}^{N} \frac{\mu_n}{\alpha_n} \left(\bar{\lambda}_1^{\alpha_n}+\bar{\lambda}_2^{\alpha_n}+\bar{\lambda}_2^{\alpha_n} - 3 \right) + \frac{\kappa}{4} \left(J^2 - 2 \ln J - 1 \right),
% \end{equation}
% where $\bar{\lambda}_i = J^{-1/3} \lambda_i$ and we get the second Piola-Kirchhoff stresses \cite{steinmann2012hyperelastic}:
% \begin{equation}
% \Sb_i = J^{-2/3} \left[\sum_{n=1}^{N} \mu_n \bar{\lambda}_i^{\alpha_{n}-2} - \sum_{j=1}^{3} \sum_{n=1}^{N} \frac{\mu_n}{3} \frac{\bar{\lambda}_j^{\alpha_n}}{\bar{\lambda}_i^{2}} \right] + \frac{\kappa}{2} \frac{1}{{\lambda}_i^{2}} (J^2-1), \quad i=1,2,3.
% \end{equation}
% and the tangent modulus:
% \begin{equation}
% \mathbb{C} = \sum_{i,j=1}^{3} \frac{1}{\lambda_j} \frac{\partial {\Sb_i}}{\partial {\lambda_j}}  \gb_i \otimes \gb_i \otimes \gb_j \otimes \gb_j + \sum_{\substack{i,j=1\\ i\ne j}}^{3} \frac{\Sb_j - \Sb_i}{\lambda_j^2 - \lambda_i^2} \left[\gb_i \otimes \gb_j \otimes \gb_i \otimes \gb_j + \gb_i \otimes \gb_j \otimes \gb_j \otimes \gb_i \right],
% \end{equation}
% where $\gb$ are the eigenvectors associated with the principal stretches.
% % \COMMENT{RJ: assocated with whatt? fiber aligned directions?}
The design parameters chosen for this RVE were again the ratio of shear moduli, $\frac{\mu_1}{\mu_2}$, and the volume fraction $\phi$. The ratio of shear moduli was further parameterized as $\log (\frac{\mu_1}{\mu_2})$ to be able to cover a wider design space. 50 sets from $\left(\phi, \log\left(\frac{\mu_1}{\mu_2}\right)\right) \in [0.15,0.35] \times [-2,2]$ were sampled via Latin Hypercube Sampling. $N_F=42\cdot 20$ deformation gradients were sampled using the same technique as for the single inclusion RVE. However, transverse isotropy was used here for filtering redundant states.
In any case, we end up with a dataset of the form $\lbrace \log (\frac{\mu_1}{\mu_2})_i, {\phi}_i, \Cb_j, \Sb_j\rbrace$ where $i=1,\ldots,50$ and $j=1,\ldots,N_F$ with the preferred direction $\bm{n}_{1} = (0,0,1)$.

\section{Results}
In this section, we present the results obtained for the different datasets generated in the previous section. We approach the problem systematically, with increasing levels of complexity. First, we consider the synthetic data obtained on the macroscale with the (an)isotropy class and preferred directions, if any, known beforehand. 
The second case also deals with macroscopic data, but with unknown anisotropy class and orientation, both of which are learned during the forward process. Then, we move on to the microscopic data, starting off with the RVE with a single spherical inclusion before presenting the results for the RVE with unidirectional fibers. The inverse problem in all these cases is to find a set of design variables $\mathcal{D}^\star$ that result in a targeted stress response, i.e., $\mathcal{T}^\star=\Sb^\star(\Cb^\star,\mathcal{D}^\star)$. Lastly, we embed the RVE with unidirectional fibers inside a beam implicitly, i.e., we follow the decoupled multiscale scheme approach and employ our framework within a finite element solver to solve an inverse problem that would give us design variables $\mathcal{D}^\star$ that lead to targeted maximum Von Mises stress $\sigma_{\text{VM}_\text{max}}$ over all elements. Therefore, in this case, $\mathcal{T}^\star=\sigma_{\text{VM}_\text{max}}(\mathcal{D}^\star)$.

\subsection{Macroscale Data}
\begin{figure}
    \begin{subfigure}{0.5\linewidth}
        \includegraphics[scale=0.35]{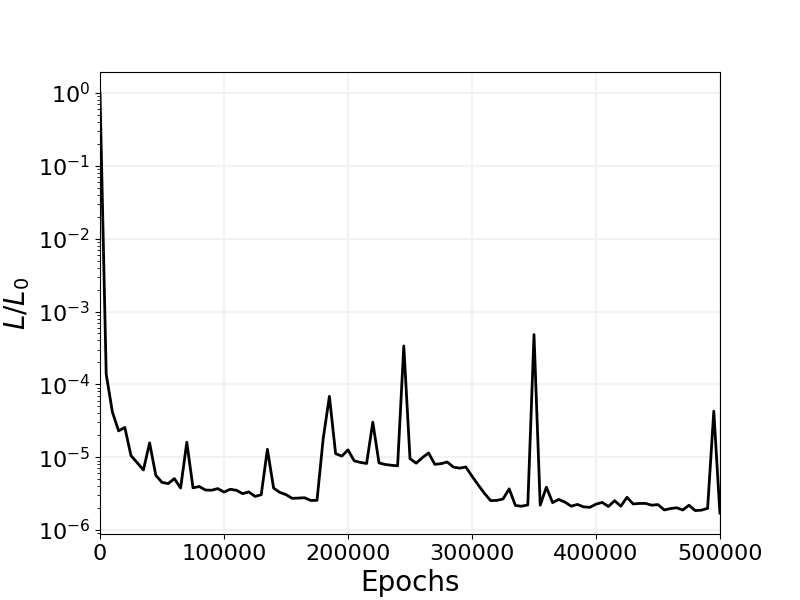}
    \caption{Loss for the forward problem}\label{fig:case1_iso_loss}
    \end{subfigure}
        \begin{subfigure}{0.5\linewidth}
        \includegraphics[scale=0.35]{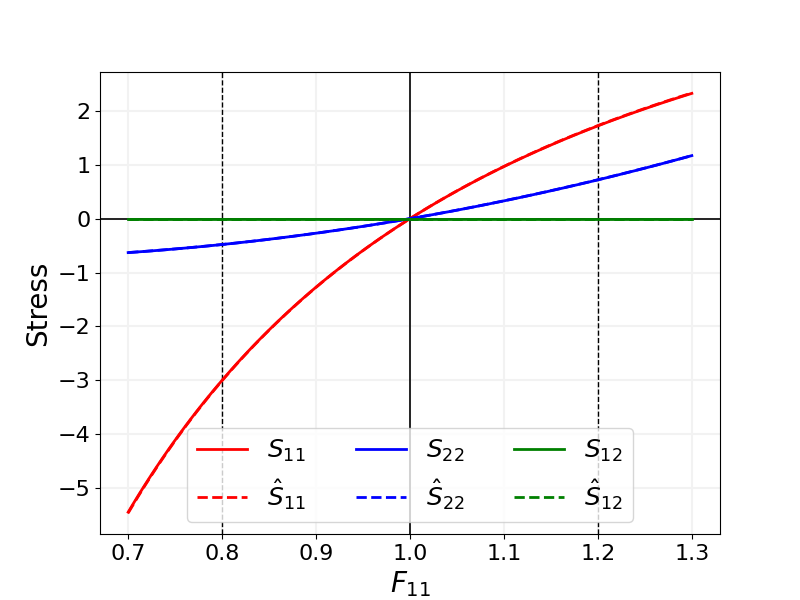}
    \caption{Stress comparison}\label{fig:case1_iso_stress}
    \end{subfigure}
    \begin{center}
            \begin{subfigure}{0.5\linewidth}
        \includegraphics[scale=0.35]{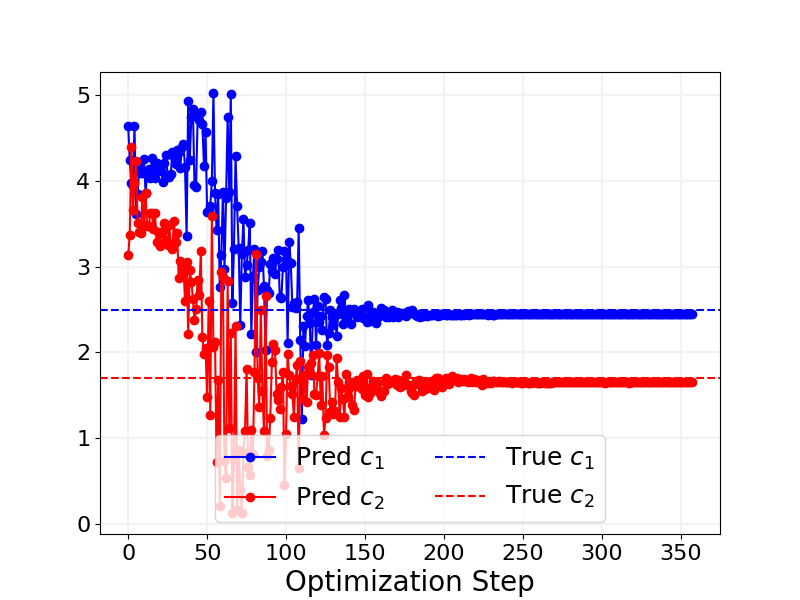}
    \caption{Inverse problem}\label{fig:case1_iso_inverse}
    \end{subfigure}
    \end{center}
    \caption{Results for data with known class and preferred direction from the \textbf{isotropic} material model of  Section \ref{sec:macrodata} (a) Training loss over epochs, (b) true (solid lines) and predicted stresses (dashed lines) over $F_{11}$ on material parameters seen during training and (c) true (dashed lines) and predicted parameters (solid lines with markers) over optimization iterations.}
    \label{iso_case1}
\end{figure}

\begin{figure}
    \begin{subfigure}{0.5\linewidth}
        \includegraphics[scale=0.35]{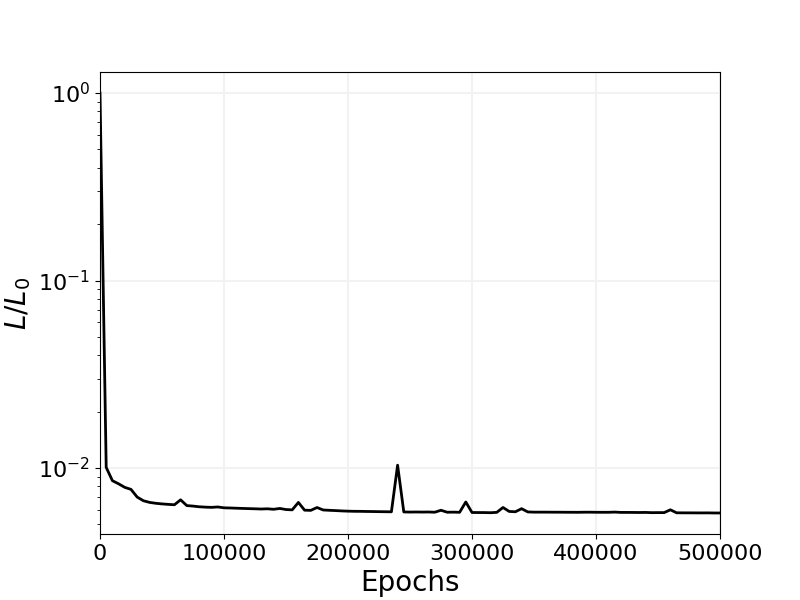}
    \caption{Loss for the forward problem}\label{fig:case1_tran_loss}
    \end{subfigure}
        \begin{subfigure}{0.5\linewidth}
        \includegraphics[scale=0.35]{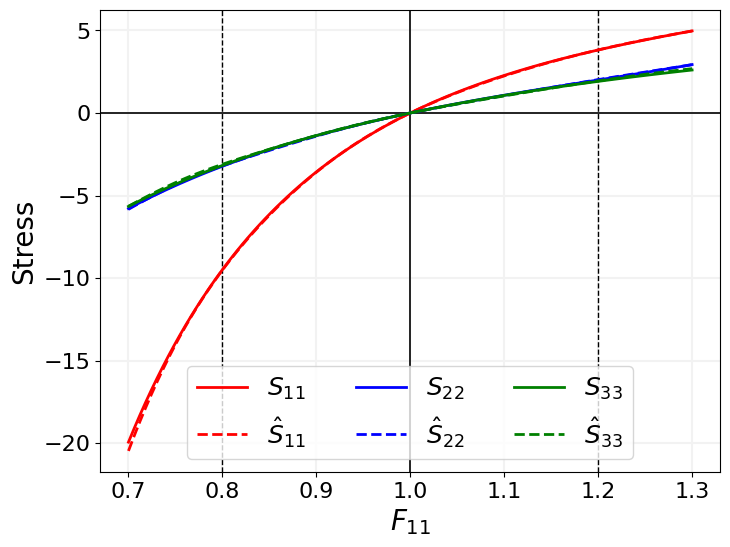}
    \caption{Stress comparison}\label{fig:case1_tran_stress}
                \end{subfigure}
            \begin{subfigure}{0.5\linewidth}
        \includegraphics[scale=0.35]{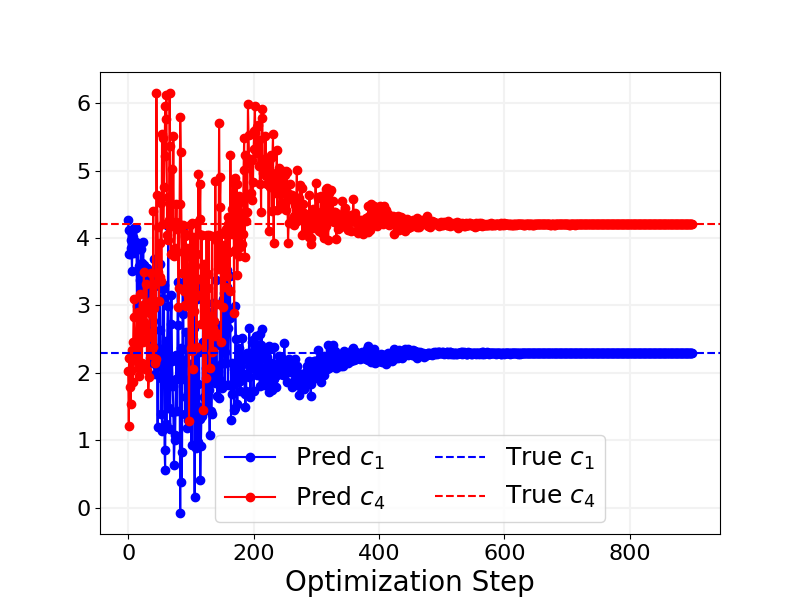}
    \caption{Inverse problem}\label{fig:case1_tran_inverse1}
    \end{subfigure}
                \begin{subfigure}{0.5\linewidth}
        \includegraphics[scale=0.35]{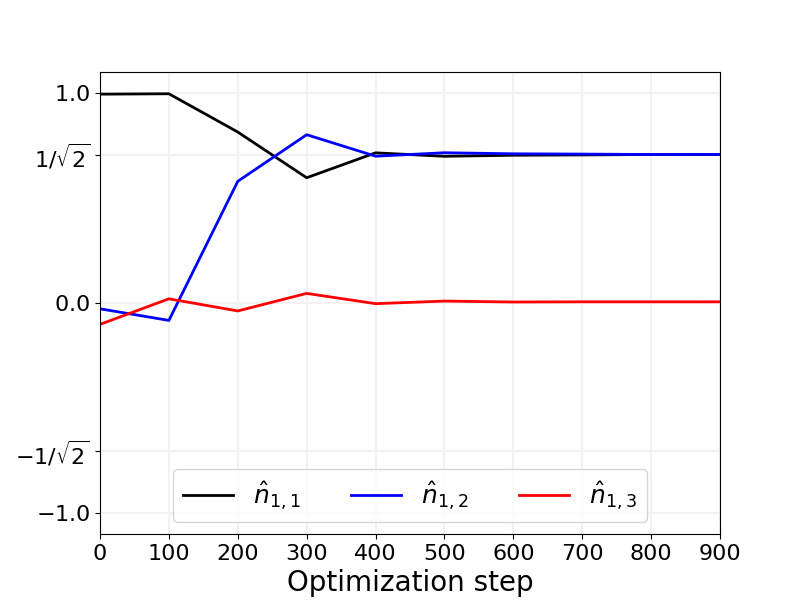}
    \caption{Inverted preferred direction}\label{fig:case1_tran_inverse2}
    \end{subfigure}

    \caption{Results for the data with known class and preferred direction from the \textbf{transversely isotropic} material model of  Section \ref{sec:macrodata} (a) Training loss over epochs, (b) true (solid lines) and predicted stresses (dashed lines) over $F_{11}$ on material parameters seen during training, (c) true (dashed lines) and predicted parameters (solid lines with markers) over optimization iterations and (d) the inverted preferred direction.}
    \label{trans_case1}
\end{figure}

\begin{figure}
    \begin{subfigure}{0.5\linewidth}
        \includegraphics[scale=0.35]{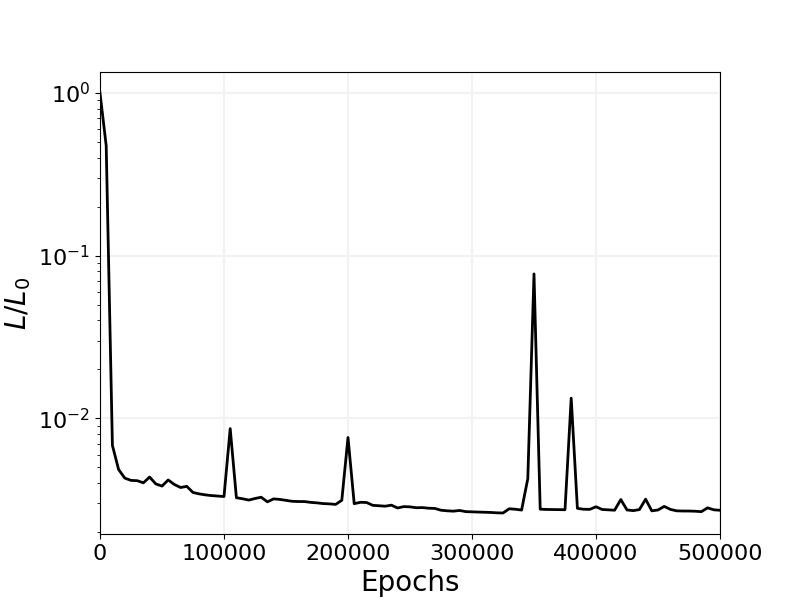}
    \caption{Loss for the forward problem}\label{fig:case1_ortho_loss}
    \end{subfigure}
        \begin{subfigure}{0.5\linewidth}
        \includegraphics[scale=0.35]{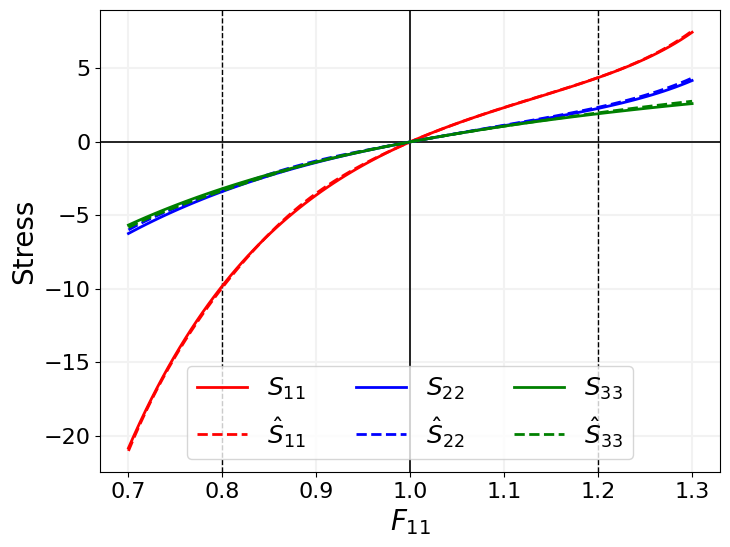}
    \caption{Stress comparison}\label{fig:case1_ortho_stress}
    \end{subfigure}
        \begin{subfigure}{0.5\linewidth}
        \includegraphics[scale=0.35]{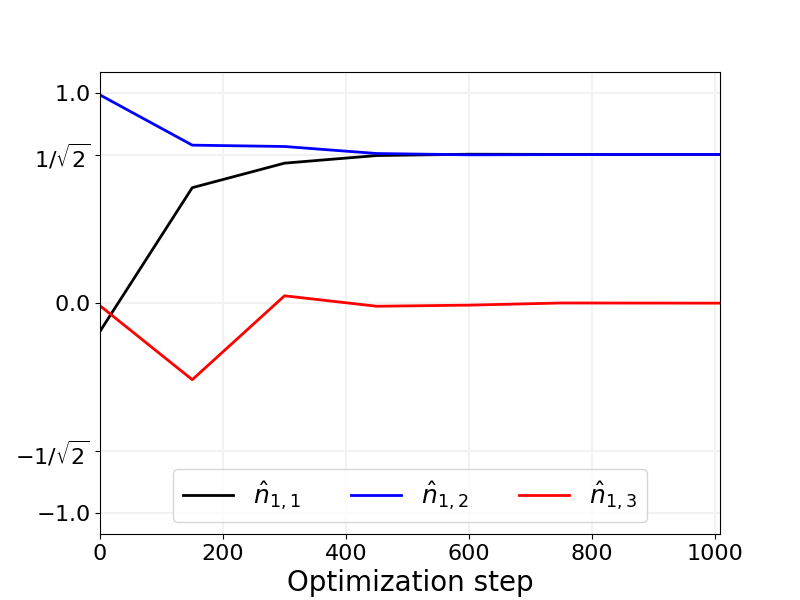}
    \caption{Inverted preferred direction 1}\label{fig:case1_ortho_invDir1}
    \end{subfigure}
        \begin{subfigure}{0.5\linewidth}
        \includegraphics[scale=0.35]{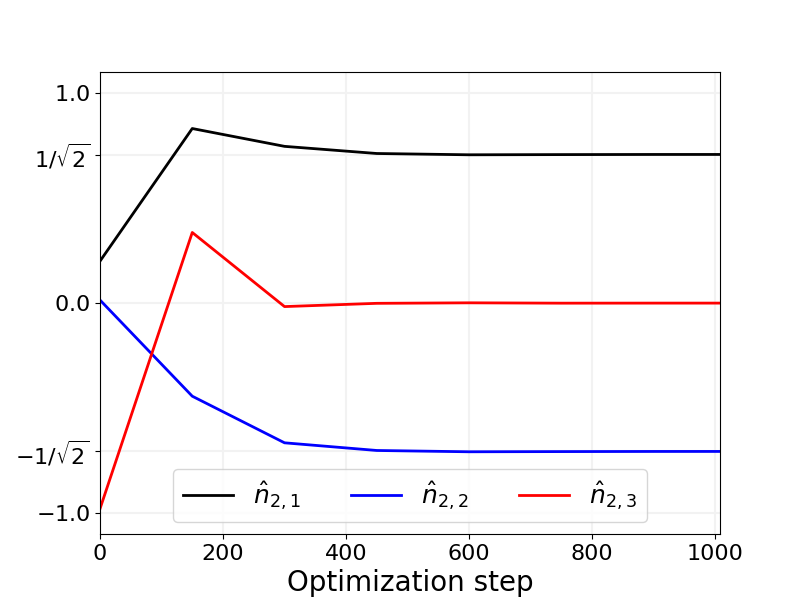}
    \caption{Inverted preferred direction 2}\label{fig:case1_ortho_invDir2}
    \end{subfigure}
    \begin{center}
            \begin{subfigure}{0.5\linewidth}
        \includegraphics[scale=0.35]{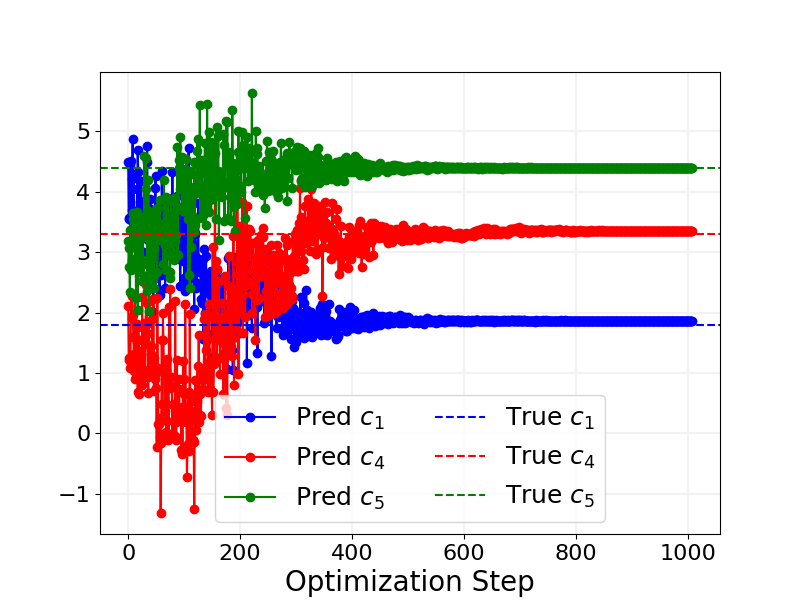}
    \caption{Inverse problem}\label{fig:case1_ortho_inverse}
    \end{subfigure}
    \end{center}
    \caption{Results for the data with known class and preferred direction from the \textbf{orthotropic} material model of  Section \ref{sec:macrodata} (a) Training loss over epochs, (b) true (solid lines) and predicted stresses (dashed lines) over $F_{11}$ on material parameters seen during training and (c) true (dashed lines) and predicted parameters (solid lines with markers) over optimization iterations.}
    \label{ortho_case1}
\end{figure}

% \COMMENT{RJ: the design problem here is to determine the model coeffients/anisotropy? I'm not sure I understand the goal other than verify that the training for the forward problem works?}
For the macroscopic data, we first consider the simple case where we assume to know \textit{a priori} the anisotropy class and orientation. Figure \ref{iso_case1} contains the results for an isotropic model, with figure \ref{fig:case1_iso_loss} showing the evolution of the loss function during training of the forward problem. Figure \ref{fig:case1_iso_stress} shows the uniaxial stress response of the trained model along with the true response. This confirms the ability of the trained model to generalize to unseen strain states. Figure \ref{fig:case1_iso_inverse} shows the results for the inverse problem for this specific material model. Dashed lines show the true design parameters, which were not seen by the network during the training process. Figures \ref{trans_case1} and \ref{ortho_case1} show the same for transversely isotropic and orthotropic datasets respectively. However, in this case, the stress-strain tuples for the inverse problem are generated with different preferred directions than the ones model was trained on, i.e., $\bm{n}_{1} = (\frac{1}{\sqrt{2}},\frac{1}{\sqrt{2}},0)$, $\bm{n}_{2} = (\frac{1}{\sqrt{2}},-\frac{1}{\sqrt{2}},0)$. Figures \ref{fig:case1_tran_inverse1} and \ref{fig:case1_tran_inverse2} show respectively the inverted material parameters and the preferred direction for the transversely isotropic case whereas Figures \ref{fig:case1_ortho_invDir1} and \ref{fig:case1_ortho_invDir2} show the inverted preferred directions for the orthotropic dataset while the inverted material parameters are shown in Figure \ref{fig:case1_ortho_inverse}. Both the forward and inverse problems give satisfactory results.

\begin{figure}
    \begin{subfigure}{0.5\linewidth}
        \includegraphics[scale=0.35]{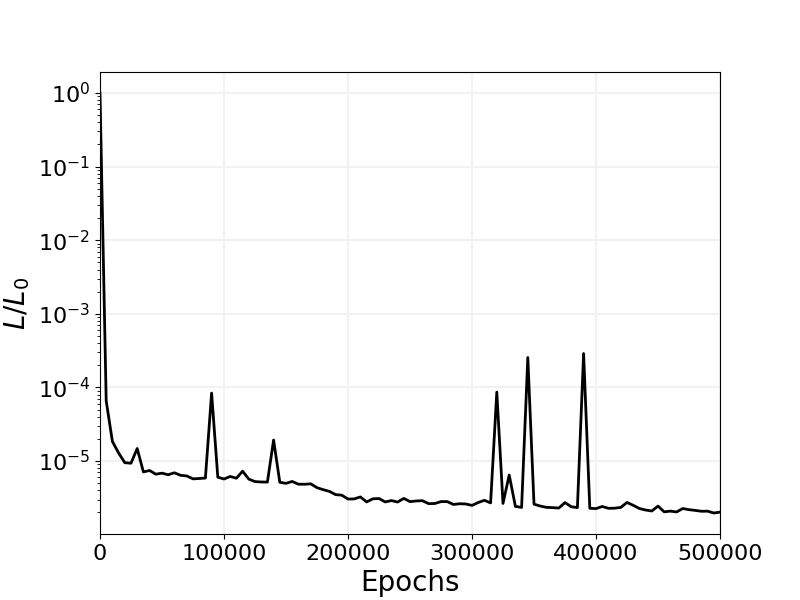}
    \caption{Loss for the forward problem}\label{fig:case2_iso_loss}
    \end{subfigure}
                \begin{subfigure}{0.5\linewidth}
        \includegraphics[scale=0.35]{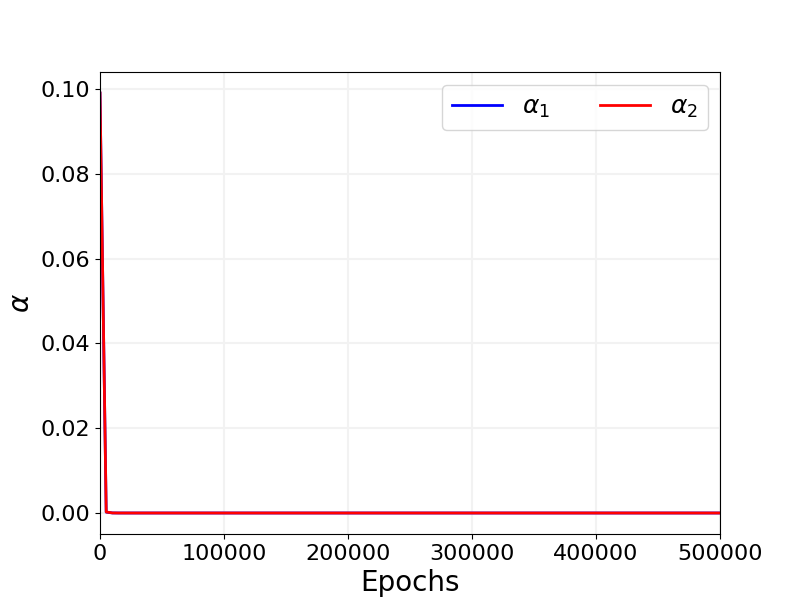}
    \caption{Anisotropic coefficients}\label{fig:case2_iso_alpha}
    \end{subfigure}
        \begin{subfigure}{0.5\linewidth}
        \includegraphics[scale=0.35]{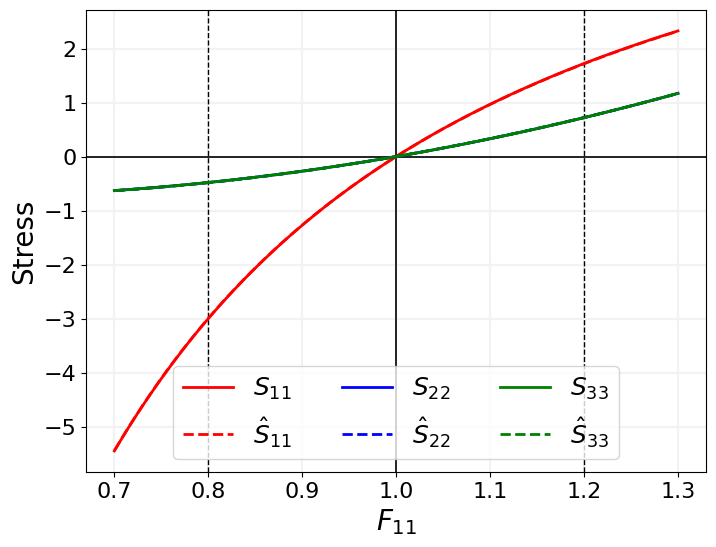}
    \caption{Stress comparison}\label{fig:case2_iso_stress}
    \end{subfigure}
            \begin{subfigure}{0.5\linewidth}
        \includegraphics[scale=0.35]{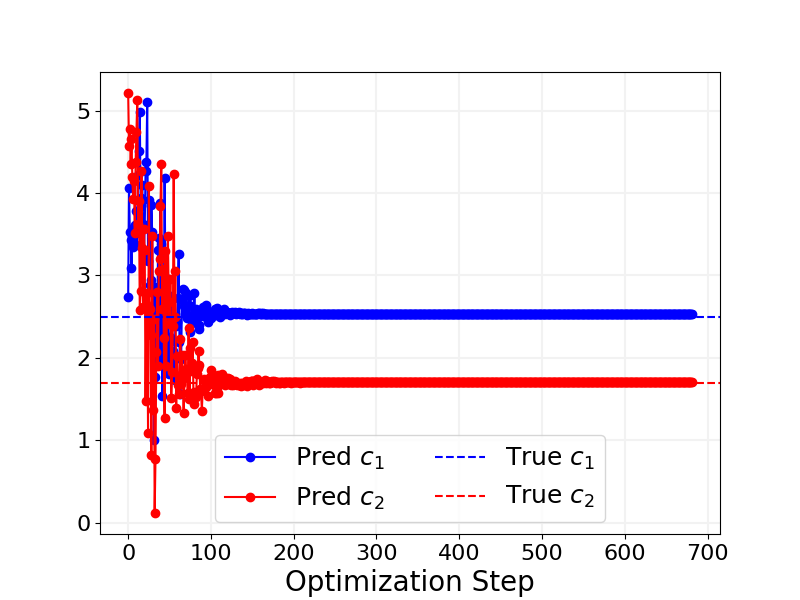}
    \caption{Inverse problem}\label{fig:case2_iso_inverse}
    \end{subfigure}

    \caption{Results for data with unknown class and preferred direction from the \textbf{isotropic} material model of  Section \ref{sec:macrodata} (a) Training loss over epochs, (b) Evolution of the anisotropic coefficients, (c) true (solid lines) and predicted stresses (dashed lines) over $F_{11}$ on material parameters seen during training and (d) true (dashed lines) and predicted parameters (solid lines with markers) over optimization iterations.}
    \label{iso_case2}
\end{figure}

\begin{figure}
    \begin{subfigure}{0.5\linewidth}
        \includegraphics[scale=0.35]{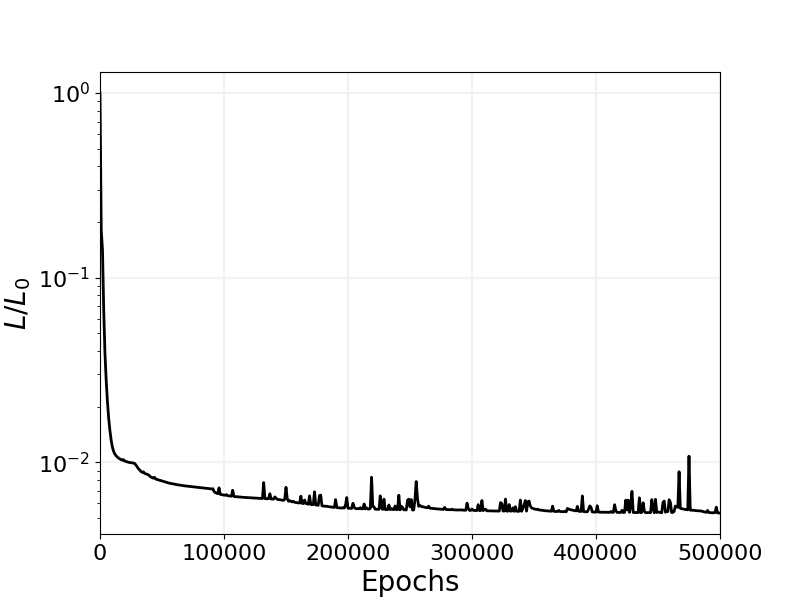}
    \caption{Loss for the forward problem}\label{fig:case2_tran_loss}
    \end{subfigure}
                \begin{subfigure}{0.5\linewidth}
        \includegraphics[scale=0.35]{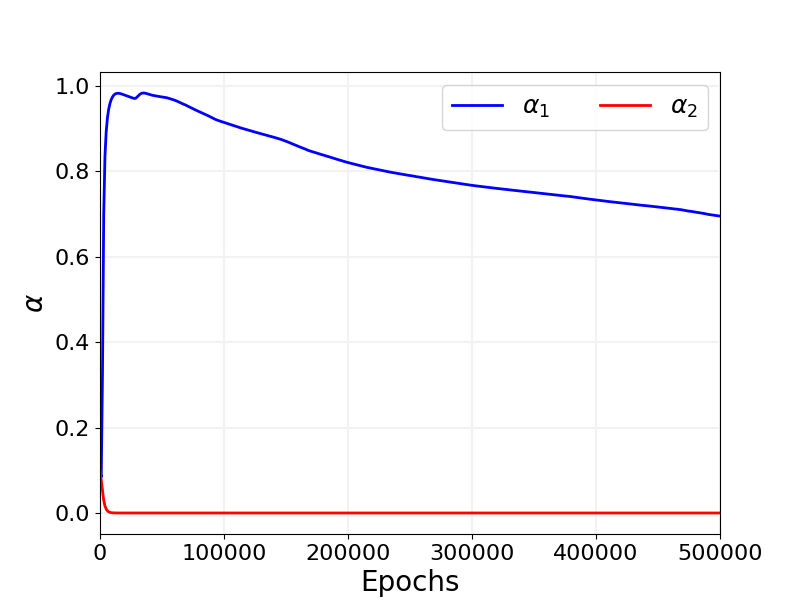}
    \caption{Anisotropic coefficients}\label{fig:case2_tran_alpha}
    \end{subfigure}
        \begin{subfigure}{0.5\linewidth}
        \includegraphics[scale=0.35]{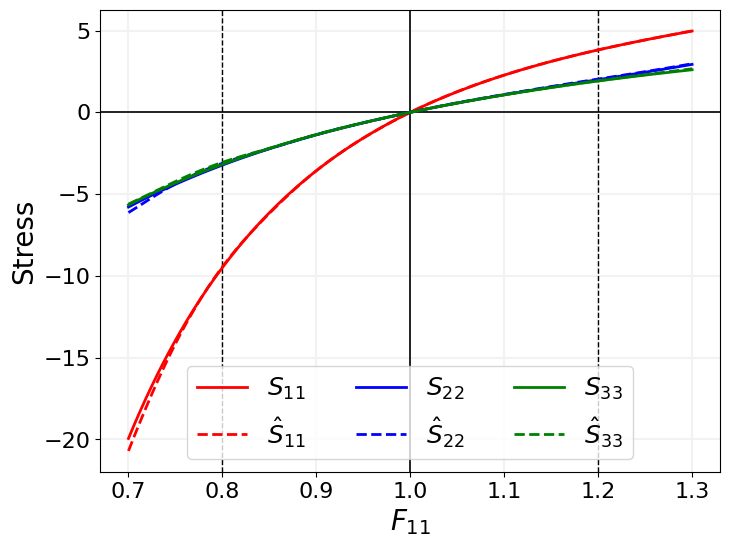}
    \caption{Stress comparison}\label{fig:case2_tran_stress}
    \end{subfigure}
                \begin{subfigure}{0.5\linewidth}
        \includegraphics[scale=0.35]{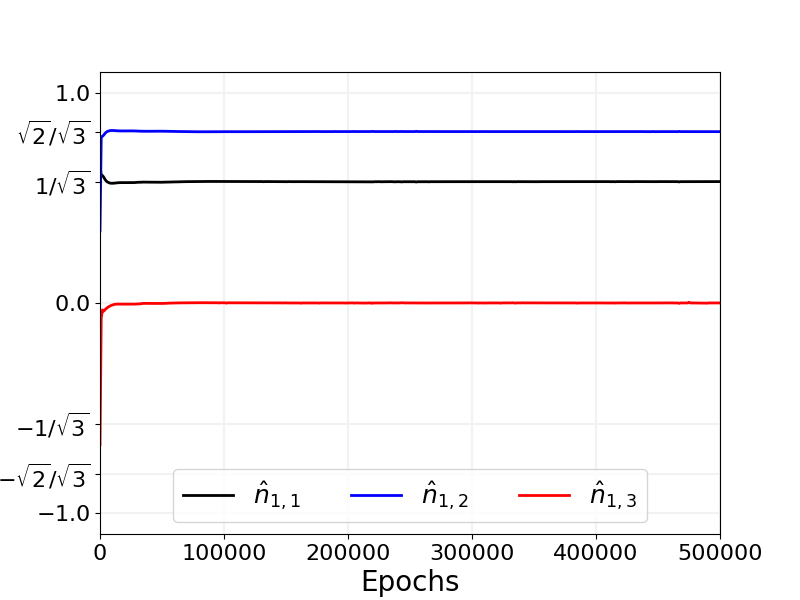}
    \caption{Preferred direction}\label{fig:case2_tran_prefdir1}
    \end{subfigure}
     \begin{subfigure}{0.5\linewidth}
        \includegraphics[scale=0.35]{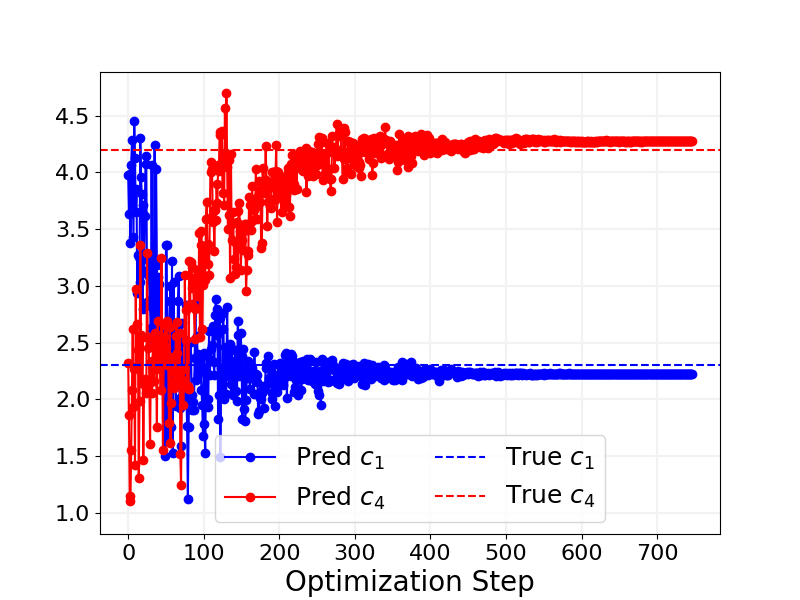}
    \caption{Inverse problem}\label{fig:case2_tran_inverse}
    \end{subfigure}
            \begin{subfigure}{0.5\linewidth}
        \includegraphics[scale=0.35]{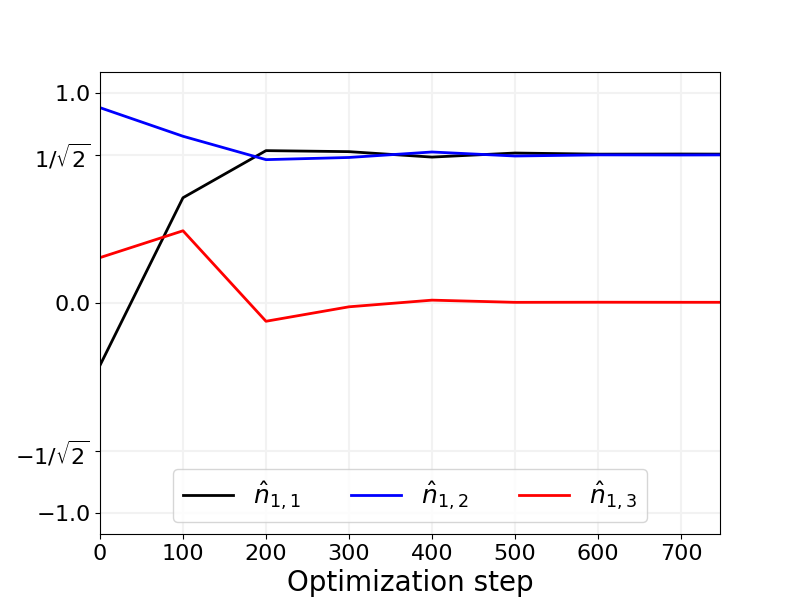}
    \caption{Inverted preferred direction}\label{fig:case2_tran_inversePrefDir}
    \end{subfigure}

    \caption{Results for data with unknown class and preferred direction from the \textbf{transversely isotropic} material model of  Section \ref{sec:macrodata} (a) Training loss over epochs, (b) Evolution of the anisotropic coefficients, (c) true (solid lines) and predicted stresses (dashed lines) over $F_{11}$ on material parameters seen during training, (d) recovered direction of anisotropy, (e) true (dashed lines) and predicted parameters (solid lines with markers) over optimization iterations and (f) the inverted preferred direction for unseen data.}
    \label{trans_case2}
\end{figure}

\begin{figure}
    \begin{subfigure}{0.5\linewidth}
        \includegraphics[scale=0.33]{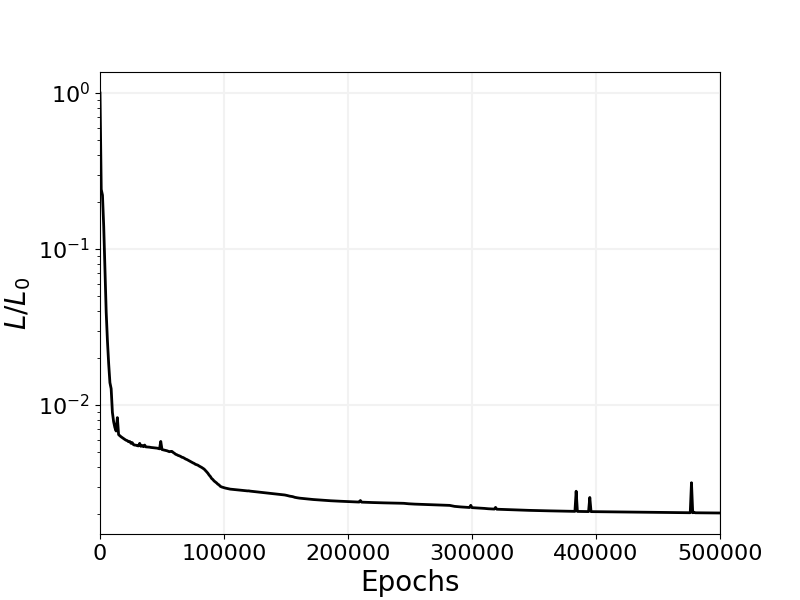}
    \caption{Loss for the forward problem}\label{fig:case2_ortho_loss}
    \end{subfigure}
                \begin{subfigure}{0.5\linewidth}
        \includegraphics[scale=0.33]{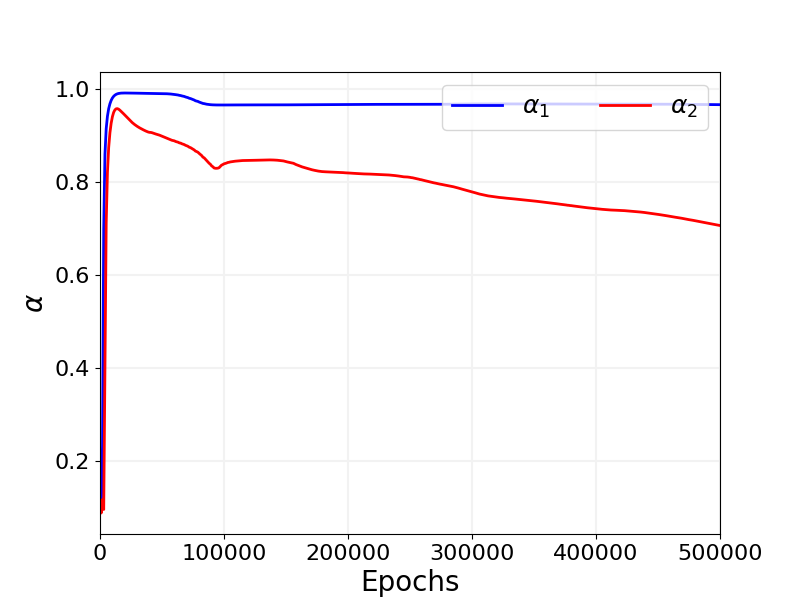}
    \caption{Anisotropic coefficients}\label{fig:case2_ortho_alpha}
    \end{subfigure}
        \begin{subfigure}{0.5\linewidth}
        \includegraphics[scale=0.33]{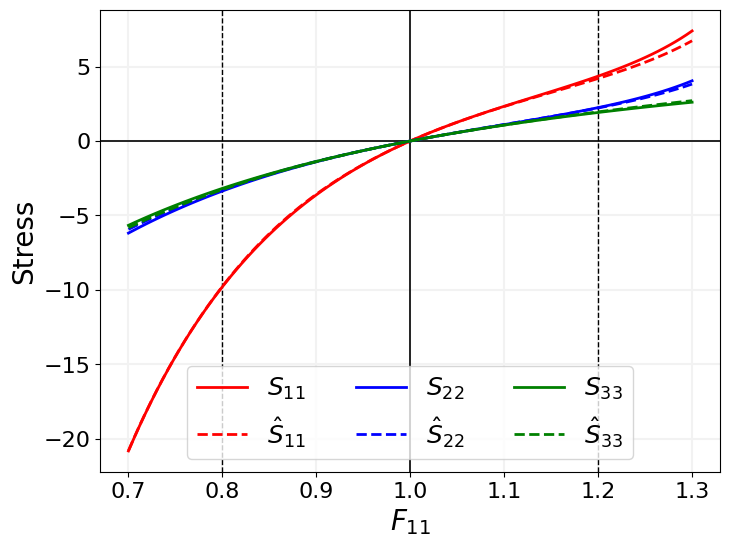}
    \caption{Stress comparison}\label{fig:case2_ortho_stress}
    \end{subfigure}
                \begin{subfigure}{0.5\linewidth}
        \includegraphics[scale=0.33]{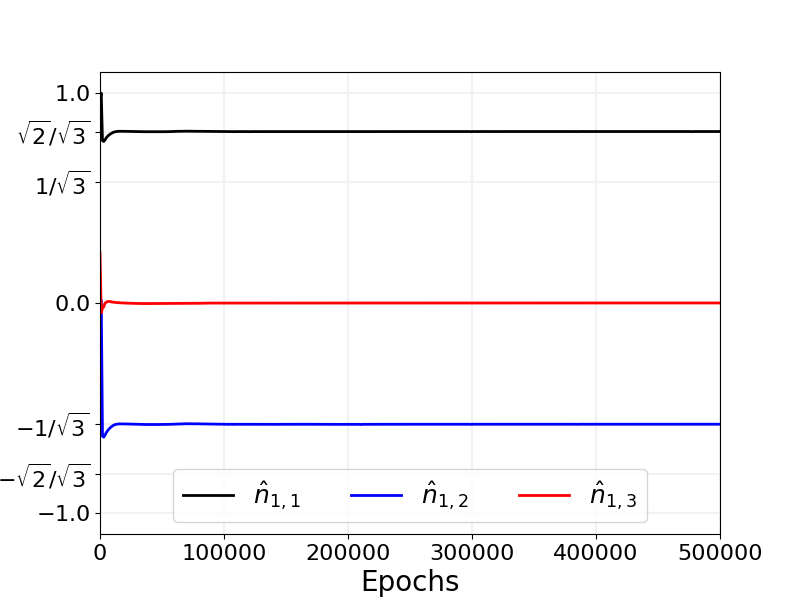}
    \caption{Preferred direction 1}\label{fig:case2_ortho_prefdir1}
    \end{subfigure}
    \begin{center}

                    \begin{subfigure}{0.33\linewidth}
        \includegraphics[scale=0.33]{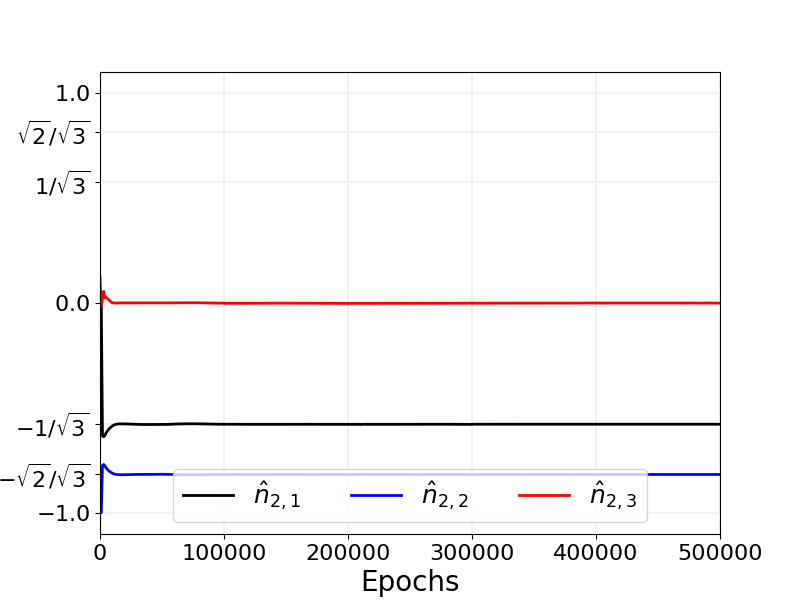}
    \caption{Preferred direction 2}\label{fig:case2_ortho_prefdir2}
    \end{subfigure}
        
    \end{center}

    \caption{Results for the forward problem for data with unknown class and preferred direction from the \textbf{orthotropic} material model of  Section \ref{sec:macrodata} (a) Training loss over epochs, (b) Evolution of the anisotropic coefficients, (c) true (solid lines) and predicted stresses (dashed lines) over $F_{11}$ on material parameters seen during training, and (d,e) recovered directions of anisotropy.}
    \label{ortho_case2_forward}
\end{figure}

\begin{figure}
                    \begin{subfigure}{0.5\linewidth}
        \includegraphics[scale=0.33]{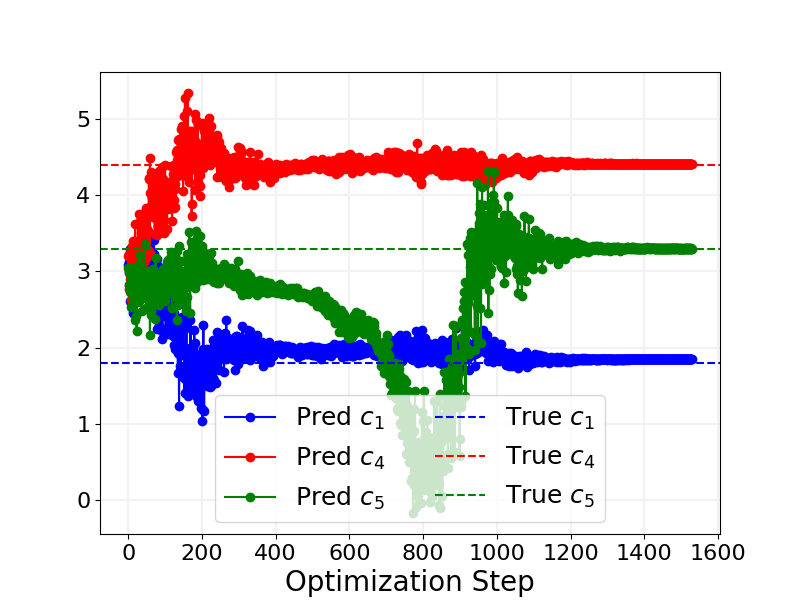}
    \caption{Inverse problem}\label{fig:case2_ortho_inverse}
    \end{subfigure}
                        \begin{subfigure}{0.5\linewidth}
        \includegraphics[scale=0.33]{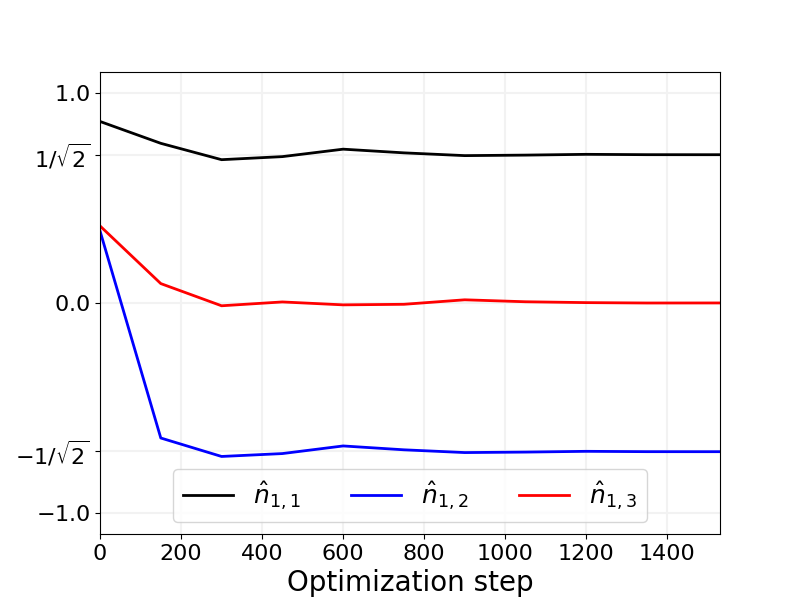}
    \caption{Inverted preferred direction 1}\label{fig:case2_ortho_invPrefDir1}
    \end{subfigure}
    \begin{center}
                    \begin{subfigure}{0.5\linewidth}
        \includegraphics[scale=0.33]{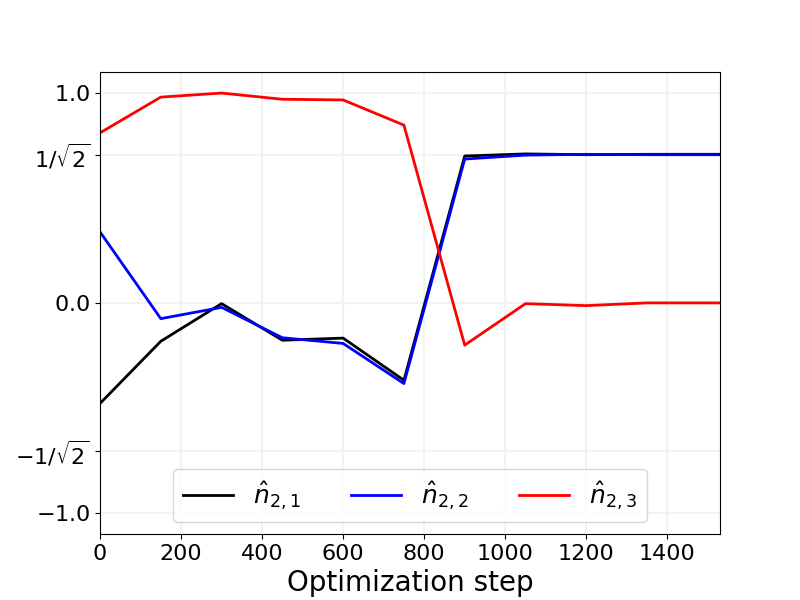}
    \caption{Inverted preferred direction 2}\label{fig:case2_ortho_invPrefDir2}
    \end{subfigure}
    \end{center}

    \caption{Results for the inverse problem for data with unknown class and preferred direction from the \textbf{orthotropic} material model of  Section \ref{sec:macrodata} (a) True (dashed lines) and predicted parameters (solid lines with markers) over optimization iterations, and (b,c) inverted preferred directions.}
    \label{ortho_case2_inverse}
\end{figure}

We then move on to the next case where we do not know the anisotropy class and the preferred direction and want to find them during the forward process. The same dataset with different preferred directions than those used in training was utilized for the inverse problems. Figure \ref{iso_case2} shows the results obtained for this case with the model trained on isotropic data. It can be seen in Figure \ref{fig:case2_iso_alpha} that both the anisotropic coefficients go to zero, signifying isotropy. Figures \ref{fig:case2_iso_stress} and \ref{fig:case2_iso_inverse} show the uniaxial stress fit and the inverted design parameters respectively. The results for the transversely isotropic dataset are presented in Figure \ref{trans_case2}. The subfigure \ref{fig:case2_tran_alpha} shows the evolution of anisotropic coefficients. It can be seen that one of them goes to zero and the other is close to 1, implying a transversely isotropic behavior as expected. Figure \ref{fig:case2_tran_prefdir1} shows the preferred direction recovered from the network parameters $\pb$ and $\varphi$ for the training dataset. The solution of the inverse problem is shown in figure \ref{fig:case2_tran_inverse} and we see that the model is able to accurately predict the design parameters and \ref{fig:case2_tran_inversePrefDir} gives the inverted preferred direction for our testing dataset. Figures \ref{ortho_case2_forward} and \ref{ortho_case2_inverse} give the results for training the model with an orthotropic dataset. Both the anisotropic coefficients attain nonzero values as can be seen in figure \ref{fig:case2_ortho_alpha} indicating an orthotropic response and the existence of two preferred directions. Figures \ref{fig:case2_ortho_prefdir1} and \ref{fig:case2_ortho_prefdir2} show these recovered preferred directions which closely align with the true directions the dataset was generated on. Figure \ref{fig:case2_ortho_inverse} shows the optimization steps while solving the inverse problem and the framework is able to correctly predict the material parameters it had not seen during training, whereas Figures \ref{fig:case2_ortho_invPrefDir1} and \ref{fig:case2_ortho_invPrefDir2} give the inverted preferred directions for the testing dataset. Note that it is possible that the network swaps the two preferred directions for the orthotropic dataset. This does not affect the response in any way. However, in such a case, the design parameters are also swapped with the preferred directions. \aref{app:ParamStudy} contains a concise study on the effect of the number of deformation gradient samples.

\subsection{Homogenized Data}
Now we present the results for the microscopic, or multiscale, dataset generated with the free energy and the corresponding stresses presented in Section \ref{sec:microData}. An RVE can exhibit anisotropic behavior upon homogenization even if all the materials are modeled individually with an isotropic hyperelastic free energy. Therefore, we try to qualitatively see if anisotropy exists in our particular RVE. In all cases, an informed guess is made regarding the (an)isotropic behavior of the RVE which will be explained subsequently.

\subsubsection{RVE with a single spherical inclusion}
\begin{figure}
    \begin{subfigure}{0.5\linewidth}
        \includegraphics[scale=0.35]{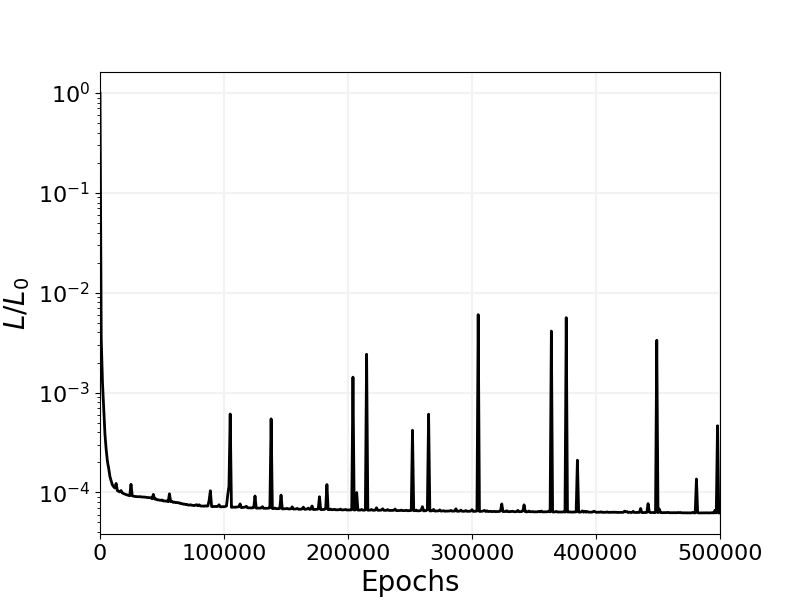}
    \caption{Loss for the forward problem}\label{fig:SingleIncRVE_loss}
    \end{subfigure}
                \begin{subfigure}{0.5\linewidth}
        \includegraphics[scale=0.35]{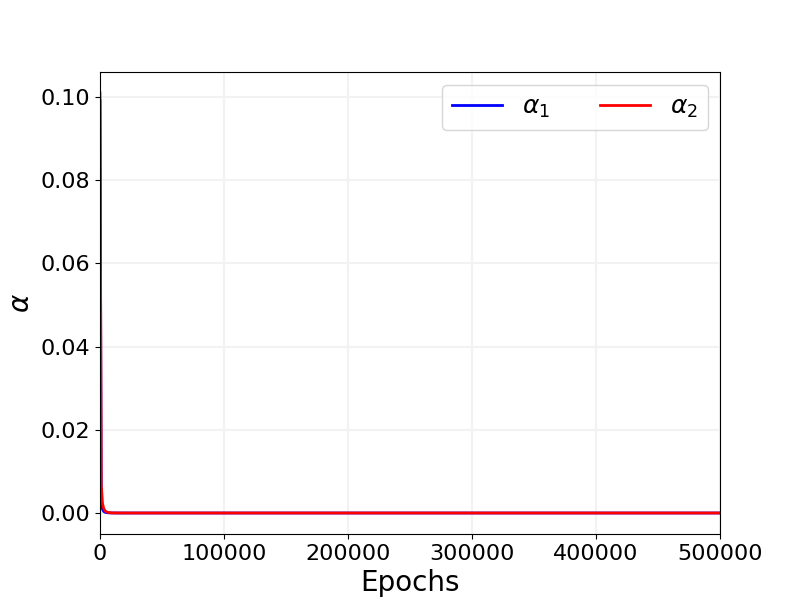}
    \caption{Anisotropic coefficients}\label{fig:SingleIncRVE_alpha}
    \end{subfigure}
        \begin{subfigure}{0.5\linewidth}
        \includegraphics[scale=0.35]{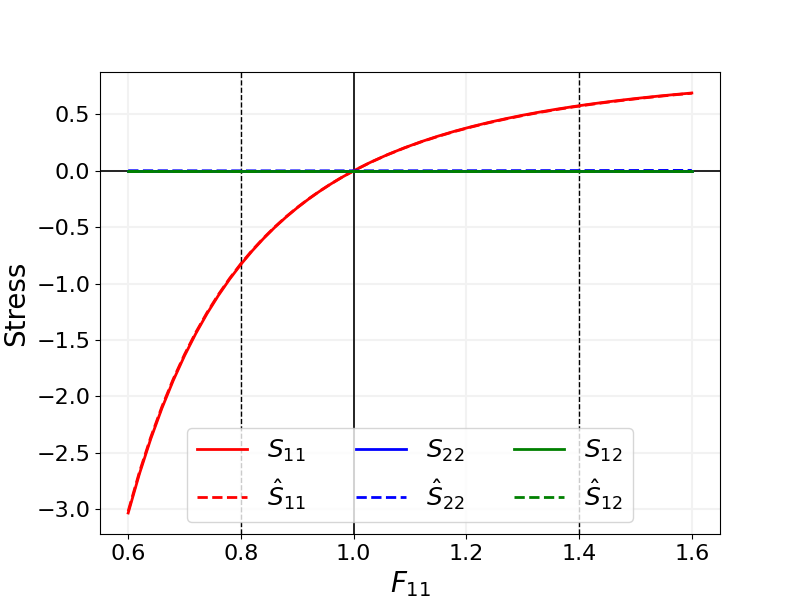}
    \caption{Stress comparison}\label{fig:SingleIncRVE_stress}
    \end{subfigure}
            \begin{subfigure}{0.5\linewidth}
        \includegraphics[scale=0.35]{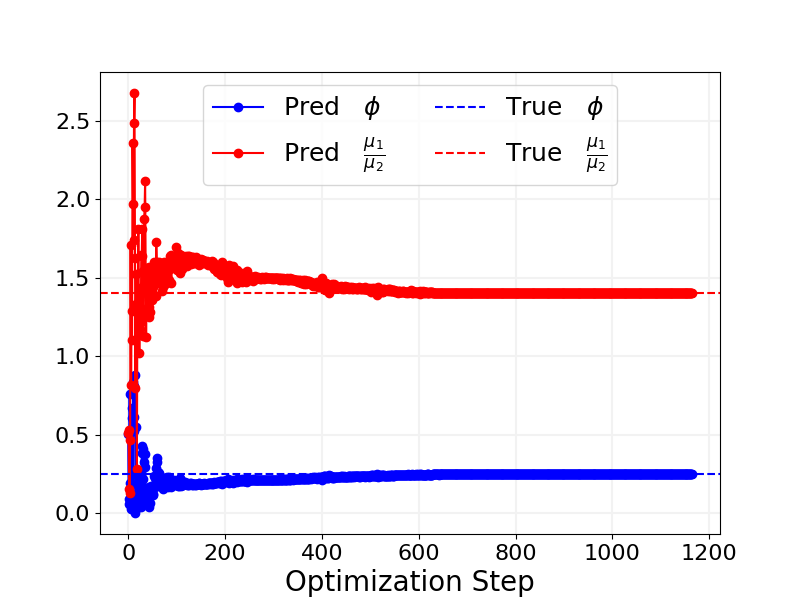}
    \caption{Inverse problem}\label{fig:SingleIncRVE_inverse}
    \end{subfigure}

    \caption{Results for the single, spherical inclusion RVE with unknown class and preferred direction from the material model of  Section \ref{sec:microData} (a) Training loss over epochs, (b) Evolution of the anisotropic coefficients, (c) true (solid lines) and predicted stresses (dashed lines) over $F_{11}$ on material parameters seen during training and (d) true (dashed lines) and predicted parameters (solid lines with markers) over optimization iterations.}
    \label{rve_singleInc}
\end{figure}

An anisotropy index termed the \emph{Zener Index} was presented in \cite{zener1948elasticity} as early as 1948. However, it was only applicable to cubic crystals. The idea was extended decades later to give a \emph{Universal Anisotropy Index} \cite{PhysRevLett.101.055504} that was applicable to all classes of anisotropy. A modified form of the Universal Anisotropy Index was presented in \cite{Sokolowski2018-fn} for the case where material exhibits no internal symmetries. However, all of these anisotropy indices are only applicable to small strain linear elasticity and do not distinguish between the secant and tangent modulus as pointed out in \cite{VlassisMD}. For the case of finite deformations and nonlinear elasticity, we get a fully populated tangent modulus which renders these anisotropy measures inapplicable. Therefore, in this work, we follow the approach presented in \cite{YangHyperelastic} where (an)isotropic response is determined by applying pure shear deformations along different directions and if the material exhibits the same response for any direction shear deformations are applied in, we can conclude that the material is isotropic. In our case, we do this for four combinations of extreme values of the design parameters: 
the radius of the inclusion $R$ and the ratio of shear moduli for the inclusion and the matrix $\frac{\mu_1}{\mu_2}$. 
%\COMMENT{RJ: are these defined?}
The resulting plots are presented in \aref{app:RVEIsotropy} and since the stress response overlaps for all cases, we can expect an isotropic response for the chosen material parameters. It should be noted that RVEs with a single inclusion often show effective cubic properties in the small strain context \cite{vazic2022mechanical}. However, it appears that isotropy holds for the ranges of material and geometric parameters we chose. Figure \ref{rve_singleInc} shows the results obtained for this RVE and the evolution of anisotropic coefficients going to zero, as can be seen in Figure \ref{fig:SingleIncRVE_alpha}. This aligns with our preliminary study that our RVE exhibits isotropic responses. The inverse problem is solved in Figure \ref{fig:SingleIncRVE_inverse} and the model is accurately able to predict both the ratio of shear moduli for the inclusion and the matrix, as well as the radius of the inclusion.

\subsubsection{RVE with unidirectional fibers}\label{sec:FiberRVEResults}
\begin{figure}
    \begin{subfigure}{0.5\linewidth}
        \includegraphics[scale=0.35]{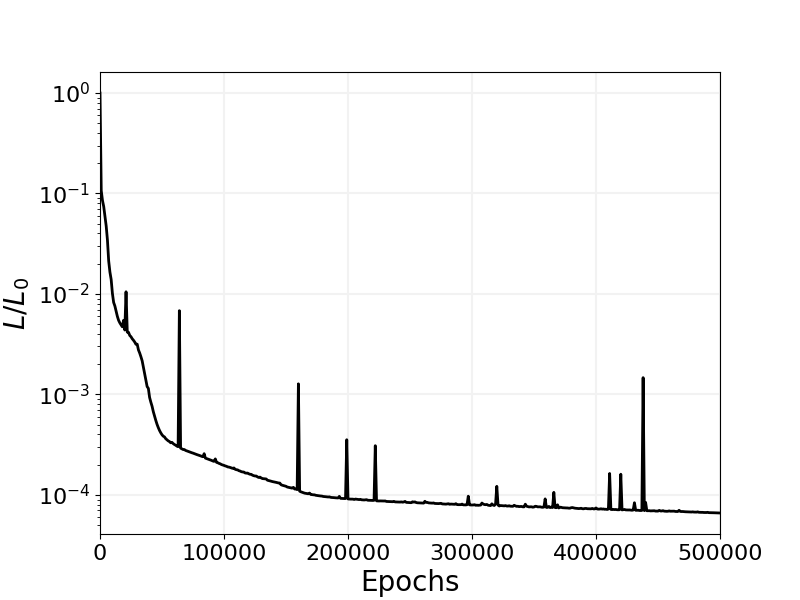}
    \caption{Loss for the forward problem}\label{fig:FiberRVE_loss}
    \end{subfigure}
    \begin{subfigure}{0.5\linewidth}
        \includegraphics[scale=0.35]{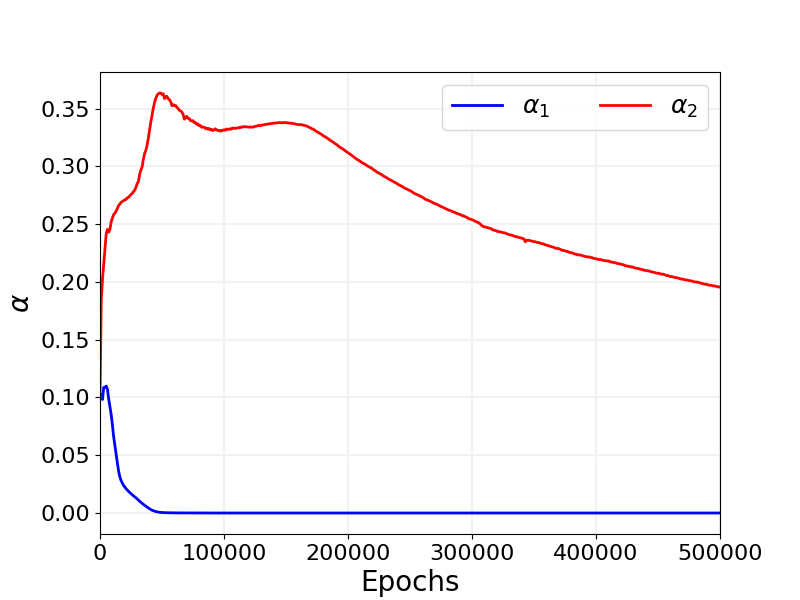}
    \caption{Anisotropic coefficients}\label{fig:FiberRVE_alpha}
    \end{subfigure}
        \begin{subfigure}{0.5\linewidth}
        \includegraphics[scale=0.35]{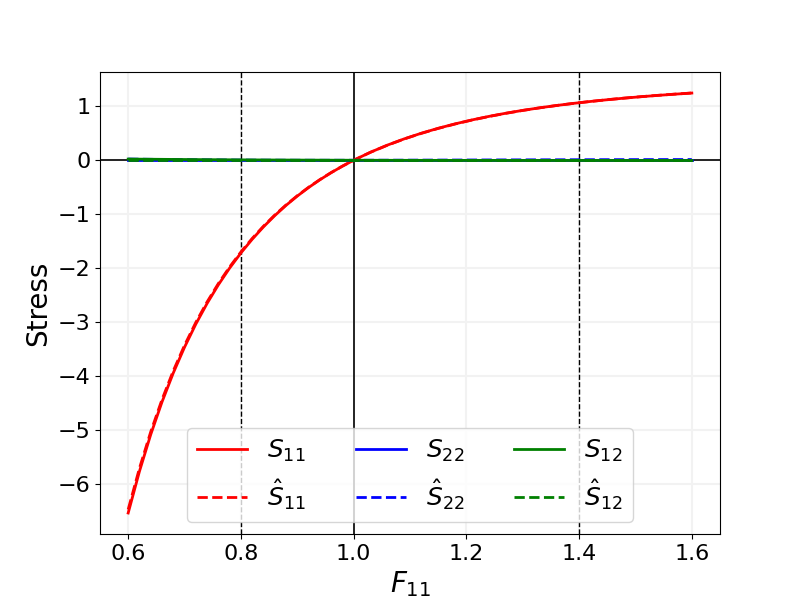}
    \caption{Stress comparison}\label{fig:FiberRVE_stress}
    \end{subfigure}
            \begin{subfigure}{0.5\linewidth}
        \includegraphics[scale=0.35]{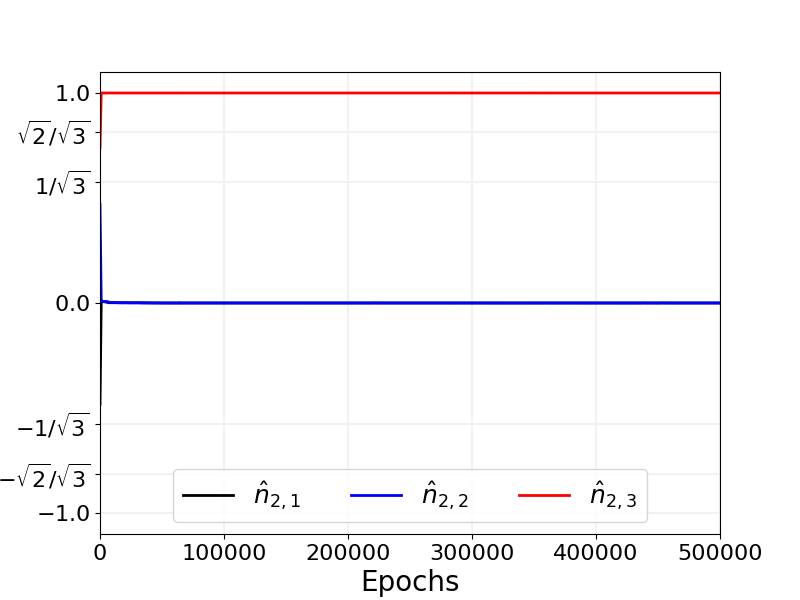}
    \caption{Preferred direction}\label{fig:FiberRVE_prefDirTrain}
    \end{subfigure}
            \begin{subfigure}{0.5\linewidth}
        \includegraphics[scale=0.35]{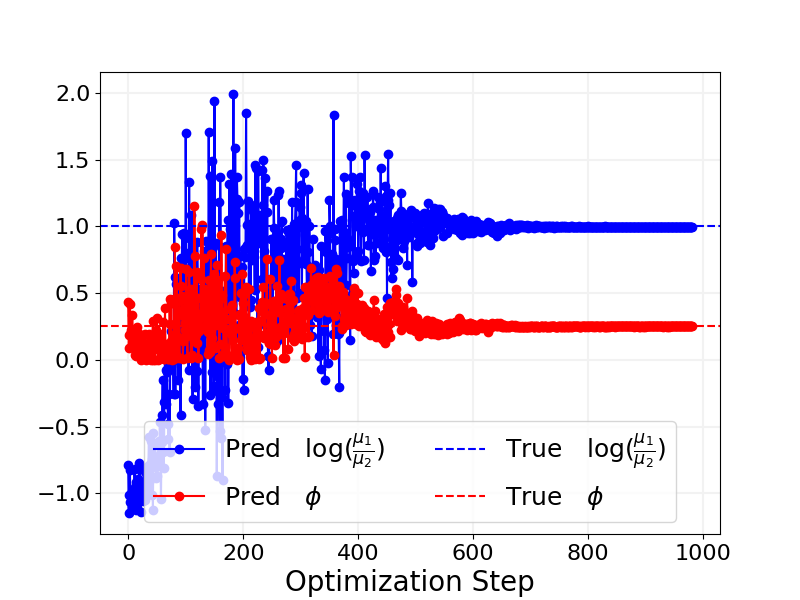}
    \caption{Inverted material parameters}\label{fig:FiberRVE_inverse}
    \end{subfigure}
                \begin{subfigure}{0.5\linewidth}
        \includegraphics[scale=0.35]{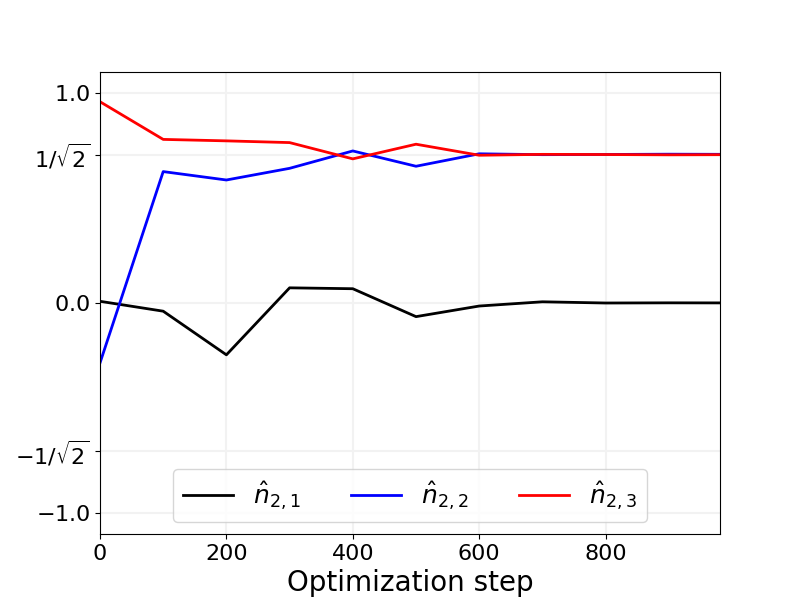}
    \caption{Inverted preferred direction}\label{fig:FiberRVE_prefDir_inverse}
    \end{subfigure}

    \caption{Results for the RVE with unidirectional fibers with unknown class and preferred direction from the material model of  Section \ref{sec:microData} (a) Training loss over epochs, (b) Evolution of the anisotropic coefficients, (c) true (solid lines) and predicted stresses (dashed lines) over $F_{11}$ on material parameters seen during training, (d) Preferred directions learned during training (e) true (dashed lines) and predicted parameters (solid lines with markers) over optimization iterations and (f) the inverted preferred direction.}
    \label{rve_fiber}
\end{figure}

We now present the results for the fiber-reinforced composite material with an RVE containing fibers aligned in one direction. Such alignment results in a transversely isotropic response with the preferred direction of the response being the same as the orientation of the fibers \cite{kalina2023fe}. As explained in Section \ref{sec:microData}, for data generation, the fibers were placed along the direction $\bm{n}_{1} = (0,0,1)$. However, training this dataset with a polyconvex NN is too restrictive as it is possible to lose polyconvexity during the homogenization process \cite{barchiesi2007loss, braides1994loss, kalina2024neuralnetworksmeetanisotropic}. Therefore, we remove the polyconvexity requirements on our network for this dataset. Two methods of doing so are discussed here. The first, and most straightforward, technique would be to remove all the constraints on the neural network, i.e., using arbitrary activation functions and allowing negative weights. Although training this network gives good results as shown in \aref{app:ArbNet}, this comes with a serious drawback. The end goal is to place these trained neural networks in a finite element solver at each integration point. These solvers extract not only the stresses but also the tangent moduli for each \textit{element} to build a global stiffness matrix. Since the solution involves inverting this stiffness matrix, it needs to be positive semi-definite. However, having no constraints on the network would mean the stiffness matrix could violate positive definiteness rendering the inverse impossible. The second method is a bit more involved as it keeps the convexity and monotonicity constraints on the network but violates polyconvexity through the normalization term. Instead of using the normalization term presented in Eq. \eqref{Finalnormalization}, we introduce a different term such that the stress normalization component of free energy reads:

\begin{equation}\label{oC_norm}
\Psi_{sn} = - \tr \left[ (\sum_i \partial_{\bar{\Ic}_i} {\Psi} \, \bar{\Bc}_i) (\Cb-\Ib)\right],
\end{equation}

where the bar represents that both the invariants and the bases are evaluated at $\Cb=\Ib$. Although this modification still results in normalized stresses, we now have a subtraction operator in the free energy and are no longer polyconvex. However, since the tangent modulus requires the second derivative with regard to $\Cb$, this new term does not appear in the tangent modulus and it preserves its positive definiteness. The forward and inverse problems for both these formulations on an anisotropic dataset are presented in \aref{app:nonPolyisotropic}. Employing the latter formulation, Figure \ref{fig:FiberRVE_loss} shows the evolution of loss during training and the fact that one of our anisotropic coefficients goes to zero whereas the other stays high, as shown in Figure \ref{fig:FiberRVE_alpha}, indicates transversely isotropic behavior. The preferred direction for the training data shown in Figure \ref{fig:FiberRVE_prefDirTrain} matches with the true preferred direction. The stress response also agrees with the true one, even for deformation gradients outside the training set as evident from Figure \ref{fig:FiberRVE_stress}. The inverse problem was run on a dataset generated with design parameters not included in the training data along with a different preferred direction, i.e., $\bm{n}_{1} = (0,\frac{1}{\sqrt{2}},\frac{1}{\sqrt{2}})$. The inverted material parameters are shown in Figure \ref{fig:FiberRVE_inverse} whereas the inverted preferred direction for this dataset is shown in Figure \ref{fig:FiberRVE_prefDir_inverse} thus highlighting the framework's capabilities to work for a wide range of data while being trained on only a small subset of it.

\subsection{Inverse problem in a finite element setting}

\begin{figure}%[h]
    \centering
    \includegraphics[width=0.8\textwidth]{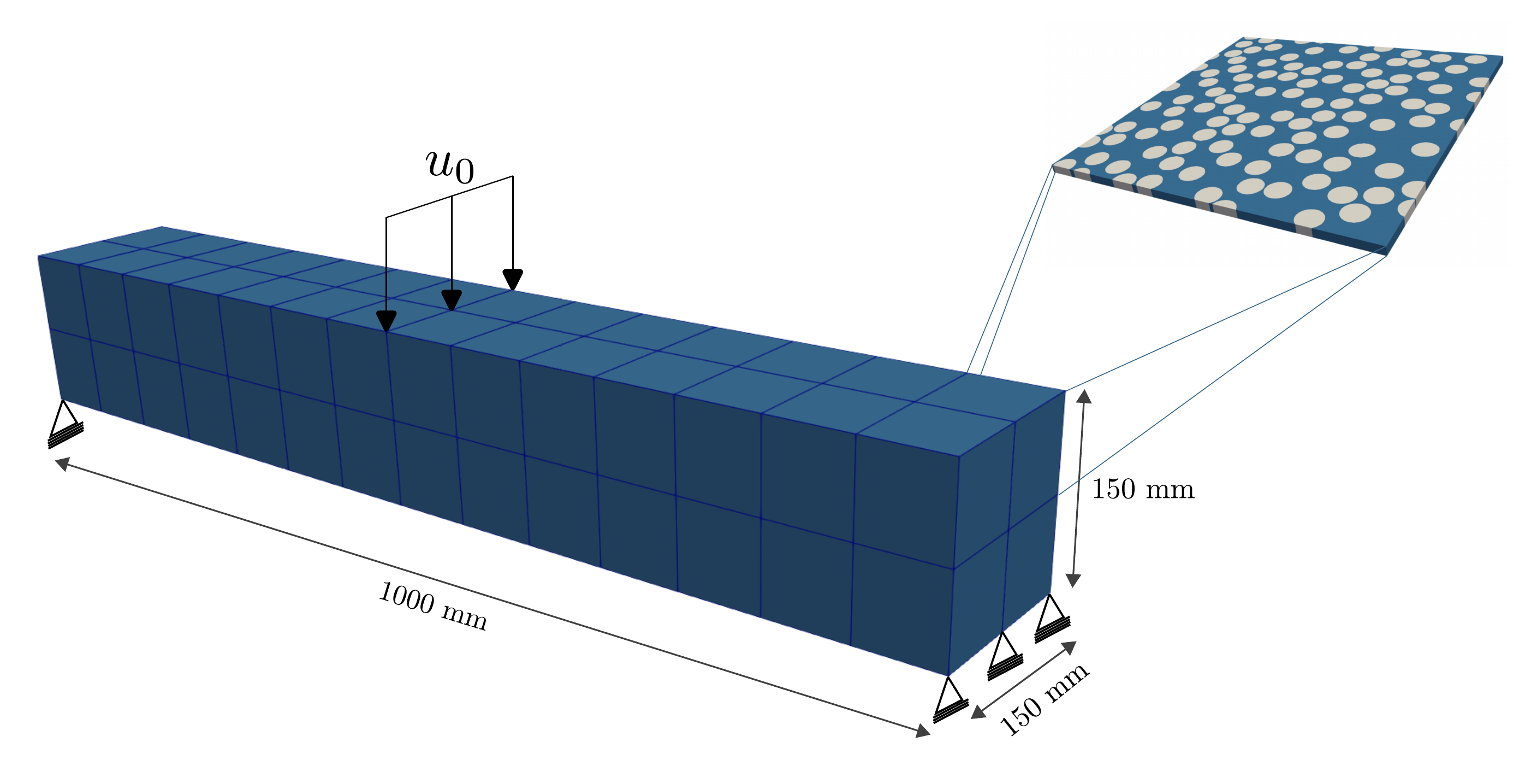}
    \caption{A simply supported beam subjected to displacement $u_0$.}
    \label{fig:FEconfig}
\end{figure}

The framework was also integrated into a finite element solver by adopting a decoupled multiscale scheme approach. This methodology involves employing the pICNNs, trained offline, to give homogenized macroscopic responses at the integration points. Doing so, we then solve the inverse problem to get a targeted macroscopic response. The problem configuration is depicted in Figure \ref{fig:FEconfig} where a simply supported beam is subjected to some displacement and the beam's microstructure is chosen as the RVE with unidirectional fibers from the previous section. Along with the stresses, we also need the tangent modulus which can be calculated as:
\begin{equation}
        \mathbb{C} = 4 \sum_i \left[ \partial_{\Ic_i^2} {\Psi} \left( \partial_{\Cb} {\Ic_i} \otimes \partial_{\Cb} {\Ic_i} \right) + \partial_{\Ic_i} {\Psi} \left( \partial_{\Cb^2} {\Ic_i} \right) \right],
\end{equation}
where the subscript $i$ represents the active invariants based on the anisotropy class learned during training.  The neural network was implemented in an in-house PyTorch-based Finite Element model. We pose the following inverse problem: Find a suitable orientation of the fibers such that it minimizes the maximum Von Mises stress across all elements. To this end, we evaluate five runs of the finite element model with fixed material parameters ($\log (R)=1.5,\, \Phi = 0.25$) and different initial fiber orientations. Our goal is to invert the fiber orientation that gives the lowest Von Mises stress.

\begin{figure}
    \begin{subfigure}{0.5\linewidth}
        \includegraphics[scale=0.35]{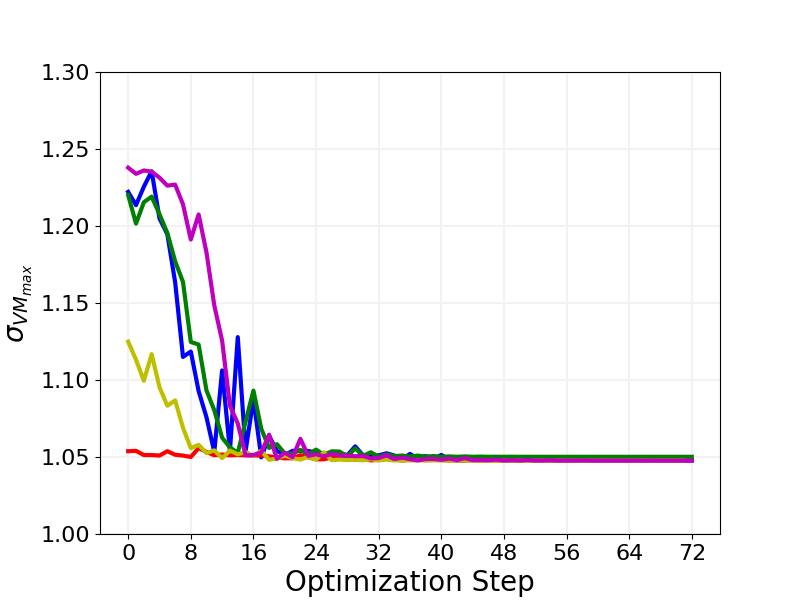}
    \caption{}
    \label{fig:FEstressEvol}
    \end{subfigure}
                \begin{subfigure}{0.5\linewidth}
        \includegraphics[scale=0.35]{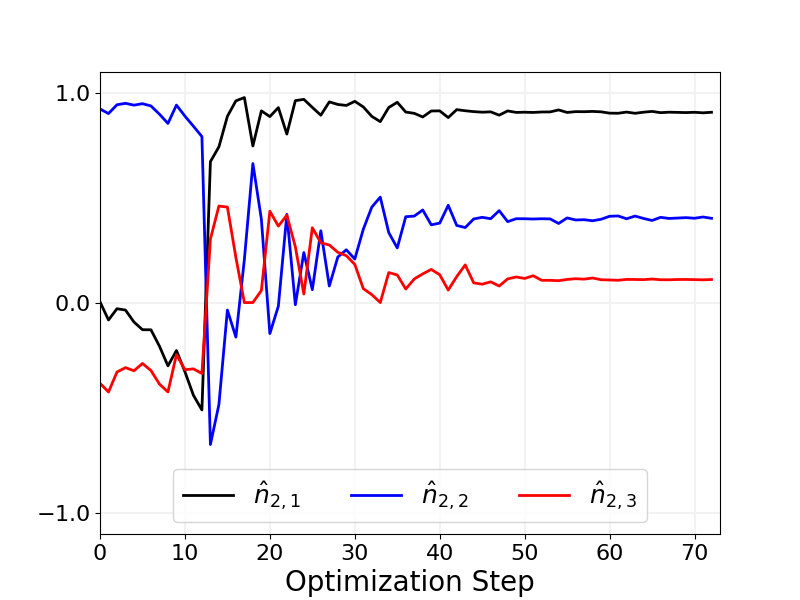}
    \caption{}
    \label{fig:FEparamEvol}
    \end{subfigure}

    \caption{Inverse problem results: (a) Evolution of the maximum Von Mises stress for five different initializations of fiber orientation and (b) the inverted fiber orientation for the model with lowest loss.}
    \label{fig:FEevol}
\end{figure}

\begin{figure}%[h]
    \centering
    \includegraphics[width=0.8\textwidth]{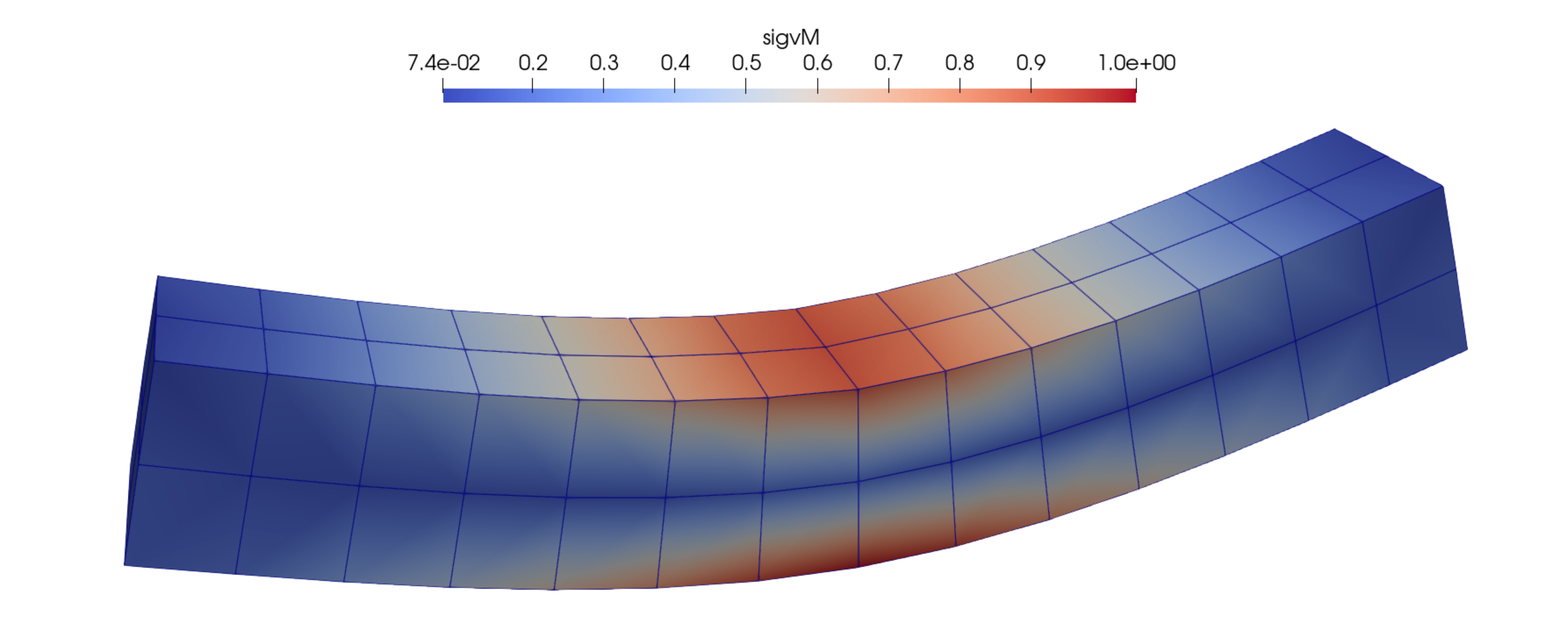}
    \caption{Stress field with the inverted fiber orientation.}
    \label{fig:FEStressField}
\end{figure}

We do this by employing a Nelder-Mead \cite{nelder1965simplex, gao2012implementing} optimization scheme that starts off with a random initial fiber orientation and then updates these parameters using the information from the loss which we now write as:
\begin{equation}
L = \| \sigma_{VM_\text{max}} - 0\|_2^2,
\end{equation}
where $\sigma_{VM_\text{max}}$ is the predicted maximum Von Mises stress. Figure \ref{fig:FEstressEvol} shows the evolution of the maximum Von Mises stress for five different initializations of the fiber orientation. We see that all five runs converged approximately to the same Von Mises stress value. Figure \ref{fig:FEparamEvol} shows the inverted preferred direction or the fiber orientation for the run with the lowest loss. Figure \ref{fig:FEStressField} depicts the stress field with the inverted preferred direction.

We remark, that out of these five runs, three converged to the same fiber orientation whereas two converged to a different one. However, even those runs converged approximately the same maximum Von Mises stress, highlighting an apparent non-uniqueness which is a common issue with design optimization and symmetry in the design space (see e.g. \cite{mahnken1996parameter, mahnken2004identification}).

% \COMMENT{RJ we could find a few refs that state this is a common issue with design opt}

\section{Discussion}
In this work, we developed a computational framework for the inverse design of microstructures in anisotropic finite strain hyperelasticity. During the forward problem, the framework combines the prowess of partially input convex and tensor basis neural networks to learn the free energy density along with the anisotropy class and the preferred direction(s) of anisotropy. The first and second derivatives of the free energy can be used to derive the stresses and the tangent modulus respectively. Once we have a trained model, we can solve the inverse problem, i.e., given a set of deformation gradients and the corresponding stresses, the framework is accurately able to invert the material and geometric parameters that would give those stresses for the respective deformations. The framework only needs to be trained for a single set of preferred direction(s) and it can invert the preferred direction for the new set of stresses and deformation gradients. The framework was first tested against synthetic macroscopic data where it was able to provide an accurate representation of the stresses, and hence the free energy, and was also able to classify the anisotropic class and direction. The solution to the inverse problem gave correct material parameters even ones outside the training data set. The framework was then extended to a microstructural setting where two different RVEs were tested. The framework performed well on both the forward and the inverse problems. Out of the two RVEs, the one exhibiting transverse isotropy was assumed to represent the microstructure of a beam problem and implemented in a finite element setting. Since the framework had already learned the macrostructural response during training, the solver was able to extract the stresses and the modulus at each integration point from the model to solve the inverse problem of finding the fiber orientation that would result in lowest maximum stress across all elements. The framework's ability to solve the forward and inverse problems effectively, while fulfilling all the physical requirements, makes it a reliable tool for the design of microstructures with required properties. In future works, we plan to extend the framework to include inelastic material behavior, higher-order anisotropy, and multiphysics, see e.g. \cite{rosenkranz2024viscoelasticty,jadoon2024automated,kalina2024neural,FUHG2024105837}. It can also prove useful for topology optimization \cite{vijayakumaran2024consistentmachinelearningtopology} and the design of functionally graded materials.

\section*{Acknowledgments}
The authors acknowledge the Texas Advanced Computing Center (TACC) at The University of Texas at Austin for providing computational resources that have contributed to the research results reported within this paper. URL: http://www.tacc.utexas.edu

The presented RVE computations were performed on a PC-Cluster at the Center for Information Services and High Performance Computing (ZIH) at TU Dresden. The authors thus thank the ZIH for generous allocations of computer time. K. A. Kalina thanks the German Research Foundation (DFG) for
the support within the Research Training Group GRK 2868 D3--Project Number 493401063.

Sandia National Laboratories is a multimission laboratory managed and operated by National Technology and Engineering Solutions of Sandia, LLC., a wholly owned subsidiary of Honeywell International, Inc., for the U.S.
Department of Energy's National Nuclear Security Administration under contract DE-NA-0003525.
This paper describes objective technical results and analysis.
Any subjective views or opinions that might be expressed in the paper do not necessarily represent the views of the U.S.  Department of Energy or the United States Government.

\section*{Data availability}
The code will be made available after the acceptance of this manuscript.

\section*{CRediT authorship contribution statement}
\textbf{AAJ:} Conceptualization, Methodology,  Writing - Original Draft, Writing - Review \& Editing, Formal analysis, \textbf{KAK:} Methodology, Data Curation, Writing - Original Draft, Writing - Review \& Editing, \textbf{MKR:} Writing - Original Draft, Writing - Review \& Editing, \textbf{RJ:} Methodology, Writing - Original Draft, Writing - Review \& Editing, \textbf{JNF:} Conceptualization, Methodology, Supervision, Funding acquisition,  Writing - Original Draft, Writing - Review \& Editing.

\clearpage
\bibliography{hyperelasticity,machine_learning}

\clearpage
\appendix
\numberwithin{equation}{section}
%%%%%%%%%%%%%%%%%%%%%%%%%%%%%%%%%%%%%%%%%%%%%%%%%%%%%%%%%%%%%%%%%%%%%%%%%%%%%%

\textbf{\large Appendix}
\section{Stress normalization} \label{app:normalization}

The procedure for stress normalization employed in the framework is outlined in this section. It closely follows the approach presented in \cite{linden2023neural} with some modification. We write the free energy as:

\begin{equation}\label{}
\Psi = \Psi^{\mathcal{N\!N}} + \Psi^{\text{gr}} + \Psi^{\text{n}} + \Psi^{\text{sn}}
\end{equation}

where $\Psi^{sn}$ denotes the free energy associated with the stress normalization. Since $\psi^{g}$, by construction, results in normalized stress and $\Psi^{n}$ is only a constant, stress normalization boils down to:

\begin{equation}\label{stress_constr}
\Sb(\Cb=\Ib) = 2 \partial_{\bar{\Cb}} \Psi^{\mathcal{N\!N}}  + 2 \partial_{\bar{\Cb}} \Psi^{sn}  = \Sb^{\mathcal{N\!N}}(\Cb=\Ib) + \Sb^{sn} = \mathbf{0},
\end{equation}

which gives:

\begin{equation}\label{normalizedStress}
\Sb^{sn}  = -\Sb^{\mathcal{N\!N}}(\Cb=\Ib),
\end{equation}

Evaluating the derivative for $\Psi^{\mathcal{N\!N}}$ at $\Cb=\Ib$, not writing the potential superscript for clarity, we can write the resulting stresses as:

\begin{equation}
\begin{aligned}
   \Sb^{\mathcal{N\!N}}(\Cb=\Ib) &= 2 \left[\partial_{\bar{I}_1}  \Psi \, \Ib + 2\partial_{\bar{I}_2} \Psi \, \Ib + \frac{1}{2}\partial_{\bar{I}_3} \Psi \, \Ib - \partial_{\bar{I}_4} \Psi \, \Ib\right] \\
   &+ 2 \left[\partial_{\bar{I}_5} \Psi \, \alpha_1 \Nb_1 + \partial_{\bar{I}_6} \Psi \,  (\tr(\alpha_1\Nb_1)\Ib - \alpha_1\Nb_1)\right] \\
   &+2 \left[ \partial_{\bar{I}_7} \Psi \, \alpha_2 \Nb_2 + \partial_{\bar{I}_8} \Psi \, (\tr(\alpha_2\Nb_2)\Ib - \alpha_2\Nb_2)  \right]. \\
\end{aligned}
\end{equation}

Rearranging,

\begin{equation} \label{S_NN_atCequalI}
\begin{aligned}
   \Sb^{\mathcal{N\!N}}(\Cb=\Ib) &= 2 \left[\partial_{\bar{I}_1}  \Psi  + 2\partial_{\bar{I}_2} \Psi + \frac{1}{2}\partial_{\bar{I}_3} \Psi - \partial_{\bar{I}_4} \Psi + \partial_{\bar{I}_6} \Psi \, \tr(\alpha_1\Nb_1) + \partial_{\bar{I}_8} \Psi \, \tr(\alpha_2\Nb_2) \right] \Ib\\
   &+ 2 \left[ \partial_{\bar{I}_5} \Psi  \,  -  \partial_{\bar{I}_6} \Psi  \, \right] \alpha_1 \Nb_1\\
   &+2 \left[  \partial_{\bar{I}_7} \Psi  \,   -  \partial_{\bar{I}_8} \Psi  \, \right] \alpha_2 \Nb_2\\
\end{aligned}
\end{equation}

From equation \eqref{normalizedStress}, we get:

\begin{equation}
\begin{aligned}
   \Sb^{sn} &= -2 \left[\partial_{\bar{I}_1}  \Psi  + 2\partial_{\bar{I}_2} \Psi + \frac{1}{2}\partial_{\bar{I}_3} \Psi - \partial_{\bar{I}_4} \Psi + \partial_{\bar{I}_6} \Psi \, \tr(\alpha_1\Nb_1) + \partial_{\bar{I}_8} \Psi \, \tr(\alpha_2\Nb_2) \right] \Ib\\
   &+ 2 \left[ \partial_{\bar{I}_6} \Psi -  \partial_{\bar{I}_5} \Psi \right] \alpha_1\Nb_1\\
   &+2 \left[  \partial_{\bar{I}_8} \Psi  -  \partial_{\bar{I}_7} \Psi \right] \alpha_2\Nb_2\\
\end{aligned}
\end{equation}

Now we need to find a form for the potential $\Psi^{sn}$ which results in these stresses while ensuring polyconvexity. Letting

\begin{subequations}
\begin{align}
\mathfrak{p} &= \partial_{\bar{I}_6} \Psi\\
\mathfrak{q} &= \partial_{\bar{I}_5} \Psi\\
\mathfrak{r} &= \partial_{\bar{I}_8} \Psi\\
\mathfrak{s} &= \partial_{\bar{I}_7} \Psi
\end{align}
\end{subequations}

and

\begin{equation}
 \bar{\mathfrak{n}} = 2 \left[\partial_{\bar{I}_1}  \Psi + 2\partial_{\bar{I}_2} \Psi + \frac{1}{2}\partial_{\bar{I}_3} \Psi - \partial_{\bar{I}_4} \Psi + \mathfrak{p}\tr(\alpha_1\Nb_1) + \mathfrak{r}\tr(\alpha_2\Nb_2) \right] \\    
\end{equation}

we can write:

% \begin{equation}
% \begin{aligned}
%    \Sb^{\mathcal{N\!N}}(\Cb=\Ib) &= 2\bar{\mathfrak{n}}\Ib + 2(\mathfrak{q}-\mathfrak{p})\alpha_1\Nb_1 + 2(\mathfrak{s}-\mathfrak{r})\alpha_2\Nb_2
% \end{aligned}
% \end{equation}

% and

\begin{equation}
\begin{aligned}
   \bar{\Sb}^{sn} &= -2\bar{\mathfrak{n}}\Ib + 2(\mathfrak{p}-\mathfrak{q})\alpha_1\Nb_1 + 2(\mathfrak{r}-\mathfrak{s})\alpha_2\Nb_2
\end{aligned}
\end{equation}

We first choose a potential of the form 

\begin{equation}\label{eq::PsiStressnorm}
\hat{\Psi}^{sn} = -\bar{\mathfrak{n}}(\Ic_3 - \Bar{\Ic_3}) + \mathfrak{p}(\Ic_5-\Bar{\Ic_5})+ \mathfrak{q}(\Ic_6-\Bar{\Ic_6})+ \mathfrak{r}(\Ic_7-\Bar{\Ic_7})+ \mathfrak{s}(\Ic_8-\Bar{\Ic_8}),
\end{equation}

where the bar represents the invariants evaluated at $\Cb=\Ib$ and are constant. These ensure that free energy is zero at the undeformed configuration but don't appear in the derivatives since they are constant. Then the stresses resulting from this potential read:

\begin{equation}
\begin{aligned}
    \hat{\Sb}^{sn} &= -\bar{\mathfrak{n}}J\Cb^{-1} + 2\mathfrak{p}\alpha_1\Nb_1+ 2\mathfrak{q}[\alpha_1\tr ((\text{Cof}\Cb) \Nb_1)\Cb^{-1} -  \alpha_1(\text{Cof}\Cb) \Nb_1\Cb^{-1} ] \\
    &+ 2\mathfrak{r}\alpha_2\Nb_2+ 2\mathfrak{s}[\alpha_2\tr ((\text{Cof}\Cb) \Nb_2)\Cb^{-1} -  \alpha_2(\text{Cof}\Cb) \Nb_2\Cb^{-1} ],
\end{aligned}
\end{equation}

At $\Cb=\Ib$,

\begin{equation}\label{}
\hat{\Sb}^{sn} = -\bar{\mathfrak{n}}\Ib + 2\mathfrak{p}\alpha_1\Nb_1+ 2\mathfrak{q}[\alpha_1\tr ( \Nb_1)\Ib -  \alpha_1\Nb_1]+ 2\mathfrak{r}\alpha_2\Nb_2+ 2\mathfrak{s}[\alpha_2\tr (\Nb_2)\Ib -  \alpha_2\Nb_2],
\end{equation}

Writing equation \eqref{S_NN_atCequalI} in terms of newly defined variables, we have

\begin{equation}\label{S_NN_atCequalI_pqrs}
\begin{aligned}
   \Sb^{\mathcal{N\!N}}(\Cb=\Ib) &= \bar{\mathfrak{n}}\Ib - 2\mathfrak{p}\alpha_1\Nb_1 + 2\mathfrak{q}\alpha_1\Nb_1 - 2\mathfrak{r}\alpha_2\Nb_2 + 2\mathfrak{s}\alpha_2\Nb_2
\end{aligned}
\end{equation}

From equation \eqref{stress_constr},

\begin{equation}
\Sb^{\mathcal{N\!N}}(\Cb=\Ib) + \hat{\Sb}^{sn} = 2\mathfrak{q}\alpha_1\tr ( \Nb_1)\Ib + 2\mathfrak{s}\alpha_2\tr ( \Nb_2)\Ib \neq \mathbf{0},
\end{equation}

Therefore, we introduce a new variable

\begin{equation}
 {\mathfrak{o}} = 2 \left[\partial_{\bar{I}_1}  \Psi + 2\partial_{\bar{I}_2} \Psi + \frac{1}{2}\partial_{\bar{I}_3} \Psi - \partial_{\bar{I}_4} \Psi + (\mathfrak{p}+\mathfrak{q})\tr(\alpha_1\Nb_1) + (\mathfrak{r}+\mathfrak{s})\tr(\alpha_2\Nb_2) \right] \\    
\end{equation}

and modify the free energy potential as:

\begin{equation}
{\Psi}^{sn} = -{\mathfrak{o}}(\Ic_3 - \Bar{\Ic_3}) + \mathfrak{p}(\Ic_5-\Bar{\Ic_5})+ \mathfrak{q}(\Ic_6-\Bar{\Ic_6})+ \mathfrak{r}(\Ic_7-\Bar{\Ic_7})+ \mathfrak{s}(\Ic_8-\Bar{\Ic_8}),
\end{equation}

This is the final form of our stress normalization part of the potential. It is evident that such a potential would fulfill the requirement from equation \eqref{stress_constr} while still being polyconvex. Since it is a linear equation in $\Ic_3$, i.e., $J$, terms with $\mathfrak{o}$ do not appear in the hessian and therefore can have the negative sign with them. Whereas all the other terms involve polyconvex invariants multiplied with some positive constants since we have a monotonically increasing neural network and therefore preserve polyconvexity of the free energy.

% and

% \begin{subequations}
% \begin{align}
%  \mathfrak{o} &= 2 \left[\partial_{I_1}  \Psi + 2\partial_{I_2} \Psi + \frac{1}{2}\partial_{I_3} \Psi - \partial_{I_4} \Psi + (\partial_{I_6}\Psi  + \mathfrak{q})\alpha_1 + (\partial_{I_8} \Psi + \mathfrak{s})\alpha_2 \right] \\
% \mathfrak{p} &= \max(0,\partial_{I_6} \Psi -  \partial_{I_5} \Psi)\\
% \mathfrak{q} &= \max(0,\partial_{I_5} \Psi -  \partial_{I_6} \Psi)\\
% \mathfrak{r} &= \max(0,\partial_{I_8} \Psi -  \partial_{I_7} \Psi)\\
% \mathfrak{s} &= \max(0,\partial_{I_7} \Psi -  \partial_{I_8} \Psi)\\
% \end{align}
% \end{subequations}

% Such a formulation leads to polyconvex, normalized free energy which gives normalized stresses with

% \begin{equation}\label{eq::PsiStressnorm}
% \Sb^{sn} = -\mathfrak{o}\Ic_3\Cb^{-1} + 2\mathfrak{p}\alpha_1\Nb_1+ 2\mathfrak{q}(\Ic_6\Cb^{-1} - \alpha_1(\text{Cof}\Cb) \Nb_1\Cb^{-1})+ \mathfrak{r}\alpha_2\Nb_2+ \mathfrak{s}(\Ic_8\Cb^{-1} - \alpha_2(\text{Cof}\Cb) \Nb_2\Cb^{-1}).
% \end{equation}

\section{Recovering parameters seen during training} \label{app:recoverCase1}
\begin{figure}[hbt!]
    \begin{subfigure}[b]{0.5\linewidth}
        \includegraphics[scale=0.3]{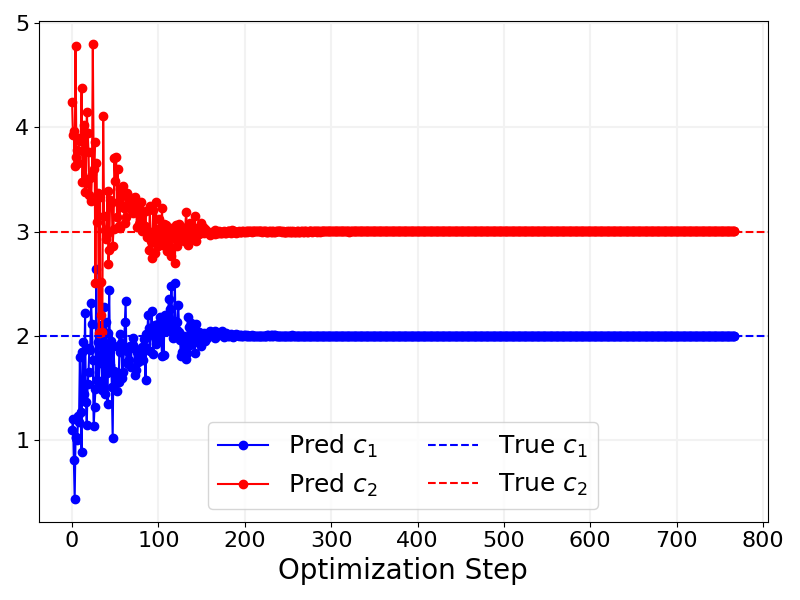}
    \caption{Isotropic}\label{fig:case1_iso_seen}
    \end{subfigure}
        \begin{subfigure}[b]{0.5\linewidth}
        \includegraphics[scale=0.3]{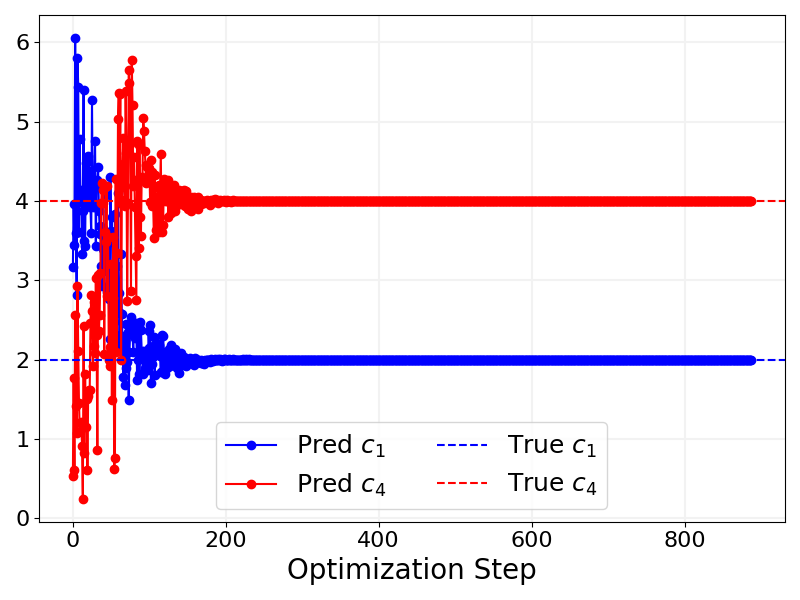}
    \caption{Transversely isotropic}\label{fig:case1_tran_seen}
    \end{subfigure}
    \begin{center}
                    \begin{subfigure}[b]{0.5\linewidth}
        \includegraphics[scale=0.3]{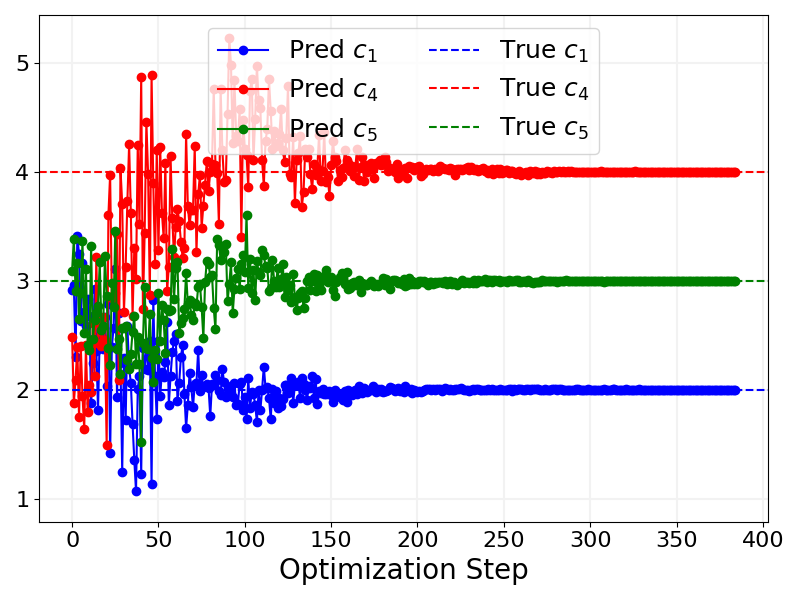}
    \caption{Orthotropic}\label{fig:case1_ortho_seen}
    \end{subfigure}
    \end{center}
    \caption{Solution of the inverse problem for material parameters already seen during training for (a) isotropic, (b) transversely isotropic and (c) orthotropic classes for known anisotropy class and preferred direction.}
    \label{paramRecovery_case1}
    
\end{figure}

\begin{figure}[hbt!]
    \begin{subfigure}[b]{0.5\linewidth}
        \includegraphics[scale=0.3]{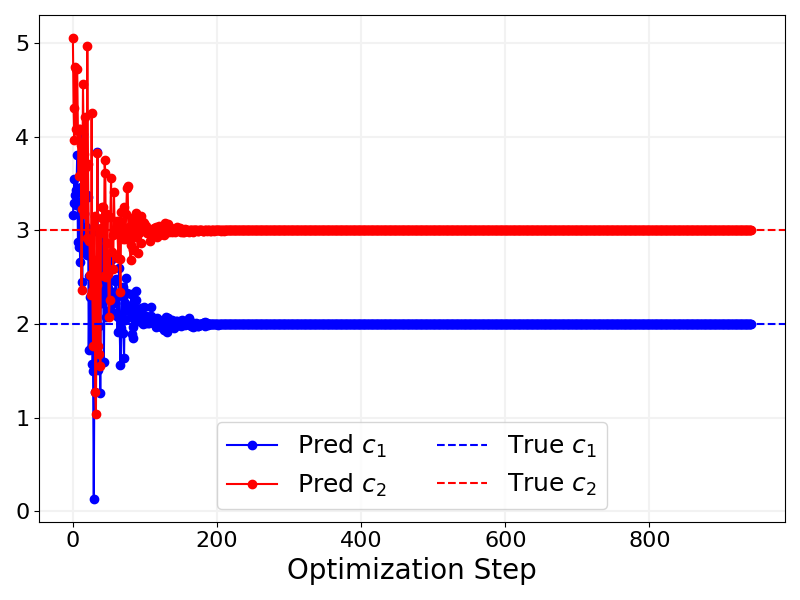}
    \caption{Isotropic}\label{fig:case2_iso_seen}
    \end{subfigure}
        \begin{subfigure}[b]{0.5\linewidth}
        \includegraphics[scale=0.3]{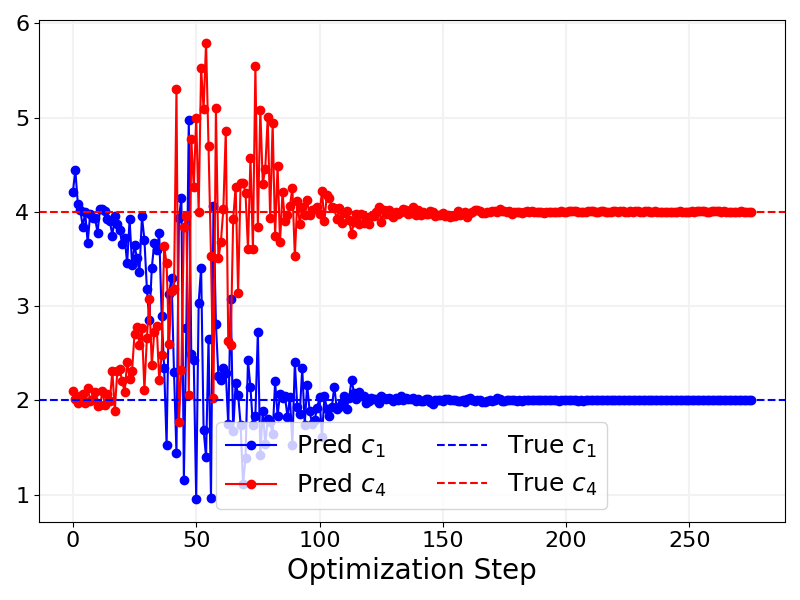}
    \caption{Transversely isotropic}\label{fig:case2_tran_seen}
    \end{subfigure}
    \begin{center}
                    \begin{subfigure}[b]{0.5\linewidth}
        \includegraphics[scale=0.3]{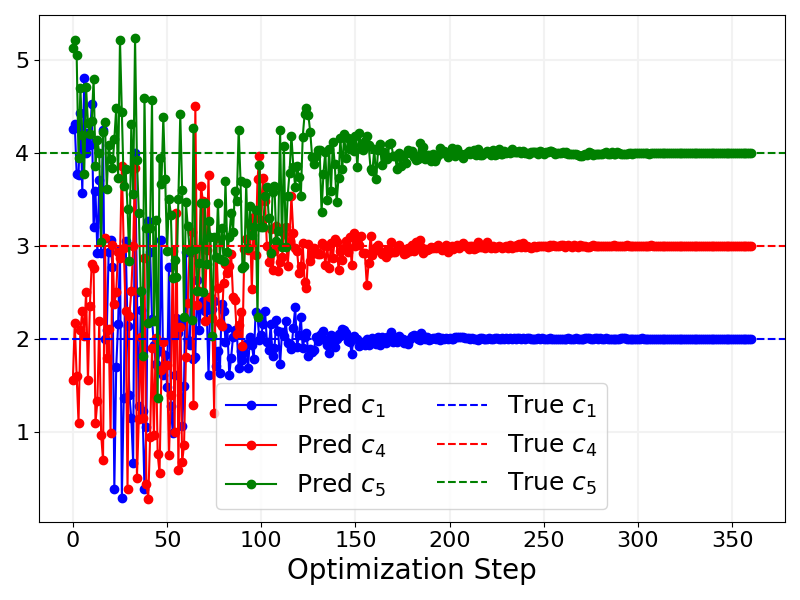}
    \caption{Orthotropic}\label{fig:case2_ortho_seen}
    \end{subfigure}
    \end{center}
    \caption{Solution of the inverse problem for material parameters already seen during training for (a) isotropic, (b) transversely isotropic and (c) orthotropic classes while learning anisotropy class and preferred direction.}
    \label{paramRecovery_case2}
    
\end{figure}

In order to check the efficacy of our inverse design framework, we run the inverse problem for material parameters seen during training as a part of the training dataset noting that the framework should, at least, be able to recover accurately the material parameters it was trained on. Figure \ref{paramRecovery_case1} shows the recovered material parameters for the case of known anisotropy class and preferred direction whereas Figure \ref{paramRecovery_case2} shows the results for the inverse problem for the case where our model learns the class and orientation of anisotropy during the training process. As evident from the figures, we are able to recover exactly the material parameters seen during training.


\section{Parameter Study} \label{app:ParamStudy}
\begin{figure}[hbt!]
    \begin{subfigure}[b]{0.5\linewidth}
        \includegraphics[scale=0.35]{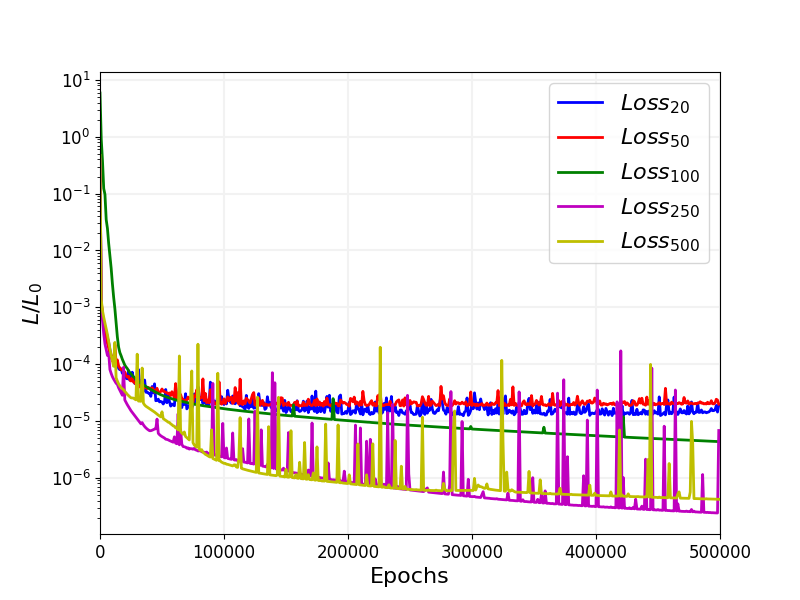}
    \caption{Isotropic}\label{fig:ParamStudy_iso}
    \end{subfigure}
        \begin{subfigure}[b]{0.5\linewidth}
        \includegraphics[scale=0.35]{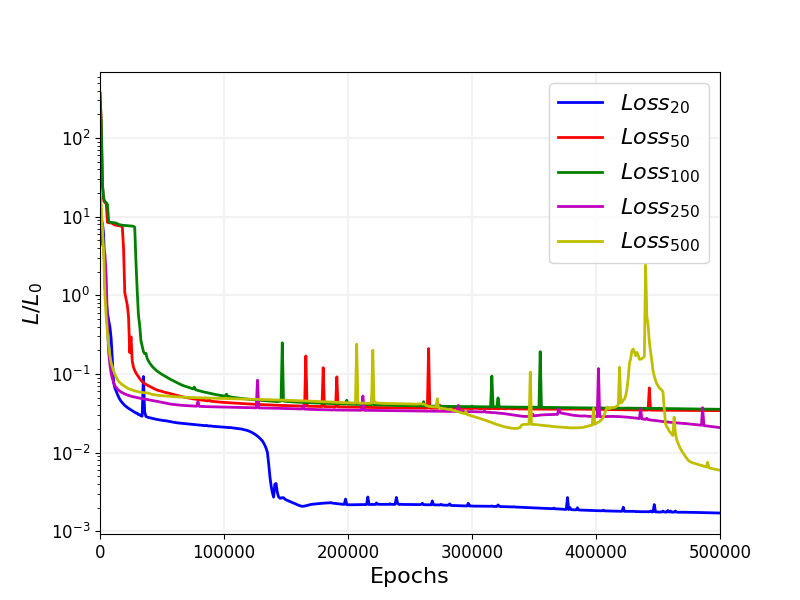}
    \caption{Transversely isotropic}\label{fig:ParamStudy_tran}
    \end{subfigure}
    \begin{center}
                    \begin{subfigure}[b]{0.5\linewidth}
        \includegraphics[scale=0.35]{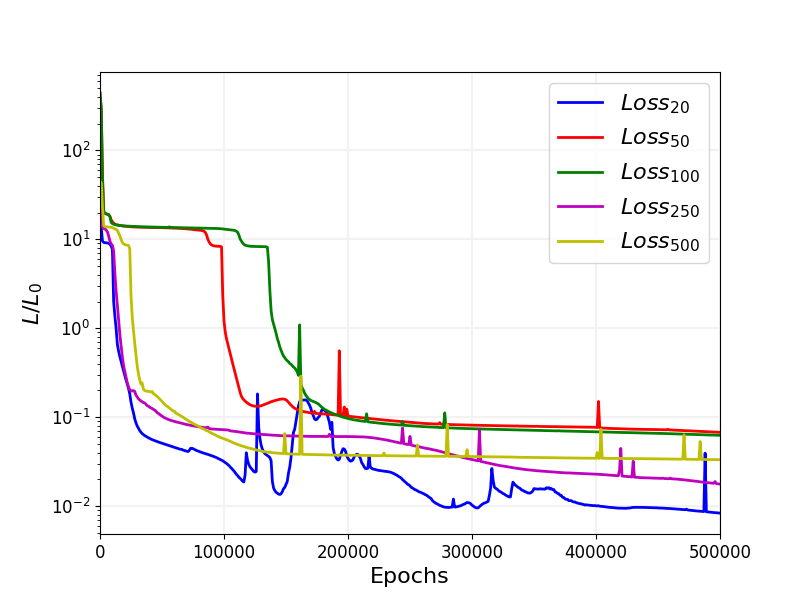}
    \caption{Orthotropic}\label{fig:ParamStudy_ortho}
    \end{subfigure}
    \end{center}
    \caption{Training loss for (a) isotropic, (b) transversely isotropic and (c) orthotropic classes with different sample sizes. The subscript in the legend denotes the number of samples chosen for each material parameter set.}
    \label{paramStudy_loss}
\end{figure}

\begin{figure}[hbt!]
    \begin{subfigure}[b]{0.5\linewidth}
        \includegraphics[scale=0.35]{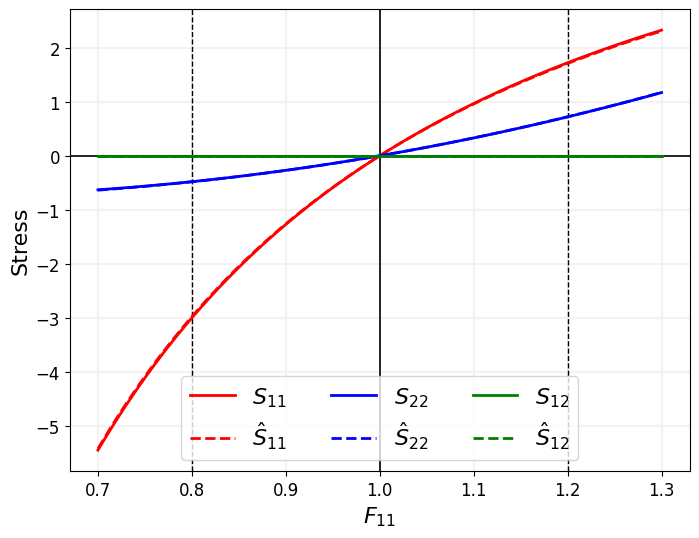}
    \caption{Isotropic}\label{fig:ParamStudy_iso_stress}
    \end{subfigure}
        \begin{subfigure}[b]{0.5\linewidth}
        \includegraphics[scale=0.35]{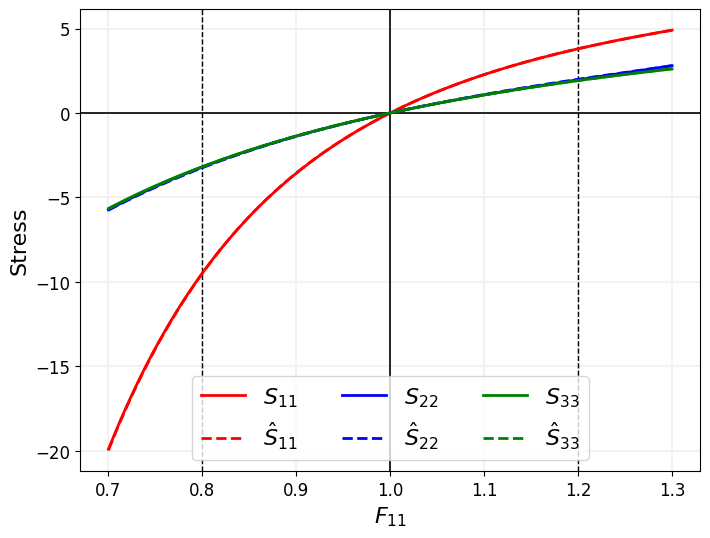}
    \caption{Transversely isotropic}\label{fig:ParamStudy_tran_stress}
    \end{subfigure}
    \begin{center}
                    \begin{subfigure}[b]{0.5\linewidth}
        \includegraphics[scale=0.35]{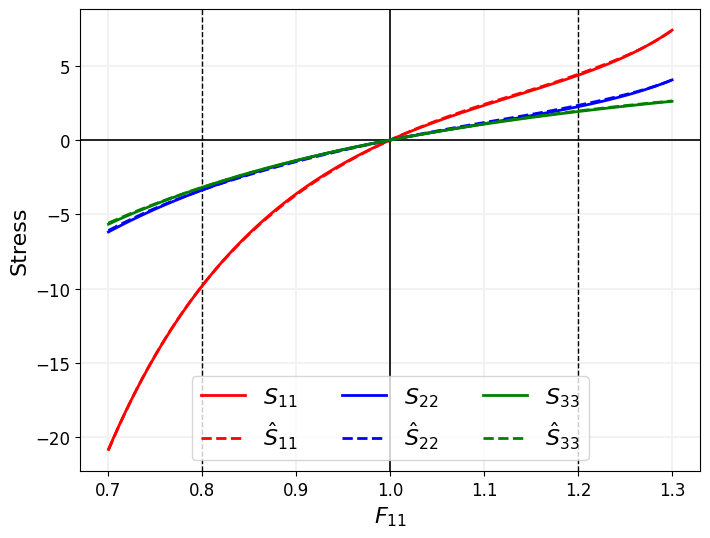}
    \caption{Orthotropic}\label{fig:ParamStudy_ortho_stress}
    \end{subfigure}
    \end{center}
    \caption{Stress fit for (a) isotropic, (b) transversely isotropic and (c) orthotropic classes for a sample size of 20 for each material parameter set.}
    \label{paramStudy_response}
\end{figure}

Efficient sampling strategies can help reduce the size of training datasets while ensuring a complete exploration of the sampling space. In our framework, we sample from the deformation gradient space using Latin Hypercube Sampling \citep{stein1987large}. Since it is a nice-dimensional space, although bounded, we would theoretically require a large number of samples. However, since our framework requires stress-strain data for different material parameters, we can reduce the required number of deformation gradients for one set of material parameters. The idea is to sample the deformation gradients independently for each material parameter set, thereby reducing the number of samples altogether while ensuring the samples are space-filling. We carried out a comparative study for different sample sizes and the training losses for the model are presented in Figure \ref{paramStudy_loss} with the subscripts in the legend denoting the number of independently sampled deformation gradients for each set of material parameters. It can be seen that the evolution of the loss function follows the same trajectory for the different sample sizes. Figure \ref{paramStudy_response} shows the stress response for a known material parameter set under uniaxial loading for the sample size of 20. The stress response from the neural network derivatives matches with the true stress response. Since the deformation gradients associated with uniaxial loading were not explicitly included in the training dataset, a good stress fit in this case confirms a good exploration of the deformation gradient space. This can be accredited to the fact that even with 20 deformation gradient samples if we have two material parameters with five samples each, we effectively have 500 deformation gradient samples.

\section{Isotropy of the RVE with a single spherical inclusion} \label{app:RVEIsotropy}
\begin{figure}
    \begin{subfigure}{0.5\linewidth}
        \includegraphics[scale=0.35]{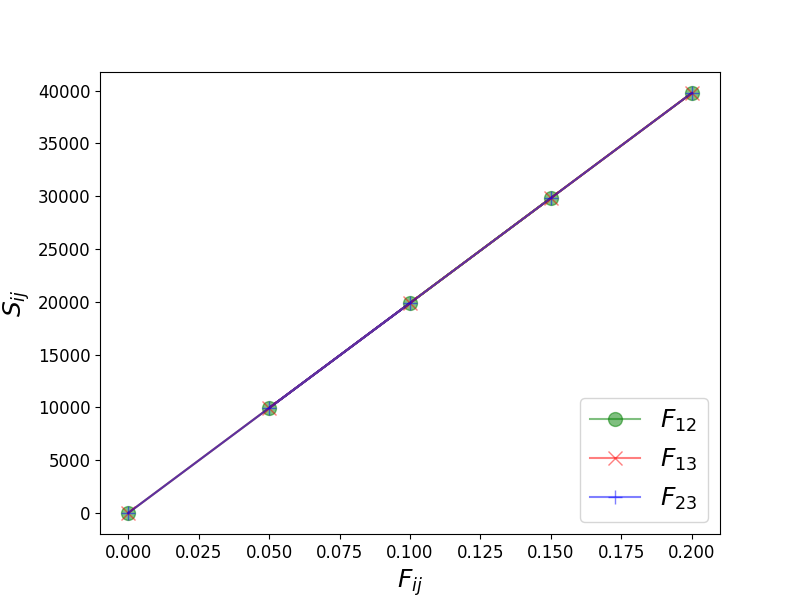}
    \caption{}\label{fig:SingleIncRVE_isotropy_path1}
    \end{subfigure}
                \begin{subfigure}{0.5\linewidth}
        \includegraphics[scale=0.35]{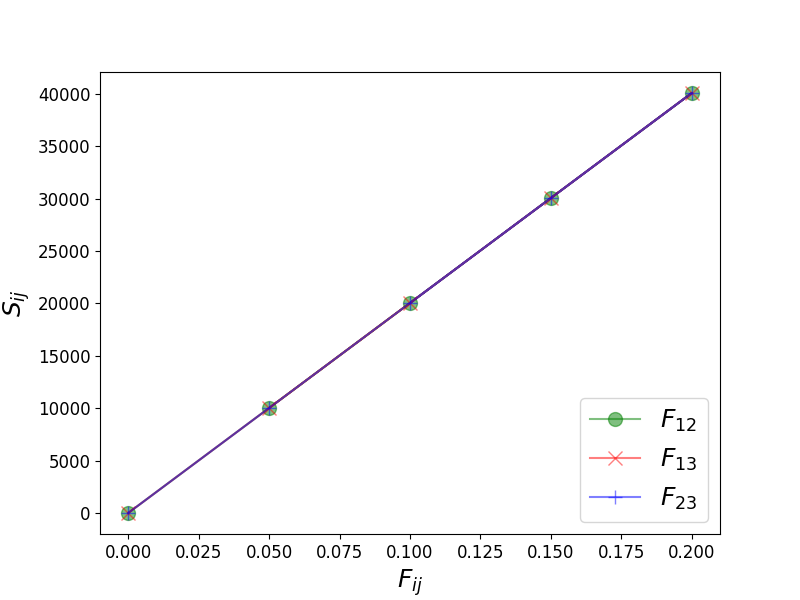}
    \caption{}\label{fig:SingleIncRVE_isotropy_path2}
    \end{subfigure}
        \begin{subfigure}{0.5\linewidth}
        \includegraphics[scale=0.35]{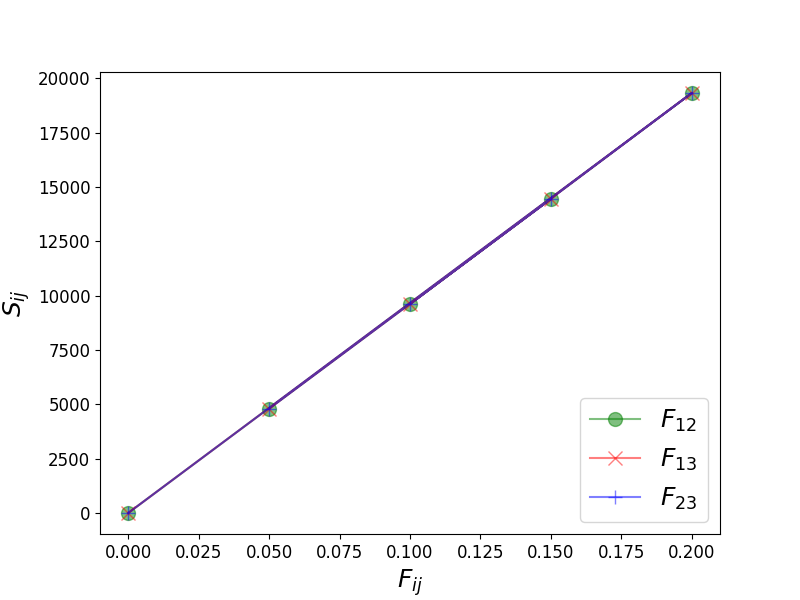}
    \caption{}\label{fig:SingleIncRVE_isotropy_path3}
    \end{subfigure}
            \begin{subfigure}{0.5\linewidth}
        \includegraphics[scale=0.35]{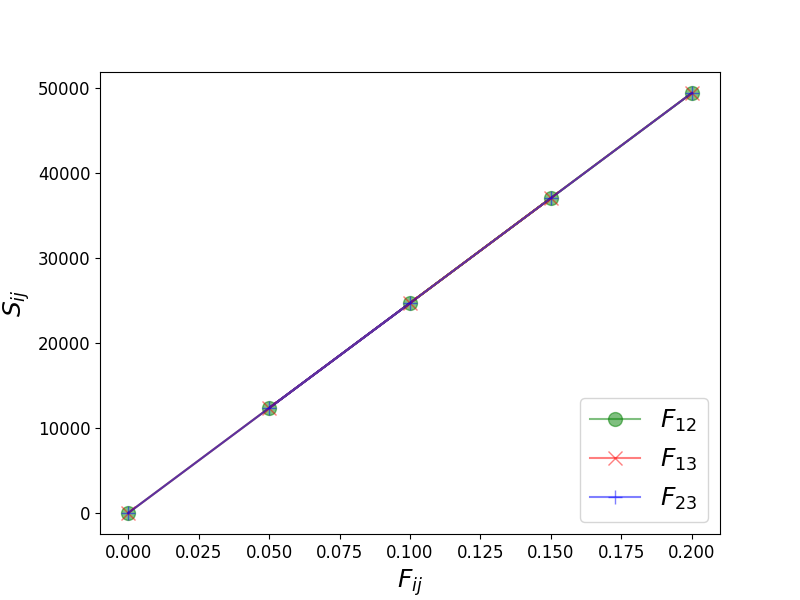}
    \caption{}\label{fig:SingleIncRVE_isotropy_path4}
    \end{subfigure}
    
    \caption{Stress response under shear loading for an RVE with a single spherical inclusion with material parameters (a) $R=0.1,  \frac{\mu_1}{\mu_2} = 0.25$, (b) $R=0.1,  \frac{\mu_1}{\mu_2} = 1.5$, (c) $R=0.5,  \frac{\mu_1}{\mu_2} = 0.25$ and (d) $R=0.5,  \frac{\mu_1}{\mu_2} = 1.5$.}
    \label{fig:SingleIncRVE_isotropy}
\end{figure}

Figure \ref{fig:SingleIncRVE_isotropy} shows the results of the numerical experiment carried out on the RVE with a single spherical inclusion to confirm an isotropic response. The stress response for shear deformations in all directions overlap with each other, implying isotropy. 

\section{Violation of polyconvexity through network constraints}\label{app:ArbNet}

\begin{figure}
    \begin{subfigure}{0.5\linewidth}
        \includegraphics[scale=0.35]{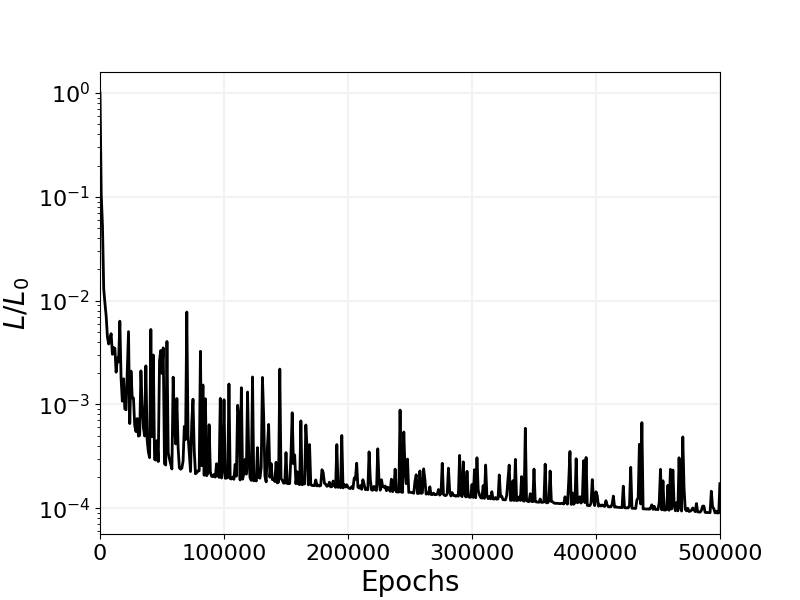}
    \caption{Loss}\label{fig:FiberRVE_loss_ArbNet}
    \end{subfigure}
    \begin{subfigure}{0.5\linewidth}
        \includegraphics[scale=0.35]{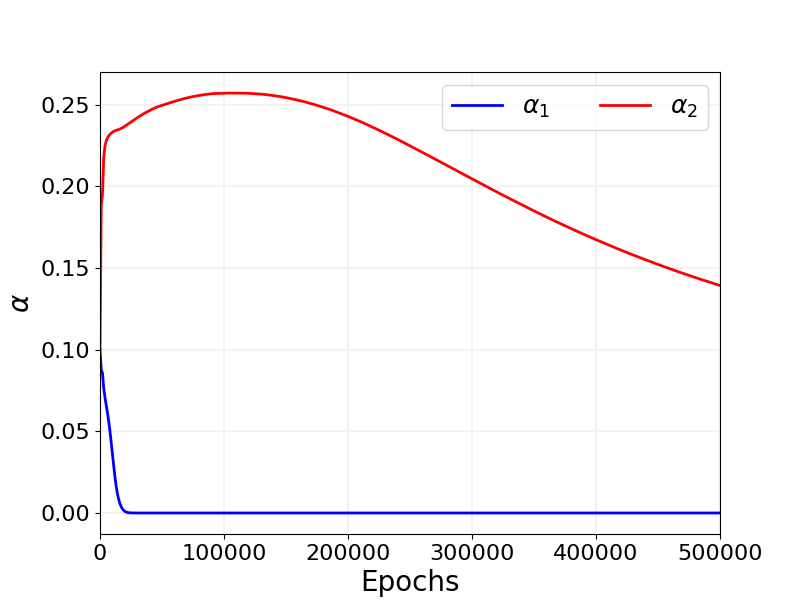}
    \caption{Anisotropic coefficients}\label{fig:FiberRVE_alpha_ArbNet}
    \end{subfigure}
        \begin{subfigure}{0.5\linewidth}
        \includegraphics[scale=0.35]{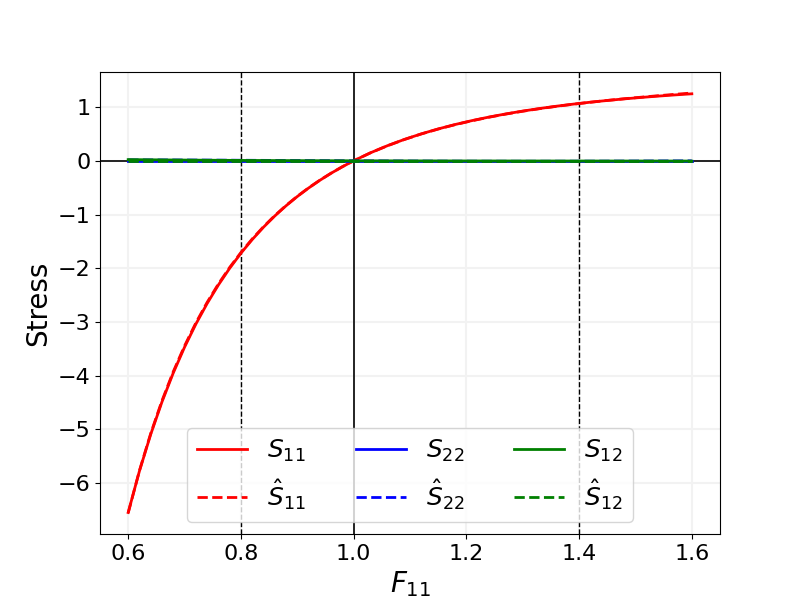}
    \caption{Stress comparison}\label{fig:FiberRVE_stress_ArbNet}
    \end{subfigure}
            \begin{subfigure}{0.5\linewidth}
        \includegraphics[scale=0.35]{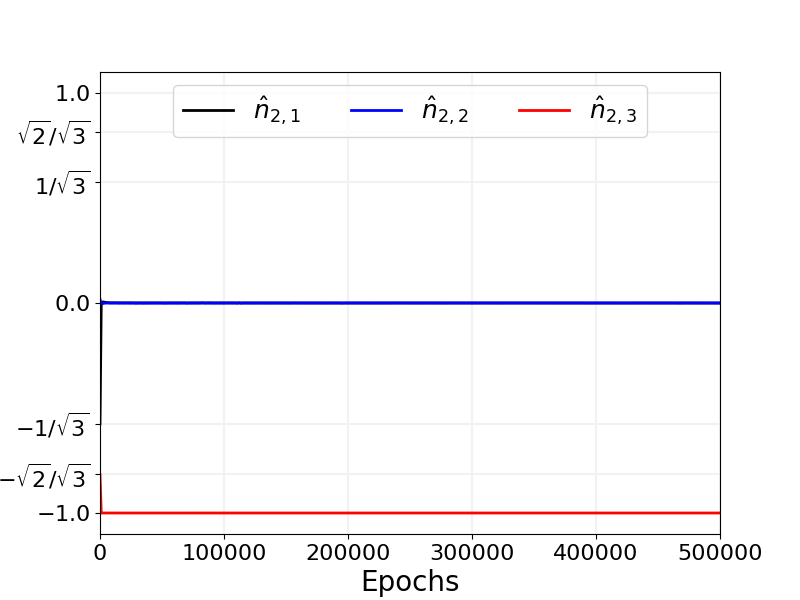}
    \caption{Preferred direction}\label{fig:FiberRVE_prefDirTrain_ArbNet}
    \end{subfigure}
            \begin{subfigure}{0.5\linewidth}
        \includegraphics[scale=0.35]{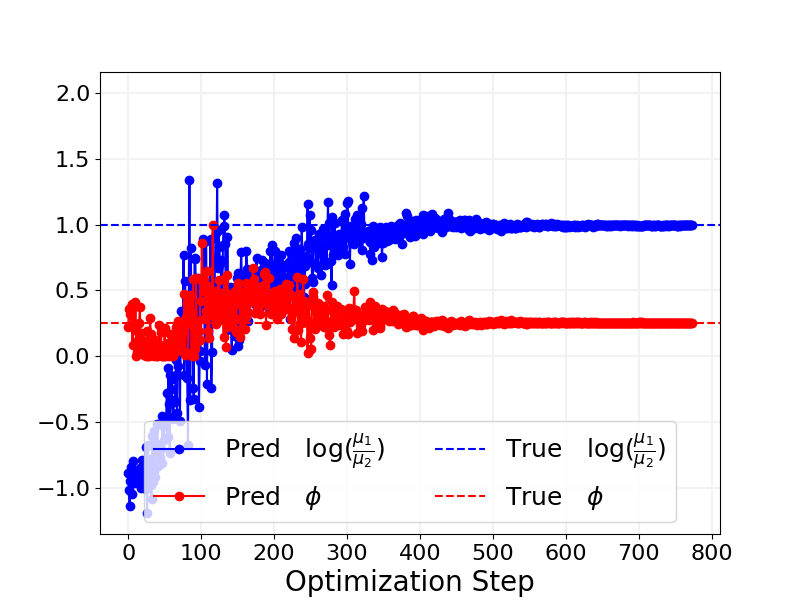}
    \caption{Inverted material parameters}\label{fig:FiberRVE_inverse_ArbNet}
    \end{subfigure}
                \begin{subfigure}{0.5\linewidth}
        \includegraphics[scale=0.35]{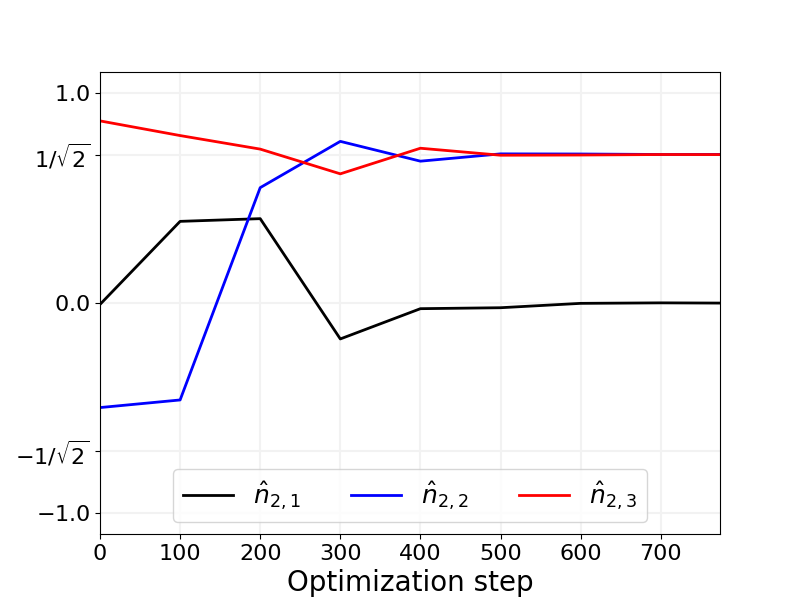}
    \caption{Inverted preferred direction}\label{fig:FiberRVE_prefDir_inverse_ArbNet}
    \end{subfigure}

    \caption{Results for the RVE with unidirectional fibers with unknown class and preferred direction from the material model of  Section \ref{sec:microData} (a) Training loss over epochs, (b) Evolution of the anisotropic coefficients, (c) true (solid lines) and predicted stresses (dashed lines) over $F_{11}$ on material parameters seen during training, (d) Preferred directions learned during training (e) true (dashed lines) and predicted parameters (solid lines with markers) over optimization iterations and (f) the inverted preferred direction.}
    \label{rve_fiber_ArbNet}
\end{figure}

In this section, we present the results for the forward and inverse problem on the unidirectional fiber RVE for an unconstrained neural network. Figure \ref{fig:FiberRVE_loss_ArbNet} shows the loss evolution during training which seems to fluctuate a bit more than the other formulation where we violate polyconvexity through the normalization term. It is still able to learn the anisotropy class and the preferred directions as shown in subfigures \ref{fig:FiberRVE_alpha_ArbNet} and \ref{fig:FiberRVE_prefDirTrain_ArbNet} with an almost perfect stress-fit under uniaxial loading as illustrated in Figure \ref{fig:FiberRVE_stress_ArbNet}. The model works well on the inverse problem as well, recovering the desired design parameters and inverting the correct preferred direction as can be seen in Figures \ref{fig:FiberRVE_inverse_ArbNet} and \ref{fig:FiberRVE_prefDir_inverse_ArbNet}.

\section{Nonpolyconvex formulations for macroscale data} \label{app:nonPolyisotropic}

\begin{figure}
    \begin{subfigure}{0.5\linewidth}
        \includegraphics[scale=0.30]{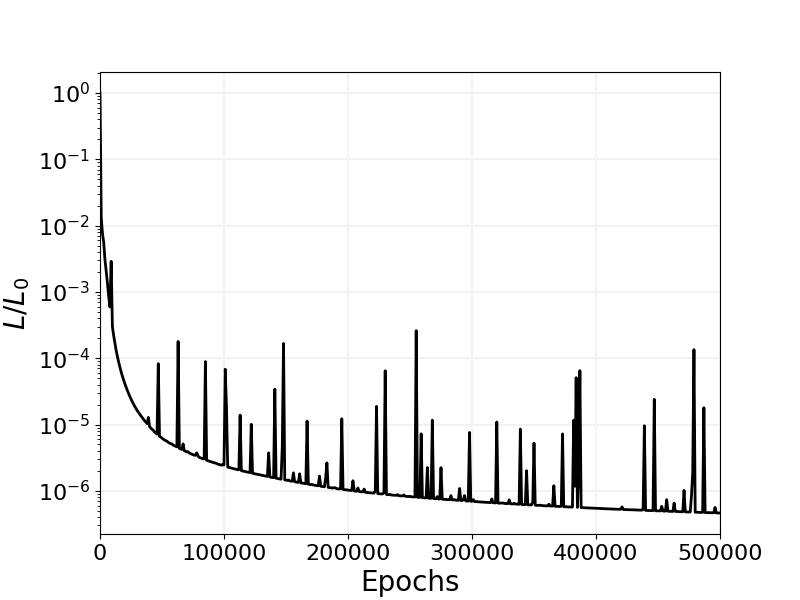}
    \caption{Loss for formulation 1}\label{fig:iso_Cnorm_loss}
    \end{subfigure}
                \begin{subfigure}{0.5\linewidth}
        \includegraphics[scale=0.30]{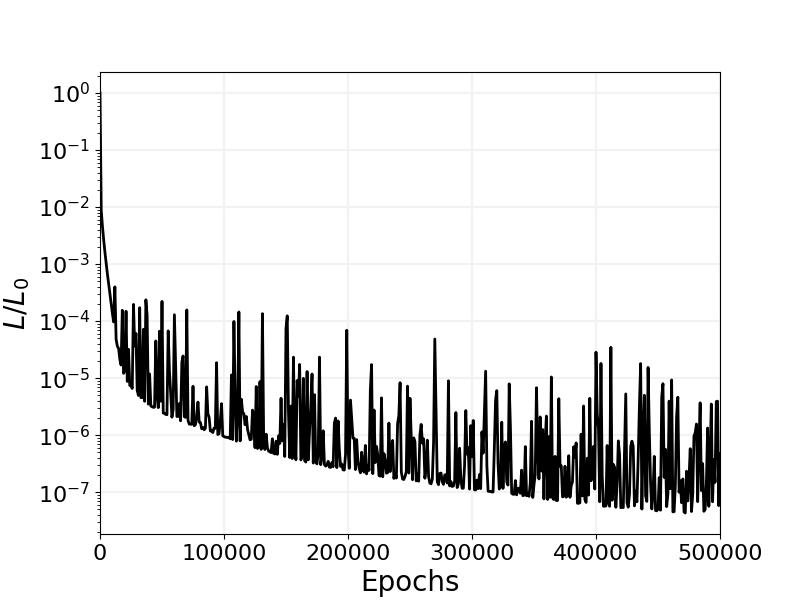}
    \caption{Loss for formulation 2}\label{fig:iso_arbnet_loss}
    \end{subfigure}
        \begin{subfigure}{0.5\linewidth}
        \includegraphics[scale=0.30]{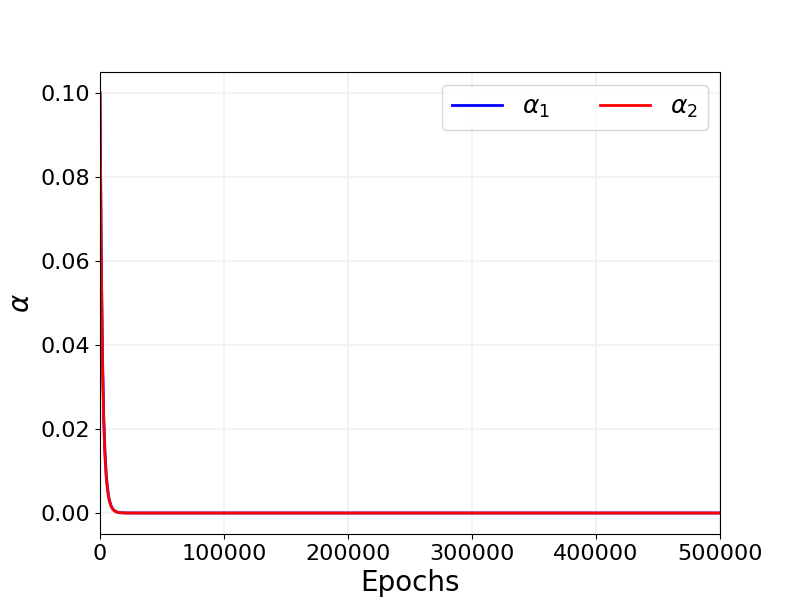}
    \caption{Anisotropic coefficients for formulation 1}\label{fig:iso_Cnorm_alpha}
    \end{subfigure}
                \begin{subfigure}{0.5\linewidth}
        \includegraphics[scale=0.30]{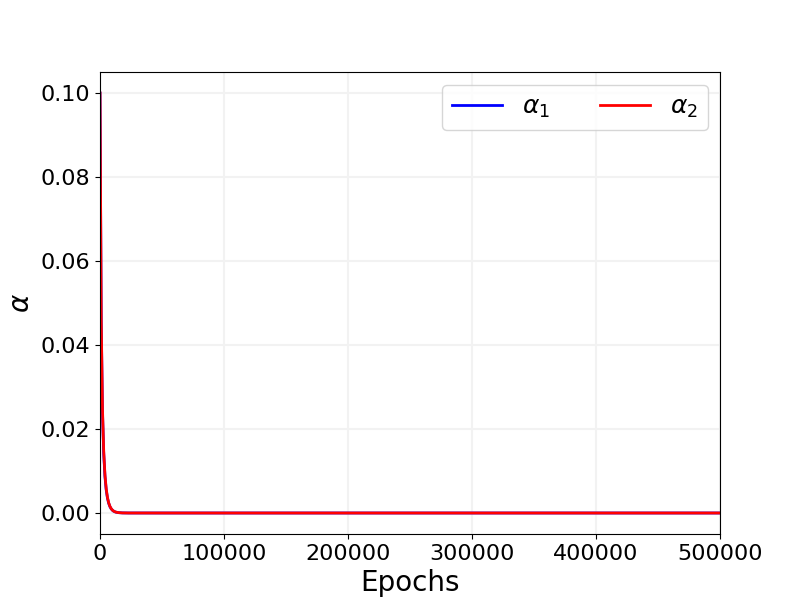}
    \caption{Anisotropic coefficients for formulation 2}\label{fig:iso_arbnet_alpha}
    \end{subfigure}
                    \begin{subfigure}{0.5\linewidth}
        \includegraphics[scale=0.30]{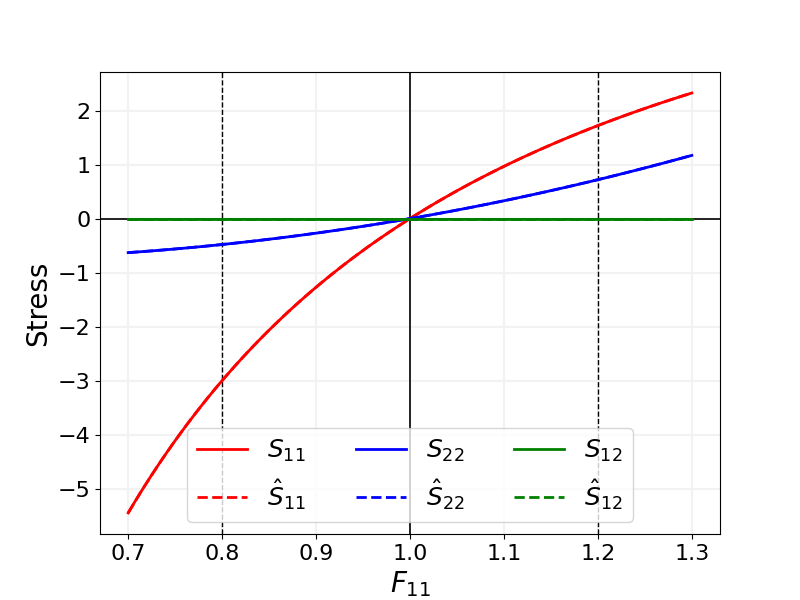}
    \caption{Stress comparison for formulation 1}\label{fig:iso_Cnorm_stress}
    \end{subfigure}
                    \begin{subfigure}{0.5\linewidth}
        \includegraphics[scale=0.30]{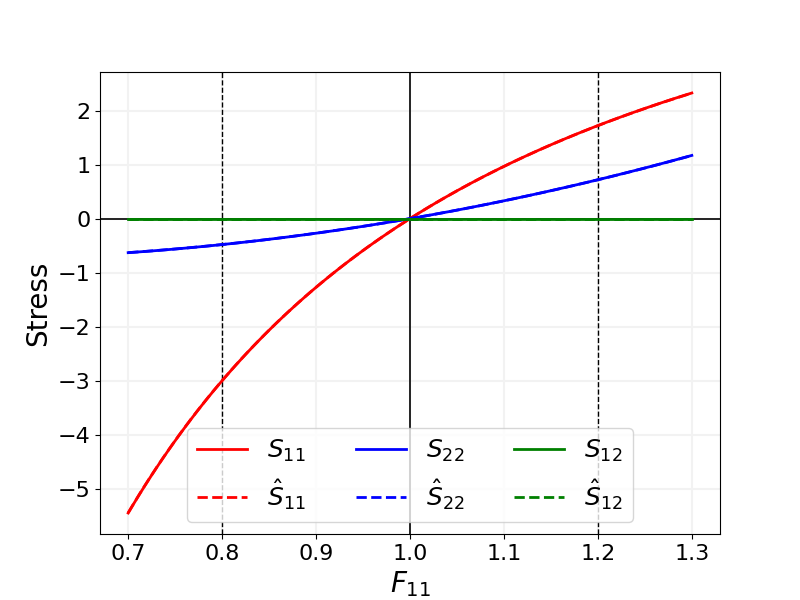}
    \caption{Stress comparison for formulation 2}\label{fig:iso_arbnet_stress}
    \end{subfigure}
                        \begin{subfigure}{0.5\linewidth}
        \includegraphics[scale=0.30]{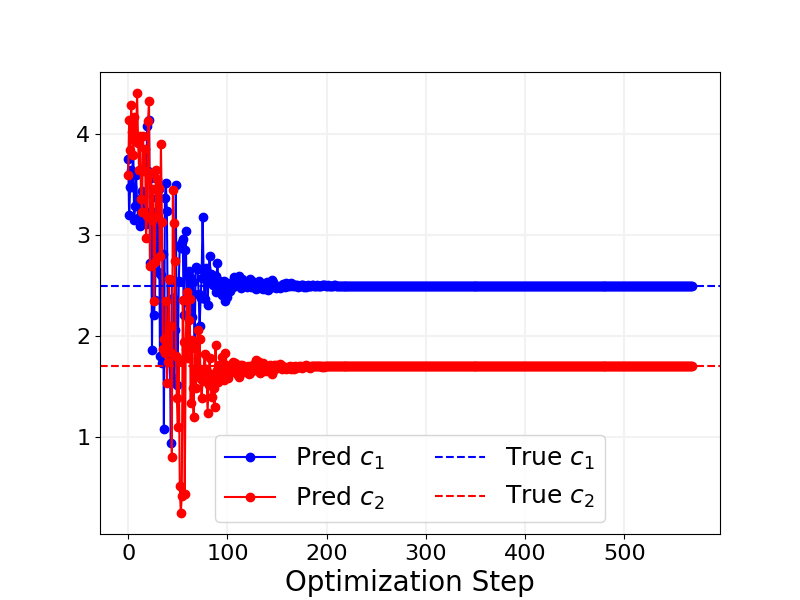}
    \caption{Inverted material parameters for formulation 1}\label{fig:iso_Cnorm_Cupdate}
    \end{subfigure}
                    \begin{subfigure}{0.5\linewidth}
        \includegraphics[scale=0.30]{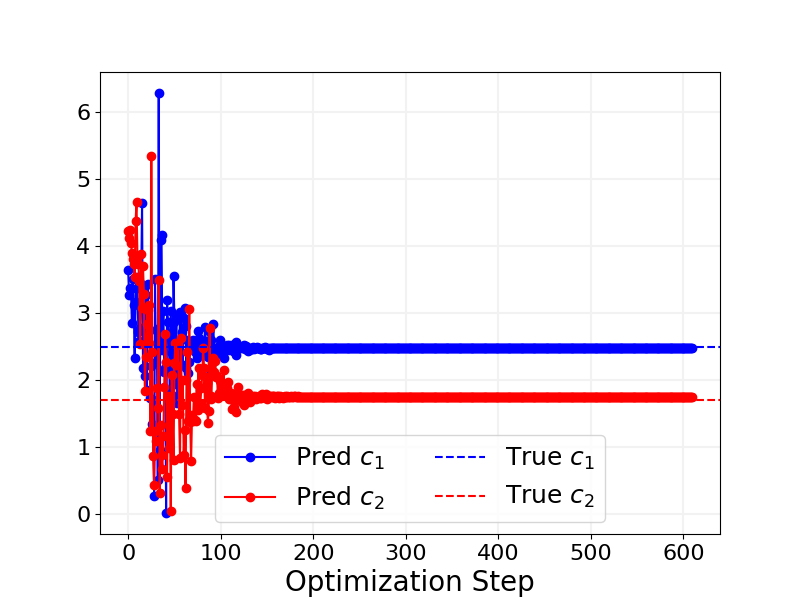}
    \caption{Inverted material parameters for formulation 2}\label{fig:iso_arbnet_Cupdate}
    \end{subfigure}

    \caption{A comparison of (a,b) loss, (c,d) anisotropic coefficient learning, (e,f) stress-fit and (g,f) inverse problem for nonpolyconvex free energy with the stress normalization term from Eq. \eqref{oC_norm} (formulation 1) and an unconstrained neural network (formulation 2).}
    \label{isotropic_nonPolyconvex}
\end{figure}

This section presents a brief comparison of two formulations outlined in Section \ref{sec:FiberRVEResults} when training a nonpolyconvex free energy function. Stress-strain tuples were generated using the isotropic material model and both formulations were tested and compared under forward and inverse problems. Here, we denote the formulation with the normalization component from Eq. \eqref{oC_norm} as formulation 1 while formulation 2 represents an unconstrained neural network. Figures \ref{fig:iso_Cnorm_loss} and \ref{fig:iso_arbnet_loss} compare the loss evolution while Figures \ref{fig:iso_Cnorm_alpha} and \ref{fig:iso_arbnet_alpha} show the anisotropic coefficient learning during the forward process for each framework. Figures \ref{fig:iso_Cnorm_stress} and \ref{fig:iso_arbnet_stress} compare the stress response for each formulation whereas Figures \ref{fig:iso_Cnorm_Cupdate} and \ref{fig:iso_arbnet_Cupdate} give the inverted design parameters for each. As noted before, both formulations work are comparable in terms of performance but the positive definiteness of the tangent modulus may necessitate the use of formulation 1.

\end{document}